{}
{}

\documentclass[11pt,a4paper]{article}

\newif\ifpublic\publictrue
\newif\iffancy\fancytrue

\usepackage[a4paper,text={160mm,247mm},centering]{geometry}
\usepackage{setspace}
\usepackage{rotating}

\usepackage{amsmath,amssymb, amsfonts, geometry}

\makeatletter
\g@addto@macro\bfseries{\boldmath}
\makeatother

\makeatletter
\providecommand*{\shuffle}{%
  \mathbin{\mathpalette\shuffle@{}}%
}
\newcommand*{\shuffle@}[2]{%
  \sbox0{$#1\vcenter{}$}%
  \kern .15\ht0 
  \rlap{\vrule height .25\ht0 depth 0pt width 2.5\ht0}%
  \raise.1\ht0\hbox to 2.5\ht0{%
    \vrule height 1.75\ht0 depth -.1\ht0 width .17\ht0 %
    \hfill
    \vrule height 1.75\ht0 depth -.1\ht0 width .17\ht0 %
    \hfill
    \vrule height 1.75\ht0 depth -.1\ht0 width .17\ht0 %
  }%
  \kern .15\ht0 
}
\makeatother

\usepackage{xparse}

\ExplSyntaxOn
\NewDocumentCommand{\omwb}{m m}
{
 \omm\left(\begin{smallmatrix}
 \omwb_print:n {#1} \\
 \omwb_print:n {#2}
 \end{smallmatrix}\right)
}
\seq_new:N \l_omwb_list_seq
\cs_new_protected:Npn \omwb_print:n #1
{
  \seq_set_split:Nnn \l_omwb_list_seq { , } { #1 }
  \seq_use:Nn \l_omwb_list_seq { , & }
}
\ExplSyntaxOff

\usepackage[pdfencoding=auto,bookmarks=true,hyperfigures=true]{hyperref}
\usepackage{graphbox}%
\usepackage{float}
\usepackage[nosort]{cite}
\usepackage{lmodern}
\usepackage{amsbsy}
\usepackage[utf8]{inputenc}

\usepackage[font=small,labelfont=bf]{caption}

\usepackage[usenames,dvipsnames]{xcolor}
\definecolor{dgreen}{rgb}{0,0.70,0.30}
\definecolor{gold}{rgb}{0.85,.66,0}
\definecolor{purple}{rgb}{1.0,0.3,0.6}

\usepackage{tikz}
\usetikzlibrary{calc} 
\usetikzlibrary{patterns} 
\usetikzlibrary{decorations.pathreplacing} 
\usetikzlibrary{decorations.markings} 
\usetikzlibrary{decorations.pathmorphing} 
\usetikzlibrary{positioning}
\usetikzlibrary{arrows.meta}

\usepackage[fonts]{mpostinl}

\makeatletter
\newsavebox{\apb@box}\newlength{\apb@width}
\newcommand{\autoparbox}[2][c]{\sbox{\apb@box}{#2}%
 \settowidth{\apb@width}{\usebox{\apb@box}}%
 \parbox[#1]{\apb@width}{\usebox{\apb@box}}}

\makeatother

\iffancy

\def\showkeysrefformat#1{{\normalfont\tiny\ttfamily#1}}
\makeatletter
\def\SK@@ref#1>#2\SK@{%
 {\@inlabelfalse\leavevmode\vbox to\z@{%
 \vss\SK@refcolor\rlap{\vrule\raise .75em%
  \hbox{\showkeysrefformat{#2}}}}}}
\makeatother
\fi

\numberwithin{equation}{section}

\newcommand{\eqn}[1]{eq.~\eqref{#1}}
\newcommand{\Eqn}[1]{Equation~\eqref{#1}}
\newcommand{\eqns}[2]{eqs.~\eqref{#1} and~\eqref{#2}}

\newcommand{\rcite}[1]{ref.~\cite{#1}}
\newcommand{\rcites}[1]{refs.~\cite{#1}}

\providecommand{\href}[2]{#2}

\makeatletter
\def\mr@ignsp#1 {\ifx\:#1\@empty\else #1\expandafter\mr@ignsp\fi}%
\newcommand{\multiref}[1]{\begingroup
\xdef\mr@no@sparg{\expandafter\mr@ignsp#1 \: }%
\def\mr@comma{}%
\@for\mr@refs:=\mr@no@sparg\do{\mr@comma\def\mr@comma{,}\ref{\mr@refs}}%
\endgroup}
\renewcommand{\eqref}[1]{(\multiref{#1})}
\makeatother

\makeatletter
\newcommand{\namedref}[2]{\hyperref[#2]{#1~\ref*{#2}}}
\newcommand{\secref}{\@ifstar{\namedref{Section}}{\namedref{section}}}
\newcommand{\subsecref}{\@ifstar{\namedref{Subsection}}{\namedref{subsection}}}
\newcommand{\appref}{\@ifstar{\namedref{Appendix}}{\namedref{appendix}}}
\newcommand{\tabref}{\@ifstar{\namedref{Table}}{\namedref{table}}}
\newcommand{\figref}{\@ifstar{\namedref{Figure}}{\namedref{figure}}}
\makeatother

\providecommand{\hypersetup}[1]{}
\providecommand{\texorpdfstring}[2]{#1}

\hypersetup{plainpages=false}
\hypersetup{pdfpagemode=UseNone}
\hypersetup{bookmarksnumbered=true}
\hypersetup{pdfstartview=FitH}
\hypersetup{colorlinks=false}
\hypersetup{citebordercolor={.5 1 .5}}
\hypersetup{urlbordercolor={.5 1 1}}
\hypersetup{linkbordercolor={1 .7 .7}}

\makeatletter
\let\@keywords\@empty
\let\@subject\@empty
\providecommand{\keywords}[1]{\gdef\@keywords{#1}}
\providecommand{\subject}[1]{\gdef\@subject{#1}}
\def\thetitle{\@title}
\def\theauthor{\@author}
\def\thesubject{\@subject}
\def\thedate{\@date}
\def\thekeywords{\@keywords}
\makeatother
\AtBeginDocument{
\hypersetup{pdftitle={\thetitle}}%
\hypersetup{pdfauthor={\theauthor}}%
\hypersetup{pdfsubject={\thesubject}}%
\hypersetup{pdfkeywords={\thekeywords}}%
}

\newif\ifnote 
\notetrue

\allowdisplaybreaks

\newcommand{\ad} {\mathrm{ad}}

\let\Re\relax\DeclareMathOperator{\Re}{Re}
\let\Im\relax\DeclareMathOperator{\Im}{Im}

\newcommand{\z}{\zeta}

\newcommand{\te}{\textrm}

\newcommand{\co}{\ , \ \ \ \ \ \ }
\newcommand{\dd}{\mathrm{d}}
\newcommand{\ee}{\mathrm{e}}

\newcommand{\ap}{\alpha'}

\newcommand{\nbeta}{b}

\newcommand{\ZR}{\mathbb R}
\newcommand{\ZC}{\mathbb C}
\newcommand{\ZE}{\mathbb E}
\newcommand{\ZG}{\mathbb G}

\newcommand{\ZN}{\mathbb N}
\newcommand{\ZZ}{\mathbb Z}
\newcommand{\ZQ}{\mathbb Q}

\newcommand{\CC}{\mathcal{C}}       
       
\newcommand{\CE}{\mathcal{E}}       
       
\newcommand{\CG}{\mathcal{G}}

\newcommand{\CO}{\mathcal{O}}      
\newcommand{\CP}{\mathcal{P}}

\def\open{\textrm{open}}
\def\closed{\textrm{closed}}

\newcommand{\sv}{\ensuremath{\text{sv}}}
\newcommand{\esv}{\ensuremath{\text{esv}}}

\newcommand{\binomial}{\binom}

\DeclareMathOperator{\nhehat}{e}

\newcommand{\nnl}{\nonumber\\}

\newcommand{\suml}{\sum\limits}

\newcommand{\intl}{\int\limits}

\DeclareMathOperator{\GL}{\Gamma}

\DeclareMathOperator{\gm}{\gamma}
\DeclareMathOperator{\gmz}{\gamma_0}
\DeclareMathOperator{\ce}{\CE}
\DeclareMathOperator{\cez}{\CE_0}
\DeclareMathOperator{\zm}{\zeta}

\DeclareMathOperator{\omm}{\omega}

\DeclareMathOperator{\dlog}{\mathrm{dlog}}

\DeclareMathOperator{\GGs}{G}
\newcommand{\GGz}[1]{\GGs^0_{#1}}
\newcommand{\GG}[1]{\GGs_{#1}}

\newcommand{\gSL}{\text{SL}}

\newcommand{\PP}{P}

\DeclareMathOperator{\EEs}{E}

\newcommand{\EE}[1]{\EEs_{#1}}

\newcommand{\GLarg}[3]{\GL\left(\begin{smallmatrix}#1\\#2\end{smallmatrix};#3\right)}
\newcommand{\GGG}[3]{\CG\left[\begin{smallmatrix}#1\\#2\end{smallmatrix};#3\right]}

\usepackage{nicefrac}
\newcommand{\tauh}{\nicefrac{\tau}{2}}

\DeclareMathOperator{\AcycLetter}{\text{\textbf{A}}}
\DeclareMathOperator{\BcycLetter}{\text{\textbf{B}}}
\DeclareMathOperator{\DcycLetter}{\text{\textbf{D}}}
\DeclareMathOperator{\hatAcycLetter}{\hat{\text{\textbf{A}}}}

\DeclareMathOperator{\smallBcycLetter}{\text{\textbf{b}}}
\DeclareMathOperator{\smallDcycLetter}{\text{\textbf{d}}}
\newcommand{\Acyc}[1]{\AcycLetter\mkern-4.2mu\Big[\mpostuse[width=0.5cm,align=c]{#1}\Big]}
\newcommand{\Bcyc}[1]{\BcycLetter\mkern-4.2mu\Big[\mpostuse[width=0.5cm,align=c]{#1}\Big]}
\newcommand{\Dcyc}[1]{\DcycLetter\mkern-4.2mu\Big[\mpostuse[width=0.5cm,align=c]{#1}\Big]}
\newcommand{\hatAcyc}[1]{\hatAcycLetter\mkern-4.2mu\Big[\mpostuse[width=0.5cm,align=c]{#1}\Big]}

\newcommand{\smallBcyc}[1]{\smallBcycLetter\mkern-4.2mu\Big[\mpostuse[width=0.5cm,align=c]{#1}\Big]}
\newcommand{\smallDcyc}[1]{\smallDcycLetter\mkern-4.2mu\Big[\mpostuse[width=0.5cm,align=c]{#1}\Big]}

\newcommand{\db}{\hspace{1pt}|\hspace{1pt}}

\vspace*{0.2cm}
\title{\textbf{From elliptic multiple zeta values \\
    to modular graph functions: \\
open and closed strings at one loop}}
\author{Johannes Broedel$^{\,\textit{a},\textit{b}}$,
Oliver Schlotterer$^{\,\textit{b},\textit{c},\textit{d}}$, Federico Zerbini$^{\,\textit{e},\textit{f}}$}
\date{\today}

\begin{document}
\pdfbookmark[1]{Title Page}{title} \thispagestyle{empty}
\begin{flushright}
  \verb!HU-EP-18/05!\\
  \verb!HU-Mathematik-2018-02!\\
  \verb!NSF-ITP-18-009!\\
  \verb!IPHT-t18/019!
\end{flushright}
\vspace*{0.4cm}
\begin{center}%
  \begingroup\LARGE\bfseries\thetitle\par\endgroup
\vspace{1.0cm}

\begingroup\large\theauthor\par\endgroup
\vspace{9mm}
\begingroup\itshape
$^{\te{a}}$Institut f\"ur Mathematik und Institut f\"ur Physik, Humboldt-Universit\"at zu Berlin\\
IRIS Adlershof, Zum Gro\ss{}en Windkanal 6, 12489 Berlin, Germany
\par\endgroup
\vspace{3mm}
\begingroup\itshape
$^{\te{b}}$KAVLI Institute for Theoretical Physics, Kohn Hall,\\
University of California, Santa Barbara, CA 93106, USA
\par\endgroup
\vspace{3mm}
\begingroup\itshape
$^{\te{c}}$Max-Planck-Institut f\"ur Gravitationsphysik, Albert-Einstein-Institut\\
Am M\"uhlenberg 1, 14476 Potsdam, Germany
\par\endgroup
\vspace{3mm}
\begingroup\itshape
$^{\te{d}}$Perimeter Institute for Theoretical Physics \\
31 Caroline St N, Waterloo, ON N2L 2Y5, Canada
\par\endgroup
\vspace{3mm}
\begingroup\itshape
$^{\te{e}}$Institut de Physique Th\'eorique (IPhT),
CEA-Saclay\\
Orme des Merisiers batiment 774, 91191 Gif-sur-Yvette, France
\par\endgroup
\vspace{3mm}
\begingroup\itshape
$^{\te{f}}$Max-Planck-Institut f\"ur Mathematik,
Vivatsgasse 7, 53111 Bonn, Germany
\par\endgroup

\vspace{1.0cm}

\begingroup\ttfamily
jbroedel@physik.hu-berlin.de, \\
olivers@aei.mpg.de,\\
fzerbini@ipht.fr
\par\endgroup

\vspace{1.2cm}

\bigskip

\textbf{Abstract}\vspace{5mm}

\begin{minipage}{13.4cm}
We relate one-loop scattering amplitudes of massless open- and closed-string
states at the level of their low-energy expansion. The modular graph functions
resulting from integration over closed-string punctures are observed to follow
from symmetrized open-string integrals through a tentative generalization 
of the single-valued projection known from genus zero.
\end{minipage}

\vspace*{4cm}

\end{center}

\newpage

\setcounter{tocdepth}{2}
\tableofcontents

\begin{mpostdef}
  xu:=0.8cm;
  ds:=3.2pt;
  ls:=0.6pt;
  pair vp[];
  path c;
  def pensize(expr s)=withpen pencircle scaled s enddef;
  def poly(expr k,offset)=
    for i=1 upto k:
    vp[i]:=0.5*xu*(dir ((i-1)*360/k+offset));
    endfor;
  enddef;
  def singleline(expr i,j)=
    draw vp[i]--vp[j] pensize(ls);
  enddef;
  def doubleline(expr i,j)=
    draw vp[i]{(vp[j]-vp[i]) rotated 25}..{(vp[j]-vp[i]) rotated -25}vp[j] pensize(ls);
    draw vp[i]{(vp[j]-vp[i]) rotated -25}..{(vp[j]-vp[i]) rotated 25}vp[j] pensize(ls);
  enddef;
  def tripleline(expr i,j)=
    draw vp[i]{(vp[j]-vp[i]) rotated 35}..{(vp[j]-vp[i]) rotated -35}vp[j] pensize(ls);
    draw vp[i]{(vp[j]-vp[i]) rotated 0}..{(vp[j]-vp[i]) rotated 0}vp[j] pensize(ls);
    draw vp[i]{(vp[j]-vp[i]) rotated -35}..{(vp[j]-vp[i]) rotated 35}vp[j] pensize(ls);
  enddef;
  def dottedline(expr i,j,num,ang)=
  c:= vp[i]{(vp[j]-vp[i]) rotated ang}..{(vp[j]-vp[i]) rotated (-1*ang)}vp[j];
    draw c pensize(ls);
    for vv=1 upto (num-1):
      draw point(vv/num) of c pensize(ds);
    endfor;
  enddef;
  def drawvert(expr i,j)=
    for vv=i upto j:
      draw vp[vv] pensize(ds);
    endfor;
  enddef;
  def drawcentralvert=
    draw (0xu,0xu) pensize(ds);
  enddef;
  def background =
    c:= (-1xu,-1xu)--(-1xu,1xu)--(1xu,1xu)--(1xu,-1xu)--cycle;
    c:=c scaled 0.52;
    fill c withcolor op*white;
  enddef;
  def framed(expr p)=fill bbox p withcolor white; draw bbox p pensize(0.8pt); draw p; enddef;
  def framedd(expr p)=fill bbox p withcolor 0.94white; draw bbox p pensize(1pt); draw p; enddef;
\end{mpostdef}

\begin{mpostfig}[label=myfig]
  poly(4,45);
  doubleline(1,2);
  tripleline(2,3);
  dottedline(3,4,3,20);
  for i=1 upto 4:
   draw vp[i] pensize(ds);
  endfor;
\end{mpostfig}

\begin{mpostfig}[label=empty]
  background;
\end{mpostfig}


\begin{mpostfig}[label=Q1label]
  background;poly(2,0);dottedline(1,2,0,0);drawvert(1,2);
  draw (0xu,0.35xu)--(0xu,-0.35xu) pensize(1.4 ls) dashed withdots scaled 0.4;
  label.rt(btex $i$ etex,(-1xu,-.2xu));
  label.lft(btex $j$ etex,(1xu,-.2xu));
\end{mpostfig}

\begin{mpostfig}[label=G2np]
  background;poly(2,0);dottedline(1,2,0,35);dottedline(1,2,0,-35);drawvert(1,2);
  draw (0xu,0.4xu)--(0xu,-0.4xu) pensize(2.2 ls) dashed withdots scaled 0.52;
\end{mpostfig}

\begin{mpostfig}[label=G3np]
  background;poly(2,0);dottedline(1,2,0,-50);dottedline(1,2,0,0);;dottedline(1,2,0,50);drawvert(1,2);
  draw (0xu,0.45xu)--(0xu,-0.45xu) pensize(2.2 ls) dashed withdots scaled 0.52;
\end{mpostfig}

\begin{mpostfig}[label=G111np]
  background;poly(3,90); singleline(1,2);singleline(1,3);singleline(2,3); drawvert(1,3);
  draw (-0.5xu,0.08xu)--(0.5xu,0.08xu) pensize(2.0 ls) dashed withdots scaled 0.52;
\end{mpostfig}

\begin{mpostfig}[label=G4np]
  background;poly(2,0);dottedline(1,2,0,-60);dottedline(1,2,0,-20);dottedline(1,2,0,20);dottedline(1,2,0,60);drawvert(1,2);
  draw (0xu,0.5xu)--(0xu,-0.5xu) pensize(2.2 ls) dashed withdots scaled 0.52;
\end{mpostfig}

\begin{mpostfig}[label=G211np1]
  background; poly(3,90); singleline(1,2); singleline(1,3); doubleline(2,3); drawvert(1,3);
  draw (-0.5xu,0.08xu)--(0.5xu,0.08xu) pensize(2.0 ls) dashed withdots scaled 0.52;
\end{mpostfig}

\begin{mpostfig}[label=G211np2]
  background; poly(3,90); doubleline(1,2); singleline(1,3); singleline(2,3); drawvert(1,3);
  draw (-0.5xu,0.08xu)--(0.5xu,0.08xu) pensize(2.0 ls) dashed withdots scaled 0.52;
\end{mpostfig}

\begin{mpostfig}[label=G1111np1]
  background;poly(4,45); singleline(1,2);singleline(2,3);singleline(3,4);singleline(4,1); drawvert(1,4);
  draw (-0.5xu,0.00xu)--(0.5xu,0.00xu) pensize(2.0 ls) dashed withdots scaled 0.52;
\end{mpostfig}

\begin{mpostfig}[label=G1111np2]
  background;poly(4,45); singleline(1,2);singleline(2,3);singleline(3,4);singleline(4,1); drawvert(1,4);
  draw (0.5xu,-0.20xu)--(-0.2xu,0.5xu) pensize(2.0 ls) dashed withdots scaled 0.52;
\end{mpostfig}

\begin{mpostfig}[label=G1111np3]
  background;poly(4,45); singleline(1,3);singleline(2,4);singleline(2,3);singleline(1,4); drawvert(1,4);
  draw (-0.55xu,0.00xu)--(0.55xu,0.00xu) pensize(2.0 ls) dashed withdots scaled 0.52;
\end{mpostfig}


\begin{mpostfig}[label=G1label]
  background;poly(2,0);dottedline(1,2,0,0);drawvert(1,2);
  label.rt(btex $i$ etex,(-1xu,-.2xu));
  label.lft(btex $j$ etex,(1xu,-.2xu));
\end{mpostfig}


\begin{mpostfig}[label=G2]
  background;poly(2,0);dottedline(1,2,0,35);dottedline(1,2,0,-35);drawvert(1,2);
\end{mpostfig}


\begin{mpostfig}[label=G3]
  background;poly(2,0);dottedline(1,2,0,-50);dottedline(1,2,0,0);;dottedline(1,2,0,50);drawvert(1,2);
\end{mpostfig}

\begin{mpostfig}[label=G111]
  background;poly(3,90); singleline(1,2);singleline(1,3);singleline(2,3); drawvert(1,3);
\end{mpostfig}


\begin{mpostfig}[label=G4]
  background;poly(2,0);dottedline(1,2,0,-60);dottedline(1,2,0,-20);dottedline(1,2,0,20);dottedline(1,2,0,60);drawvert(1,2);
\end{mpostfig}

\begin{mpostfig}[label=G211]
  background; poly(3,90); singleline(1,2); singleline(1,3); doubleline(2,3); drawvert(1,3);
\end{mpostfig}

\begin{mpostfig}[label=G1111]
  background;poly(4,45); singleline(1,2);singleline(2,3);singleline(3,4);singleline(4,1); drawvert(1,4);
\end{mpostfig}

\begin{mpostfig}[label=GC211]
  background; poly(2,0);drawvert(1,2); dottedline(2,1,1,70); dottedline(2,1,1,-10); dottedline(2,1,2,-70); \end{mpostfig}


\begin{mpostfig}[label=G5]
  background;poly(2,0);dottedline(1,2,0,-60);dottedline(1,2,0,-30);dottedline(1,2,0,0);dottedline(1,2,0,30);dottedline(1,2,0,60);drawvert(1,2);
\end{mpostfig}

\begin{mpostfig}[label=G221]
  background; poly(3,90); doubleline(1,2); doubleline(1,3); singleline(2,3); drawvert(1,3);
\end{mpostfig}

\begin{mpostfig}[label=G311]
  background; poly(3,90); singleline(1,2); singleline(1,3); tripleline(2,3); drawvert(1,3);
\end{mpostfig}

\begin{mpostfig}[label=G2111]
  background; poly(4,45); singleline(1,2); doubleline(2,3); singleline(3,4); singleline(4,1); drawvert(1,4);
\end{mpostfig}

\begin{mpostfig}[label=Gp1111]
  background; poly(4,45);drawvert(1,4); singleline(1,2);singleline(2,3);singleline(3,4);singleline(4,1);singleline(1,3);
\end{mpostfig}

\begin{mpostfig}[label=G11111]
  background; poly(5,18);drawvert(1,5); singleline(1,2); singleline(2,3); singleline(3,4); singleline(4,5); singleline(5,1); 
\end{mpostfig}

\begin{mpostfig}[label=GC221]
  background;
  poly(2,0);drawvert(1,2);
  dottedline(2,1,2,70); 
  dottedline(2,1,2,-10); 
  dottedline(2,1,1,-70); 
\end{mpostfig}


\begin{mpostfig}[label=G6]
  background;poly(2,0);dottedline(1,2,0,-75);dottedline(1,2,0,-50);dottedline(1,2,0,-18);dottedline(1,2,0,18);dottedline(1,2,0,50);dottedline(1,2,0,75);drawvert(1,2);
\end{mpostfig}

\begin{mpostfig}[label=G411]
  background; poly(3,90); dottedline(3,2,0,-45);dottedline(3,2,0,-15);dottedline(3,2,0,15);dottedline(3,2,0,45); singleline(1,2); singleline(1,3); drawvert(1,3);
\end{mpostfig}

\begin{mpostfig}[label=G321]
  background; poly(3,90); tripleline(1,2); doubleline(2,3); singleline(3,1); drawvert(1,3);
\end{mpostfig}

\begin{mpostfig}[label=G222]
  background;
  poly(3,90);drawvert(1,3);
  doubleline(1,2);doubleline(1,3);doubleline(2,3);
\end{mpostfig}

\begin{mpostfig}[label=G3111]
  background; poly(4,45); singleline(1,2); tripleline(2,3); singleline(3,4); singleline(4,1); drawvert(1,4);
\end{mpostfig}

\begin{mpostfig}[label=G2211]
  background; poly(4,45);drawvert(1,4); singleline(1,2);doubleline(2,3); doubleline(3,4); singleline(4,1);
\end{mpostfig}

\begin{mpostfig}[label=Gp2111]
  background; poly(4,45);drawvert(1,4); singleline(1,2);doubleline(2,3);singleline(3,4);singleline(4,1);singleline(1,3);
\end{mpostfig}

\begin{mpostfig}[label=Gpp1111]
  background; poly(4,45);drawvert(1,4); singleline(1,2);singleline(2,3);singleline(3,4);singleline(4,1);doubleline(1,3);
\end{mpostfig}

\begin{mpostfig}[label=Gx1111]
  background; poly(4,45);drawvert(1,4); singleline(1,2);singleline(2,3);singleline(3,4);singleline(4,1);singleline(1,3);singleline(2,4);
\end{mpostfig}

\begin{mpostfig}[label=G21111]
  background; poly(5,18);drawvert(1,5); singleline(1,2); singleline(2,3); doubleline(3,4); singleline(4,5); singleline(5,1); 
\end{mpostfig}

\begin{mpostfig}[label=Gp11111]
  background; poly(5,18);drawvert(1,5); singleline(1,2);singleline(2,3);singleline(3,4);singleline(4,5);singleline(5,1);singleline(1,3);
\end{mpostfig}

\begin{mpostfig}[label=G111111]
  background; poly(6,0);drawvert(1,6); singleline(1,2); singleline(2,3); singleline(3,4); singleline(4,5); singleline(5,6); singleline(6,1); 
\end{mpostfig}

\begin{mpostfig}[label=GC411]
  background; poly(2,0);drawvert(1,2); dottedline(2,1,1,70); dottedline(2,1,1,0); dottedline(2,1,4,-70); 
\end{mpostfig}

\begin{mpostfig}[label=GC321]
  background;
  poly(2,0);drawvert(1,2);
  dottedline(2,1,1,70);
  dottedline(2,1,2,-10); 
  dottedline(2,1,3,-70); 
\end{mpostfig}

\begin{mpostfig}[label=GC222]
  background; poly(2,0);drawvert(1,2); dottedline(2,1,2,70); dottedline(2,1,2,-10); dottedline(2,1,2,-70); \end{mpostfig}

\begin{mpostfig}[label=GC2211]
  background; poly(2,0);drawvert(1,2); dottedline(2,1,1,70); dottedline(2,1,1,27); dottedline(2,1,2,-27); dottedline(2,1,2,-70); 
\end{mpostfig}

\begin{mpostfig}[label=GC3111]
  background; poly(2,0);drawvert(1,2); dottedline(2,1,0,70); dottedline(2,1,0,27); dottedline(2,1,0,-27); dottedline(2,1,3,-70); 
\end{mpostfig}


\begin{mpostfig}[label=G511]
  background; poly(3,90); 
  dottedline(2,3,0,-60);dottedline(2,3,0,-30);dottedline(2,3,0,0);dottedline(2,3,0,30);dottedline(2,3,0,60);
  singleline(1,2); singleline(1,3); drawvert(1,3);
\end{mpostfig}


\begin{mpostfig}[label=tetrahedral]
  background;poly(3,90); 
  vp[4]:=(0xu,0xu);
  singleline(1,2);singleline(1,3);singleline(2,3);drawvert(1,4);
  singleline(1,4);singleline(2,4);singleline(3,4); 
\end{mpostfig}

\begin{mpostfig}[label=test]
  background;
  poly(10,0);drawvert(1,10);
  doubleline(8,9);
  tripleline(9,10); 
  dottedline(1,6,4,-30);
\end{mpostfig}

\begin{mpostdef}
xu:=1cm;
psf:=0.85;
ahlength:=5pt;
string fftext;
qcdpos:=3;
nefpos:=1.25;
ostpos:=-0.65;
grpos:=-2.7;
\end{mpostdef}

\begin{mpostfig}[label=overview]


framed(thelabel (btex \begin{minipage}{2.8cm}\begin{center}\textbf{$A$-cycle graph function}\end{center}\end{minipage}etex,(3.0xu,4.0xu )));
framed(thelabel (btex \begin{minipage}{2.8cm}\begin{center}\textbf{$B$-cycle graph function}\end{center}\end{minipage}etex,(10.0xu,2.8xu )));
framed(thelabel (btex \begin{minipage}{2.8cm}\begin{center}\textbf{modular graph function}\end{center}\end{minipage}etex,(4.0xu,0.0xu )));

drawarrow((4.7xu,4.0xu){dir 0}..{dir 0}(8.3xu,2.8xu)) pensize(1.0pt);
drawarrow((10.0xu,2.2xu){dir 270}..{dir 180}(5.7xu,0.0xu)) pensize(1.0pt);
drawarrow((3.0xu,3.4xu){dir 270}..{dir 270}(4.0xu,0.6xu)) pensize(1.0pt);

label(btex \begin{minipage}{2.8cm}\begin{center}\begin{small}modular\\[-2pt]transformation\\[-2pt]$\tau\to-\frac{1}{\tau}$\end{small}\end{center}\end{minipage}etex,(7.0xu,4.2xu ));
label(btex \begin{minipage}{1.8cm}\begin{center}\begin{small}esv\\[-2pt] projection\end{small}\end{center}\end{minipage}etex,(10.8xu,1.0xu ));
label(btex \begin{minipage}{2.8cm}\begin{center}\begin{small}compare\\[-2pt]$\tau$-derivative\\[-2pt]and\\[-2pt]Cauchy--Riemann\\[-2pt]derivative\end{small}\end{center}\end{minipage}etex,(2.2xu,1.9xu ));

\end{mpostfig}

\begin{mpostfig}[label=overview2]
framed(thelabel (btex \begin{minipage}{2.8cm}\begin{center}\textbf{$A$-cycle graph function}\\$\omega$-representation\end{center}\end{minipage}etex,(4.0xu,6.0xu )));
framed(thelabel (btex \begin{minipage}{2.8cm}\begin{center}\textbf{$A$-cycle graph function}\\$\cal{E}$-representation\end{center}\end{minipage}etex,(8.0xu,6.0xu )));
framed(thelabel (btex \begin{minipage}{2.8cm}\begin{center}\textbf{$B$-cycle graph function}\\$\omega$-representation\end{center}\end{minipage}etex,(4.0xu,4.0xu )));
framed(thelabel (btex \begin{minipage}{2.8cm}\begin{center}\textbf{$B$-cycle graph function}\\$\cal{E}$-representation\end{center}\end{minipage}etex,(8.0xu,4.0xu )));
\end{mpostfig}

\begin{mpostfig}[label=openclosed]
c := ((0xu,0xu)--(3xu,0xu)--(3xu,5xu)--(0xu,5xu)--cycle);
fill c withcolor 0.8 white;
fill c shifted (6.0xu,0.0xu) withcolor 0.8 white;
label(btex \textbf{open string} etex, (1.5xu,4.2 xu));
label(btex \textbf{closed string} etex, (7.5xu,4.2 xu));
framed(thelabel (btex \begin{minipage}{2.2cm}\begin{center}tree-level\\\textbf{MZV}\end{center}\end{minipage}etex,(1.5xu,3.0xu )));
framed(thelabel (btex \begin{minipage}{2.2cm}\begin{center}tree-level\\\textbf{sv(MZV)}\end{center}\end{minipage}etex,(7.5xu,3.0xu )));
framed(thelabel (btex \begin{minipage}{2.2cm}\begin{center}one-loop level\\\textbf{eMZV}\end{center}\end{minipage}etex,(1.5xu,1.0xu )));
framed(thelabel (btex \begin{minipage}{2.2cm}\begin{center}one-loop level\\\textbf{esv(eMZV)}\end{center}\end{minipage}etex,(7.5xu,1.0xu )));
drawarrow((2.85xu,3.0xu){dir 15}..(6.15xu,3.0xu)) pensize(0.6pt);
label(btex \begin{minipage}{2.2cm}\begin{center}single-valued\\[-2pt]projection\end{center}\end{minipage} etex, (4.5xu,3.73xu));
drawarrow((2.85xu,1.0xu){dir -15}..(6.15xu,1.0xu)) pensize(0.6pt);
label(btex \begin{minipage}{2.2cm}\begin{center}elliptic single-valued projection?\end{center}\end{minipage} etex, (4.5xu,1.60xu));
\end{mpostfig}

\begin{mpostfig}[label=elliptic]
  picture pic;
  draw (-2xu,0xu) ..controls (-2xu,0.75xu) and (2xu,0.75xu) .. ( 2xu,0xu) dashed evenly pensize (0.6pt) withcolor blue;
  draw (0xu,-0.15xu) .. controls (0.2xu,-0.18xu) and (0.2xu,-0.92xu) .. ( 0xu,-0.95xu) dashed evenly pensize (0.6pt) withcolor red;
  draw fullcircle xscaled 4xu yscaled 1.9xu pensize (1pt);
  draw (-1xu,0.03xu){dir -20}..(1xu,0.03xu) pensize (0.8pt);
  draw (-0.8xu,-0.02xu){dir 25}..(0.8xu,-0.02xu) pensize (0.8pt);
  draw (-2xu,0xu) ..controls (-2xu,-0.75xu) and (2xu,-0.75xu) .. ( 2xu,0xu) pensize (0.6pt) withcolor blue;
  draw (0xu,-0.15xu) .. controls (-0.2xu,-0.18xu) and (-0.2xu,-0.92xu) .. ( 0xu,-0.95xu) pensize (0.6pt) withcolor red;
  pic:=currentpicture;
  currentpicture := nullpicture;
  pickup pencircle scaled 1pt;
  drawarrow (-0.15xu,0xu)--(5.2xu,0xu) withcolor black;
  drawarrow (0xu,-0.15xu)--(0xu,2.6xu) withcolor black;
  draw (0xu,0xu)--(3.9xu, 0xu) withcolor red;
  draw (4.9 xu, 2 xu)--(1 xu, 2 xu) withcolor red;
  draw (3.9 xu, 0 xu)--(4.9 xu, 2 xu) withcolor blue;
  draw (0xu, 0xu)--(1 xu, 2 xu) withcolor blue;
  label.bot (btex $0$ etex, (0,-.1xu));
  label.top (btex $\tau$ etex, (0.9xu,2xu));
  label.top (btex $\tau+1$ etex, (4.8xu,2xu));
  label.bot(btex $1$ etex, (3.9xu,-.1xu));
  label.lft(btex Im$(z)$ etex, (0xu,2.6xu));
  label.top(btex Re$(z)$ etex, (5.2xu,0xu));
  draw pic shifted (-4xu,0.8xu) scaled (1.2);
\end{mpostfig}


\section{Introduction}

Modular graph functions are building blocks for one-loop scattering amplitudes
in closed-string theories at the one-loop level. They have been thoroughly
investigated by D'Hoker, Green, Vanhove and other authors during the last
couple of years \cite{Green:1999pv, Green:2008uj, DHoker:2015gmr,
DHoker:2015sve, Basu:2015ayg, DHoker:2015wxz, DHoker:2016mwo, Basu:2016xrt,
Basu:2016kli, Basu:2016mmk, DHoker:2016quv, Kleinschmidt:2017ege, DHoker:2017zhq} and arise
from Feynman graphs of certain conformal scalar fields on the torus.  Each
modular graph function depends on the modular parameter of the torus and its
modular invariance is inherited from the underlying closed-string setup. While
the computation of their asymptotic expansion\footnote{As the modular parameter 
$\tau$ tends to $i\infty$ such that a homology cycle of the Riemann surfaces pinches.} 
is itself cumbersome, they exhibit a variety of mathematical structures:
modular graph functions are related by a network of algebraic identities and
related to holomorphic Eisenstein series by differential equations with respect
to the modular parameter. Even more, they satisfy certain eigenvalue equations
involving the modular invariant Laplace operator.

Most interestingly for the purpose of this article, however, a first connection
between elliptic multiple polylogarithms (as defined in \rcites{Zagier, Levin,
BrownLev}) and modular graph functions was established in
\rcite{DHoker:2015wxz}: The latter were written as special values of infinite
sums of single-valued multiple polylogarithms, and these infinite sums are
proposed in the reference to be a single-valued analogue of elliptic multiple
polylogarithms\footnote{It is not demonstrated that the infinite sums studied
in \rcite{DHoker:2015wxz} can be called single-valued elliptic multiple
polylogarithms in the usual mathematical sense. This would be true if one can
write them as finite linear combinations of products of elliptic multiple
polylogarithms and their complex conjugates.}.  This connection extends an
observation made for genus-zero (tree-level) open- and closed-string
amplitudes: closed-string tree amplitudes are conjectured to be obtained by
acting with the so-called single-valued projection on the multiple zeta values
appearing in their open-string counterparts \cite{Schlotterer:2012ny,
Stieberger:2013wea, Stieberger:2014hba}.  The single-valued projection maps
generic multiple zeta values to those instances which descend from
single-valued polylogarithms at genus zero \cite{Schnetz:2013hqa,
Brown:2013gia}.

At genus one (one-loop level), Enriquez's elliptic multiple zeta values
\cite{Enriquez:Emzv} were shown to capture the low-energy expansion of the open
superstring \cite{Broedel:2014vla, Broedel:2015hia, Broedel:2017jdo}. The
results of \cite{DHoker:2015wxz} suggest to expect that modular graph functions
are single-valued versions of Enriquez's elliptic multiple zeta values. However,
the precise matching and thus the relation between open- and closed-string
results at one-loop level is an open problem: First, the closed-string
\cite{DHoker:2015wxz} and open-string literature \cite{Broedel:2014vla,
Broedel:2015hia, Broedel:2017jdo} use different notions of elliptic
polylogarithms.  Second, the dependence of modular graph functions and elliptic
multiple zeta values on the modular parameters of the respective genus-one
surface is realized in rather different languages.

In the current article we are going to bridge the leftover gap between one-loop
open- and closed-string amplitudes before integration over the respective
modular parameters. We propose a setup which allows to relate certain building
blocks of open-string amplitudes with modular graph functions. This accumulates
evidence for a conjectural elliptic generalization of the single-valued projection 
known from genus zero. Simultaneously, this leads to a conjectural
formalism to explicitly construct modular graph functions starting from
open-string quantities.  The results thus obtained pass a variety of consistency checks
and match previous partial expressions.

The main idea is to define open-string graph functions within an abelian
version of one-loop open-string amplitudes. Despite the fact that the
permissible string spectrum of Type-I open-superstring theory does not contain
an abelian gauge boson \cite{Green:1984ed}, we will consider a kinematical
building block of the putative amplitude, which is non-trivial and well-defined
for auxiliary abelian particles. 
In order to implement the abelian character of the auxiliary particles, the
integration regions for open-string punctures are symmetrized in a convenient
manner.  The symmetrized open-string integrals of the abelian setup are the key
to lining up the properties of the open-string genus-one Green function with
its closed-string counterpart. In particular, the graphical organization of the
low-energy expansion of open- and closed-string amplitudes in terms of
open-string and modular graph functions agrees, which allows for direct
comparison between constituents. This includes
a matching of the respective differential equations in the modular parameter
on the open- and closed-string side.
  
\begin{figure}[h]
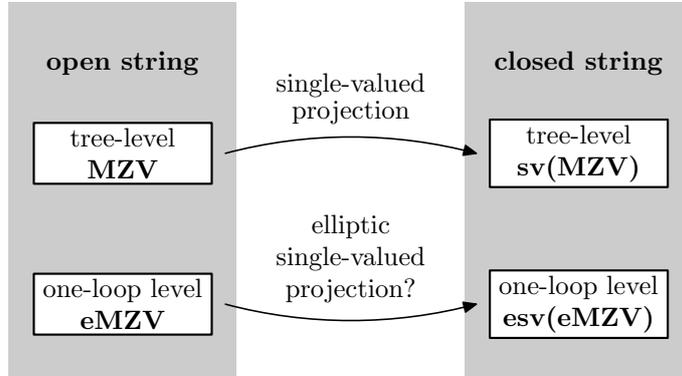

\begin{center}
\mpostuse{openclosed}
\end{center}
\caption{Context of a tentative generalization ``esv'' of the single-valued
projection to genus one.}
\label{contextesv}
\end{figure}


\subsection{Summary of results}

The notion of a \textit{single-valued projection} applies to a variety of
situations \cite{Francislecture}. The 
most common examples are multiple zeta values (MZVs), 
\begin{equation}
\zeta_{n_1,n_2,\ldots, n_r} := \sum_{0<k_1<k_2<\ldots <k_r}^{\infty} k_1^{-n_1} k_2^{-n_2} \ldots k_r^{-n_r} 
\ , \ \ \ \ n_i\in\ZN^+,\,\, n_r \geq 2 \,,
\label{tree05}
\end{equation}
of weight $n_1{+}n_2{+}\ldots{+}n_r$ and depth $r$, which can be represented as
multiple polylogarithms evaluated at unit argument. In contrast, single-valued
MZVs\footnote{While the concept of single-valuedness is well defined for a
  function, the notion is -- by slight abuse of nomenclature  -- also used for
MZVs which are numbers.} descend from single-valued multiple polylogarithms at unit
argument \cite{Brown:2013gia}.  As explained in the
reference, the single-valued projection formally denoted as
\begin{equation}
  \label{tree07a}
  \sv(\zm_{n_1,\ldots,n_r}) = \zm^\sv_{n_1,\ldots,n_r}
\end{equation}
maps generic MZVs \eqref{tree05} to their single-valued counterparts, e.g.
\begin{align}
  \zeta^\sv_{2k} &= 0 \,, \quad  \ \
 \zeta^\sv_{2k+1} = 2 \zeta_{2k+1} \,,\quad k\in \ZN^+  \label{tree07} \\
 \zeta^\sv_{3,5} &= -10 \zeta_3 \zeta_5 \, , \quad  \ \
 \zeta^\sv_{3,5,3}=2 \zeta_{3,5,3}-2 \zeta_3 \zeta_{3,5}-10 \zeta_3^2 \zeta_5 \,.
\notag
\end{align}
As will be reviewed in the next section, the single-valued projection of MZVs
appears naturally in relating tree-level scattering amplitudes of open and
closed strings:  the single-valued map acts on the MZVs arising in the
low-energy expansion of open-string disk integrals and yields the closed-string
integral over a punctured sphere. Correspondingly, it would be desirable to
identify a similar map called ``\esv'' for the elliptic version of multiple
zeta values $\omega$ (to be defined and discussed below) 
\begin{equation}
  \label{tree314}
  \esv(\omm({n_1,\ldots,n_r \db \tau})) = \omm^\esv(n_1,\ldots,n_r \db \tau) 
\end{equation}
at the one-loop level.  As will be shown in this article, one-loop open- and
closed-string amplitudes -- expressed as open-string and modular graph
functions, respectively -- can be taken as a starting point to propose an
analogous single-valued projection of elliptic multiple zeta values (eMZVs),
see \figref{contextesv}.  Accordingly, we are going to describe operations 
on open-string graph functions in suitable
presentations, which conjecturally yield modular graph functions as their
one-loop closed-string counterparts,
\begin{equation}
  \label{tree3141}
 \esv  \left( \begin{array}{c}  \te{open-string} \\ \te{graph function} \end{array} \right) = 
\left( \begin{array}{c}  \te{closed-string mo-} \\ \te{dular graph function} \end{array} \right)\,.
\end{equation}
As will be detailed below, the present formulation of the operations in $\esv$
is in general ill-defined, as it is not compatible with the shuffle
multiplication law of certain iterated Eisenstein integrals. Still, the
conjecture \eqn{tree3141} is well defined for the leading terms in the
expansion of both sides around the cusp, and for the complete expansions of
modular graph functions that evaluate to non-holomorphic Eisenstein series.
Moreover, it is highly non-trivial that there seem to exist rather natural
representations of the open-string input, which yield modular graph functions
beyond non-holomorphic Eisenstein series under the map $\esv$.

We will provide examples of the correspondence \eqn{tree3141}, up to and
including the seventh subleading order in the low-energy expansion. In
particular, starting from \eqn{tree3141}, we will establish a new connection
between building blocks of open- and closed-string four-point amplitudes
\begin{equation}
\esv \ M^{\rm open}_{4}(s_{ij}\db-\tfrac{1}{\tau}) = M_4^\closed(s_{ij}\db\tau) \ .
\label{punchline}
\end{equation}
These functions of the modular parameters $\tau$ of the underlying Riemann
surfaces result from integrating over the open- and closed-string punctures and
yield the respective building blocks for amplitudes upon integration over
$\tau$.  We will furthermore provide evidence that the planar open-string
integral on the left-hand side can be replaced by any of its non-planar
counterparts, irrespective on how the four state insertions are distributed
over the boundaries of the worldsheet.

It is important to mention that a way to produce a single-valued projection of
eMZVs (and therefore of open-string graph functions)
already exists in the literature: it is based on their representation in terms
of iterated integrals of Eisenstein series (as will be explained later in
\secref{sec:basics}), followed by the construction given in Francis Brown's
papers \cite{Brown:2017qwo} and \cite{Brown:2017qwo2}. Brown's construction
maps iterated integrals of Eisenstein series to certain modular-invariant
real-analytic functions whose coefficients are single-valued MZVs. So far,
however, it remains conjectural that modular graph functions are contained in
the image of this elliptic single-valued projection. We postpone the
investigation of the relation between our single-valued projection and Brown's
map to a sequel of the present work. 


\subsection{Outline}

Several techniques and previous results entering the construction of this work
are reviewed in \secref{sec:basics}. First, a short review is given on the
single-valued projection in the context of regular multiple zeta values, which
appear at string tree level. Second, $A$- and $B$-cycle versions of eMZVs will
be reviewed.  As it will turn out, modular transformations are facilitated by
representing $A$- and $B$-cycle elliptic multiple zeta values in the language
of iterated integrals over Eisenstein series. Modular graph functions including
some of their properties are introduced briefly. 

In \secref{sec:opensetup}, open-string graph functions are introduced. While
starting from the so-called $A$-cycle graph functions, it will turn out that
finally $B$-cycle functions are the objects necessary for the construction of
modular graph functions.

Once open-string graph functions are properly introduced, the comparison with
modular graph functions can happen, and it is presented in
\secref{sec:openVSclosed}. Using several examples, we will finally arrive at a
set of rules relating open-string graph functions to modular graph functions.
This is first of all done at the level of the relations and differential equations in the
modular parameter satisfied by the respective graph functions, see \subsecref{sec:ADrels}
and \subsecref{sec:modezero}. From the resulting conjectures, modular graph functions 
can be obtained from their open-string counterparts up to integration constants.
Moreover, since eMZVs are related to what is 
believed to be a single-valued version thereof in \subsecref{sec:funrules}, the construction is
believed to constitute a representation of an elliptic single-valued projection.
Still, in view of the issues with the shuffle multiplication of iterated Eisenstein integrals
detailed in \subsecref{sec:funrules}, parts of the operations in the tentative elliptic single-valued projection
await a reformulation in the future. 

Finally, non-planar analogues of the above open-string graph functions are
introduced in \secref{sec:npcyc}, generalizing our main result \eqn{punchline}
to admit the integrals for arbitrary non-planar four-point open-string amplitudes 
on the left-hand side.

Various details and examples can be found in the appendices.
In \appref{app:translate} we provide a table allowing to translate our
graphical notation to different notations for modular graph functions appearing
in earlier articles on the subject.


\section{Basics}
\label{sec:basics}

\subsection{Single-valued projection at tree level}
\label{ssec:sv}

In this section, we provide a brief review of the tree-level relations between
open- and closed-string amplitudes and identify them as the single-valued projection
in \eqn{tree07a}. 

Tree amplitudes among $n$ massless open-string states can be represented by
moduli-space integrals over punctured disks accompanied by partial amplitudes
of the Yang--Mills field theory \cite{Mafra:2011nv, Broedel:2013tta}. 
The moduli-space integrals read
\begin{equation}
Z(\rho(1,2,\ldots,n) \db \sigma(1,2,\ldots,n)) :=
\! \! \! \! \! \! \!
\int \limits_{D(\rho(1),\rho(2),\ldots,\rho(n))}\! \! {\dd z_1\, \dd z_2\cdots \dd z_n \over {\rm vol}(\gSL(2,\ZR))}
\frac{ \prod_{i<j}^n |z_{ij}|^{ -s_{ij}} }{\sigma( z_{12} z_{23}\ldots z_{n-1,n} z_{n,1})}\,,
\label{tree01}
\end{equation}
where $z_i$ are the positions of the punctures on the boundary of a disk.  The
integral $Z(\cdot \db \cdot)$ in \eqn{tree01} is labeled by two permutations
$\sigma,\rho \in S_{n}$ of the external legs $1,2,\ldots,n$ which govern the
cyclic product of $z_{ij} := z_{i} - z_j$ in the denominator 
(with $\sigma(z_{ij})=z_{\sigma(i),\sigma(j)}$) and the integration domains 
\begin{equation}
  D(1,2,\ldots,n) = \{ (z_1,z_2,\ldots,z_n) \in \ZR^n, \ -\infty < z_{1} < z_{2} < \ldots < z_{n} < \infty \} \,.
\label{tree02}
\end{equation}
The division by the inverse volume ${\rm vol}(\gSL(2,\ZR))$ of the
conformal Killing group can be implemented by dropping any three integrations,
fixing the respective
positions such as $(z_{1},z_{{n-1}},z_{n})=(0,1,\infty)$, and inserting the
compensating Jacobian $|z_{1,n-1}z_{1,n}z_{n-1,n}|$. Finally, the disk integrals
\eqn{tree01} depend on the lightlike momenta $k_j$ of the external states
$j=1,2,\ldots,n$ subject to momentum conservation
$\sum_{j=1}^n k_j=0$ through the dimensionless Mandelstam
variables\footnote{Throughout this work, we will follow the normalization
  convention for $\alpha'$ which is tailored to the closed-string setup. The
  fully accurate normalization of open-string quantities can be restored by
rescaling $\ap \rightarrow 4\ap$ \cite{Polchinski:1998rq}.}
\begin{equation}
s_{ij}  = - \frac{ \alpha' }{2}\,\,  k_{i} \cdot k_j  
\label{tree03}
\end{equation}
involving the inverse string tension $\alpha'$.

Tree-level amplitudes among massless closed-string states, in turn, comprise
moduli-space integrals over punctured spheres,
\begin{equation}
W(\rho(1,2,\ldots,n) \db \sigma(1,2,\ldots,n)) := \pi^{3-n} \int \limits_{\ZC^n} 
 {\dd^2 z_1\, \dd^2 z_2\cdots \dd^2 z_n \over {\rm vol}(\gSL(2,\ZC))}
\frac{ \prod_{i<j}^n |z_{ij}|^{-2s_{ij}} }{\sigma( z_{12} z_{23}\ldots   z_{n,1}) \ \rho( \bar z_{12} \bar z_{23}\ldots   \bar z_{n,1})} \,,
\label{tree04}
\end{equation}
where both permutations $\rho,\sigma \in S_{n}$ label a
cyclic product of $z_{ij}$ or their complex conjugates. The inverse volume $
{\rm vol}(\gSL(2,\ZC))$ suppresses three complex integrations and the
normalization factor $\pi^{3-n}$ is chosen for later convenience.

The low-energy regime of string amplitudes is encoded in the Taylor expansion
of the disk and sphere integrals around small values of the inverse string
tension $\ap$ and thus small values of the Mandelstam variables \eqref{tree03}.
The $w$'th order in the low-energy expansion beyond the respective field-theory
amplitudes gives rise to MZVs \eqn{tree05} of weight $w$
\cite{Terasoma, Brown:0606}, for instance
\begin{align}
s_{12} Z(1,2,3,4 \db 1,2,4,3) &=  \exp \Big( \sum_{n=2}^{\infty} \frac{ \zeta_n}{n}  \big[
s_{12}^n +s_{23}^n - (s_{12}+s_{23})^n
\big]
\Big) \label{tree05a} \\
s_{12} W(1,2,3,4 \db 1,2,4,3) &=  \exp \Big(2 \sum_{k=1}^{\infty} \frac{ \zeta_{2k+1}}{2k+1}  \big[
s_{12}^{2k+1} +s_{23}^{2k+1} - (s_{12}+s_{23})^{2k+1}
\big]
\Big)  \,.  \label{tree05b} 
\end{align}
Generic examples of multiplicity $n\geq 5$ also involve MZVs
of higher depth $r\geq 2$ \cite{Stieberger:2009rr,
Schlotterer:2012ny}, and the explicit polynomial dependence on the Mandelstam
invariants can for instance be computed\footnote{Earlier work on
$\ap$-expansions at $n=5,6,7$ points include \cite{Barreiro:2005hv,
Oprisa:2005wu, Stieberger:2006te, Stieberger:2007jv}, and the representation of
five-point integrals as hypergeometric functions has been exploited in the
all-order methods of \rcites{Boels:2013jua, Puhlfuerst:2015gta}.} via
polylogarithm manipulations \cite{Broedel:2013tta}, the Drinfeld associator
\cite{Broedel:2013aza} or a Berends--Giele recursion for a putative effective
field theory of bi-colored scalars \cite{Mafra:2016mcc}. A machine-readable
form of such results is available for download on the website
\cite{MZVWebsite}.

Closed-string integrals (\ref{tree04}) can in principle be assembled from
squares of open-string integrals (\ref{tree01}) through the
Kawai--Lewellen--Tye (KLT) relations \cite{Kawai:1985xq}. However, the KLT
formula obscures the cancellation of various MZVs from the open-string
constituents: From the all-order conjectures of \rcite{Schlotterer:2012ny},
closed-string integrals (\ref{tree04}) are expected to be single-valued
open-string integrals \cite{Stieberger:2013wea, Stieberger:2014hba},
\begin{equation}
W(\rho(1,2,\ldots,n) \db \sigma(1,2,\ldots,n)) = \sv \, Z(\rho(1,2,\ldots,n) \db \sigma(1,2,\ldots,n)) \,.
\label{tree06}
\end{equation}
The MZVs in the image of the single-valued projection \sv$(\ldots)$ are precisely the
single-valued MZVs described in \eqns{tree07a}{tree07} above -- in
agreement with the four-point examples \eqns{tree05a}{tree05b}. As can be seen
from  \eqn{tree06}, the \sv-projection trades the integration domain of
the disk integral \eqn{tree01} for an antiholomorphic cyclic denominator of a
sphere integral (\ref{tree04}).

\subsection{\texorpdfstring{$A$}{$A$}- and \texorpdfstring{$B$}{$B$}-cycle eMZVs and iterated Eisenstein integrals}
\label{ssec:eMZV}

Several versions of eMZVs have been used in different
contexts: when represented as special values of multiple elliptic
polylogarithms (defined by Brown and Levin in \cite{BrownLev}), they have made
an appearance in the evaluation of the sunrise integral, see for instance
\cite{Bloch:2013tra, Bloch:2014qca, Adams:2014vja, Adams:2015gva,
Adams:2017ejb, Remiddi:2013joa, Remiddi:2016gno, Remiddi:2017har,
Broedel:2017kkb,Broedel:2017siw
}, while when
represented as the coefficients of the elliptic associator (defined by Enriquez
in \cite{Enriquez:EllAss}), they have made an appearance in the one-loop
open-string amplitudes. The latter is the context that we consider in this
article; therefore our conventions are inspired by the string-theory setup in
\rcites{Broedel:2014vla, Broedel:2015hia, Broedel:2017jdo}. A further
comprehensive reference on eMZVs is Matthes's PhD thesis
\cite{Matthes:Thesis}.
\begin{figure}[h]
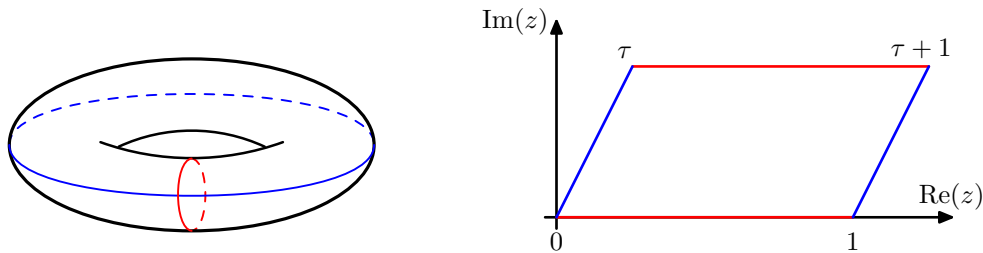

\begin{center}
\mpostuse{elliptic}
\caption{Parametrization of a torus as a lattice $\mathbb C/(\mathbb Z{+}\tau
  \mathbb Z)$ with modular parameter $\tau$ in the upper half plane and complex
  coordinate $z\cong z{+}1\cong z{+}\tau$. The homology cycle drawn in red is
  mapped to the unit interval $(0,1)$ and referred  to as the $A$-cycle.
  Accordingly, the second homology cycle mapped to the path from $0$ to $\tau$
is known as the $B$-cycle.}
\label{figureone}
\end{center}
\end{figure}
$A$-cycle eMZVs are defined as iterated integrals over the unit interval
\begin{align}
  \label{eqn:defommA}
\omm_A(n_1,n_2,\ldots,n_r \db \tau) &:= \! \! \! \! \! \! \! \! \! \int \limits_{0 \leq z_1\leq z_2\leq \ldots\leq z_r\leq  1} \! \! \! \! \! \! \! \! \!  f^{(n_1)}(z_1,\tau) \, \dd z_1 \, f^{(n_2)}(z_2,\tau) \, \dd z_2 \, \ldots \, f^{(n_r)}(z_r,\tau) \, \dd z_r \,,
\end{align}
where the integration path is taken to be on the real
line\footnote{Homotopy-invariant completions of the integrands in
\eqn{eqn:defommA} are known from \rcite{BrownLev}.}.  Using the parametrization
of the torus in \figref{figureone}, the integration domain in \eqn{eqn:defommA}
corresponds to the $A$-cycle and justifies the term ``$A$-cycle eMZVs''.
Accordingly, iterated integrals along the $B$-cycle connecting the points $0$
and $\tau$ in \figref{figureone} give rise\footnote{We think of
\eqn{eqn:defommB} as an integral over the straight path
$[0,\tau]\subset\mathbb{C}$. Again, these integrals are not homotopy invariant,
and their relation with the homotopy invariant version known from
\rcite{BrownLev} is more subtle than in the $A$-cycle case. The interested reader is
referred to \cite{Matthes:Thesis}.} to $B$-cycle eMZVs
\begin{align}
  \label{eqn:defommB}
\omm_B(n_1,n_2,\ldots,n_r \db \tau) &:= \! \! \! \! \! \! \! \! \! \! \! \int \limits_{0 \leq z_1\leq z_2\leq \ldots\leq z_r\leq  \tau} 
\! \! \! \! \! \! \! \! \! \! \!  f^{(n_1)}(z_1,\tau) \, \dd z_1 \, f^{(n_2)}(z_2,\tau) \, \dd z_2 \, \ldots \, f^{(n_r)}(z_r,\tau)\, \dd z_r \, .
\end{align}
The doubly-periodic integration kernels $f^{(n)}$ in
\eqns{eqn:defommA}{eqn:defommB} are defined by their generating series
\cite{Broedel:2014vla, Broedel:2015hia},
\begin{equation}
 \exp \bigg( 2\pi i \alpha \frac{ \Im(z)}{\Im(\tau)} \bigg) \frac{ \theta'(0,\tau) \theta(z+\alpha,\tau)}{\theta(z,\tau) \theta(\alpha,\tau)} = \sum_{n=0}^{\infty} \alpha^{n-1} f^{(n)}(z,\tau) \, ,
 \label{eqn:Kron}
\end{equation}
where $\theta(z,\tau)$ denotes the odd Jacobi theta function, and the simplest
instances are $f^{(0)}(z,\tau) = 1$ as well as $f^{(1)}(z,\tau) = \partial_z
\log \theta(z,\tau) + 2\pi i \frac{ \Im (z) }{\Im(\tau)}$.
We refer to the number $r$ of entries of eMZVs and the quantity
$n_1+n_2+\ldots+n_r$ as their {\it length} and {\it weight}, respectively.
Furthermore, the number of non-zero entries $n_j \neq 0$ of eMZVs will be
referred to as their {\it depth}.

$B$-cycle eMZVs can be obtained from $A$-cycle eMZVs by the modular
$S$-transformation, which sends $\tau\to-\frac{1}{\tau}$,
\begin{equation}
  \label{eqn:fromAtoB}
  \omm_A(n_1,n_2,\ldots,n_r \db\!-\tfrac{1}{\tau}) = \tau^{n_1+n_2+\ldots+n_r - r} \omm_B(n_1,n_2,\ldots,n_r \db \tau) \,.
\end{equation}
Since the restriction of the kernels $f^{(n)}$ to the real line admits a
Fourier-expansion in $q = \ee^{2\pi i \tau}$ spelt out in subsection 3.3.3 of
\rcite{Broedel:2014vla}, the same is true for $A$-cycle eMZVs in
\eqn{eqn:defommA}, and one can prove that the coefficients are given by
$\ZQ[(2\pi i)^{\pm 1}]$-linear combinations of MZVs \cite{Enriquez:Emzv}.

By contrast, $B$-cycle eMZVs have the more complicated behavior near the
cusp $\tau\to i\infty$ (or $q\to 0$) \cite{Enriquez:Emzv, ZerbiniThesis},
\begin{equation}\label{B-cycleAsymptExp}
\omm_B(n_1,n_2,\ldots,n_r \db \tau) =
\sum_{l=1-n_1-\cdots -n_r}^r \! \! \!  \tau^l \  \sum_{k=0}^{\infty} b_{k,l}(n_1,n_2,\ldots,n_r) \, q^k  \ , \ \ \ \ \ \ 
n_1,n_r\neq 1 \ ,
\end{equation}
where the coefficients $b_{k,l}(n_1,n_2,\ldots,n_r)$ are $\ZQ[(2\pi
i)^{\pm 1}]$-linear combinations of MZVs. In the resulting
expansion for $S$-transformed $A$-cycle eMZVs
\begin{equation}\label{Texpansion}
\omm_A(n_1,n_2,\ldots,n_r \db\!-\tfrac{1}{\tau})=\sum_{l=1-r}^{n_1+\cdots +n_r}(2\pi i \tau)^l\sum_{k= 0}^\infty c_{k,l}(n_1,n_2,\ldots ,n_r)q^k  \ , \ \ \ \ \ \ 
n_1,n_r\neq 1  \ ,
\end{equation}
it is crucial for later purposes to note that the coefficients
$c_{k,l}(n_1,n_2,\ldots,n_r)$ are $\ZQ$-linear (rather than $\ZQ[(2\pi i)^{\pm
1}]$-linear) combinations of MZVs. As will be proven in \appref{app:prf}, all
extra powers of $\pi$ can been absorbed into powers of $2\pi i \tau$ in
\eqn{Texpansion}. 


\subsubsection{Elliptic iterated integrals}

In the same way as MZVs descend from multiple polylogarithms at unit argument,
$A$-cycle eMZVs defined in \eqn{eqn:defommA} are special cases of elliptic
iterated integrals subject to the recursive definition \cite{Broedel:2014vla}
\begin{equation}
  \GLarg{n_1 &n_2 &\ldots &n_r}{a_1 &a_2 &\ldots &a_r}{z \db \tau} :=
  \int^z_0 \dd t \, f^{(n_1)}(t-a_1,\tau) \,
  \GLarg{n_2 &\ldots &n_r}{a_2 &\ldots &a_r}{t \db \tau}
\label{defgam}
\end{equation}
with initial condition $\Gamma\,(;z \db \tau)=1$, integration path along the real
line and real upper limit $z$. Accordingly,
\begin{equation}
\omm_A(n_1,n_2,\ldots,n_r \db \tau) =  \GLarg{n_r &\ldots &n_2 &n_1}{0 &\ldots &0 &0}{1 \db \tau} \,.
 \label{toomega}
\end{equation}
The integrals defined in \eqn{defgam} above are not homotopy invariant.
However, as discussed in \rcite{BrownLev} (see also subsection 3.1 of
\rcite{Broedel:2014vla}), every integral $\GLarg{n_1 &n_2 &\ldots &n_r}{a_1
						     &a_2 &\ldots
						     &a_r}{z\db \tau}$ can be
lifted to a homotopy invariant integral. Thus, despite the lack of homotopy
invariance, various manipulations are still allowed for the integrals defined
in \eqn{defgam}. In particular, as will become important for later
computations, differential equations in $a_i$ acting on the iterated elliptic
integrals defined in \eqn{defgam} can be used to eliminate any additional
occurrences of the argument $z$ on the left of the semicolon
\cite{Broedel:2014vla}, for instance
\begin{align}
  \GLarg{n}{z}{z \db \tau}&=(-1)^n\GLarg{n}{0}{z \db \tau} \\
\GLarg{1 &0 &1}{z &0 &0 }{z\db \tau}&=
2\GLarg{0 &0 &2}{0 &0 &0 }{z\db \tau}
+\GLarg{0 &2 &0}{0 &0 &0}{z\db \tau}
-2\GLarg{0 &1 &1}{0 &0 &0 }{z\db \tau}
+\zeta_2 \GLarg{0}{0}{z \db \tau}  \,.
\label{appG2}
\end{align}

\subsubsection{Iterated Eisenstein integrals}\label{SectionItEisInt}

Given that the differential equation in \appref{app:prf.2} allows to relate
eMZVs to Eisenstein series, it is natural to represent them in terms of
iterated integrals in $\tau$ (or $q=\ee^{2\pi i \tau}$), see
\rcite{Broedel:2015hia} for the detailed formalism of iterated Eisenstein
integrals\footnote{In \rcite{Broedel:2015hia}, a slightly different convention
  for iterated Eisenstein integrals has been employed. Named $\gm$ and $\gmz$,
  they differ from the objects $\ce$ and $\cez$ defined in
  \eqns{eqn:iteis}{eqn:iteis0} by powers of $2\pi i$ and can be related via
\begin{align*}
  \gm(k_1,k_2,\ldots,k_r;\tau) &= (2 \pi i)^{k_1+\cdots+k_r-2r} \ce(k_1,k_2,\ldots,k_r;\tau)\,\nnl
  \gmz(k_1,k_2,\ldots,k_r;\tau) &= (2 \pi i)^{k_1+\cdots+k_r-2r} \cez(k_1,k_2,\ldots,k_r;\tau)\,.
\end{align*}
Please see \appref{app:conv} for further details of our conventions.},
\begin{align}
  \ce(k_1,k_2,\ldots,k_r;\tau) &:= 2\pi i \int^{i\infty}_{\tau} \dd \tau_r \  \frac{\GG{k_r}(\tau_r)}{(2\pi i)^{k_r}}  \ce(k_1,k_2,\ldots,k_{r-1};\tau_r) 
 \notag \\ 
 &\phantom{:} = -\int^q_0 \dlog q_r \, \frac{\GG{k_r}(q_r)}{(2\pi i)^{k_r}}   \, \ce(k_1,k_2,\ldots,k_{r-1};q_r)  \label{eqn:iteis} \\
	&\phantom{:}=  (-1)^r \intl_{0\leq q_1\leq q_2\leq \ldots \leq q_r\leq q} 
  \dlog q_1\cdots \dlog q_r \, \frac{\GG{k_1}(q_1)}{(2 \pi i)^{k_1}}\cdots\frac{\GG{k_r}(q_r)}{(2 \pi i)^{k_r}} \ \nnl
  \cez(k_1,k_2,\ldots,k_r;\tau) & :=
  2\pi i \int^{i\infty}_{\tau} \dd \tau_r \  \frac{\GGz{k_r}(\tau_r)}{(2\pi i)^{k_r}}  \ce(k_1,k_2,\ldots,k_{r-1};\tau_r) 
  \notag \\
  &\phantom{:}= -\int^q_0 \dlog q_r \,  \frac{\GGz{k_r}(q_r)}{(2\pi i)^{k_r}}  \, \cez(k_1,k_2,\ldots,k_{r-1};q_r)   \label{eqn:iteis0}\\
	&\phantom{:}=  (-1)^r \intl_{0\leq q_1\leq q_2\leq \ldots \leq q_r\leq q}  
  \dlog q_1\cdots \dlog q_r \, \frac{\GGz{k_1}(q_1)}{(2 \pi i)^{k_1}}\cdots\frac{\GGz{k_r}(q_r)}{(2 \pi i)^{k_r}}  \ .  \notag
\end{align}
The recursion starts with $\ce(;\tau)=\cez(;\tau)=1$, and the non-constant
parts of Eisenstein series are defined as
 \begin{equation}
 \GGz{2n}(\tau)= \GG{2n}(\tau) - 2
\zm_{2n} \,, \ \ \ \ \ \ 
\GG{0}(\tau)=\GGz{0}(\tau)=-1 \, 
\label{GGk0}
\end{equation}
with $n\in \ZN^+$. Our conventions for Eisenstein series $\GG{k}$ are listed in
\appref{app:conv}, and we will interchangeably refer to the argument of
$\GG{k}$, $\GGz{k}$ and their iterated integrals by $\tau$ or $q$.  For both
$\ce(k_1,k_2,\ldots,k_r;\tau)$ and $\cez(k_1,k_2,\ldots,k_r;\tau)$ in
\eqns{eqn:iteis}{eqn:iteis0}, we will refer to the number of non-zero entries
($k_j\neq 0$) as the \textit{depth} of the respective iterated Eisenstein
integral (similar to the terminology for eMZVs). 

Throughout this article, the endpoint divergences of the above integrals as
$q_1\rightarrow 0$ are understood to be shuffle-regularized through the
tangential-basepoint prescription described in \rcite{Brown:mmv} with the net
effect $\int^q_0 \frac{ \dd q_1}{q_1} = \log q$. The iterated Eisenstein
integrals $\cez(k_1,\ldots,k_r;q)$ with $k_1\neq 0$ do not need to be regularized
and have the following Fourier-expansion (cf.~eq.~(4.62) of
\rcite{Broedel:2015hia}):
\begin{align}
&\cez(k_1,0^{p_1-1},k_2,0^{p_2-1},\ldots,k_r,0^{p_r-1};q) =(-2)^r
  \bigg(  \prod_{j=1}^{r} \frac{ 1 }{(k_j-1)!} \bigg)
\label{qgamma1}\\
& \ \ \ \ \ \ \ \times \sum_{m_i,n_i=1}^{\infty} \frac{m_1^{k_1-1} m_2^{k_2-1} \ldots m_r^{k_r-1}  q^{m_1n_1+m_2n_2+\ldots +m_rn_r}}{(m_1 n_1)^{p_1} (m_1n_1+m_2n_2)^{p_2} \ldots (m_1n_1+m_2n_2+\ldots +m_rn_r)^{p_r}}  \,, \notag
\end{align}
where $k_j\neq 0$ and $0^p$ is a shorthand for $p$ successive zeros $0,0,\ldots,0$. 
The conversion of $A$-cycle eMZVs to iterated Eisenstein
integrals therefore provides an easy way to find  their functional dependence
on $q$ and, by the linear independence of $\ce$ with different labels
\cite{Nilsnewarticle, Brown:2017qwo2}, exposes their relations
\cite{Broedel:2015hia}. 

The iterated Eisenstein integrals in \eqn{eqn:iteis} are linear combinations of
products of powers of $\tau$ and the objects
\begin{equation}
\GGG{j_1 &j_2 &\ldots &j_r}{k_1 &k_2 &\ldots &k_r}{\tau} := 
\int_{\tau}^{i\infty} \dd \tau_r \ \tau_r^{j_r} \, \GG{k_r}(\tau_r)  \, \GGG{j_1 &\ldots &j_{r-1}}{k_1 &\ldots &k_{r-1}}{\tau_r} \, ,
\label{prevBrown}
\end{equation}
where $k_i$ are even positive integers, $j_i$ are non-negative integers and 
$\GGG{}{}{\tau}=1$. The results of Brown
\cite{Brown:mmv} on the integrals \eqn{prevBrown} will be used to express 
the modular $S$-transformations $\ce(k_1,k_2,\ldots,k_r;{-}\frac{1}{\tau})$
in terms of iterated Eisenstein integrals at argument $\tau$, powers of $\tau$ 
and $\ZQ[(2\pi i)^{\pm 1}]$-linear combinations of MZVs. For $k_i\neq 0$, 
one recovers 
\begin{equation}
\GGG{0 &0 &\ldots &0}{k_1 &k_2 &\ldots &k_r}{\tau}
=  \prod_{j=1}^r (2\pi i)^{k_j-1} \ce(k_1,k_2,\ldots,k_r;\tau) \, ,
\end{equation}
and the general dictionary between the two types of iterated Eisenstein integrals
\eqns{eqn:iteis}{prevBrown} is described in \secref{sec:evalB} below. The
number $r$ of integrations in \eqn{prevBrown} will be referred to as the
\textit{depth} of Brown's iterated Eisenstein integrals, and it is compatible
with the notion of depth in their representation via $\ce$ in \eqn{eqn:iteis}.

Given a suitable regularization scheme, all objects defined as iterated
integrals naturally satisfy shuffle relations. This applies in
particular to eMZVs, elliptic iterated integrals and iterated Eisenstein
integrals. Shuffle relations can be neatly explained by reorganizing
the higher-dimensional integration domains and read for the example 
of iterated Eisenstein integrals:
\begin{equation}
  \label{shuffle}
\ce(0,0;\tau) \ce(4;\tau) = \ce(0,0,4;\tau)+\ce(0,4,0;\tau) +\ce(4,0,0;\tau)\, .
\end{equation}
\subsection{Modular graph functions}
\label{ssec:modular}

The definition of modular graph functions \cite{DHoker:2015wxz} is motivated by
the low-energy expansion of the modular invariant integral 
\begin{equation}
  M_n^\closed(s_{ij}\db\tau) := \int \dd\mu_n^\closed(\tau) \exp \left( \sum_{i<j}^n s_{ij} G_{ij}(\tau) \right) \, ,
\label{Acyc1}
\end{equation}
which appears in one-loop amplitudes of the closed superstring
\cite{Green:1982sw, Richards:2008jg} and gives rise to the right-hand side of
the correspondence in \eqn{tree3141}. The Green function $G_{ij}(\tau) :=
G(z_i,z_j;\tau)$ on the torus is defined below, and the integration measure for
$n$ external states reads
\begin{equation}
  \int \dd \mu_n^\closed(\tau) =  \frac{1}{\Im(\tau)^{n-1}} \int_{T(\tau)} \dd^2 z_2 \int_{T(\tau)} \dd^2 z_3\ldots \int_{T(\tau)} \dd^2 z_n 
\label{Acyc12}
\end{equation}
with $z_1=0$. The  $z_j$ are to be integrated over a torus $T(\tau)$ of modular parameter
$\tau$, and the above measure is normalized such that $\int_{T(\tau)} \dd^2
z=\Im(\tau)$. The Green function is only defined up to an additive function of $\tau$,
and we will employ the representative
\begin{equation}
G_{ij}(\tau) = - \log \left| \frac{\theta_1 (z_{ij},\tau) }{\eta(\tau)}
\right|^2 - \frac{\pi}{2\Im(\tau)} (z_{ij}-\bar z_{ij})^2 \, ,\quad \ \ z_{ij}=z_i-z_j
\label{Acyc2}
\end{equation}
which vanishes upon integration over the torus
\begin{equation}
\int_{T(\tau)} \dd^2 z_i \ G_{ij}(\tau) = 0 \,.
\label{Acyc3}
\end{equation}
The low-energy expansion of \eqn{Acyc1} can be conveniently represented
graphically. After expanding the exponential in the integrand as a power series
and exchanging integration and summation, one can associate a graph to every
summand in the following way: each integration variable of \eqn{Acyc12} is
represented by a vertex, and each Green function $G_{ij}$ between vertices $i$
and $j$ is visualized by an edge \cite{Green:1999pv, Green:2008uj}
\begin{equation} 
  G_{ij} (\tau) = \,\,\mpostuse[align=c]{G1label} \,.
\label{mgf1}
\end{equation}
Then, property \eqn{Acyc3} implies the vanishing of one-particle reducible
graphs\footnote{One-particle reducible graphs are those which can be
disconnected by removing an edge.}, so the simplest contributions to the
low-energy expansion of \eqn{Acyc1} stem from two-vertex graphs with multiple
edges. The associated modular graph functions are given by
\begin{align}
  \Dcyc{G2}&= \int \dd \mu_2^\closed \ G_{12}^2,\quad \ \ 
  \Dcyc{G3}=\int  \dd \mu_2^\closed \, G_{12}^3,\quad \ \ 
  \Dcyc{G4} =\int  \dd \mu_2^\closed \, G_{12}^4  \, ,
  \label{mgf2}
\end{align}
and we will employ a graphical labeling for their generalizations to one-particle 
irreducible graphs with multiple vertices, e.g.
\begin{align}
\Dcyc{G111} &= \int  \dd \mu_3^\closed\, G_{12} G_{13} G_{23} \, ,\quad \ \ 
\Dcyc{G211} = \int  \dd \mu_3^\closed\, G_{12}^2 G_{13} G_{23}   \, , \label{mgf3}\\
\Dcyc{G1111}& = \int  \dd \mu_4^\closed\, G_{12} G_{23} G_{34} G_{41} \, , \quad \ \ 
\Dcyc{GC321} = \int  \dd \mu_5^\closed\, G_{13}G_{34}G_{42} G_{15} G_{52} G_{12} \, . \notag 
\end{align}
We suppress the dependence on $\tau$ in \eqns{mgf2}{mgf3} as well as in later
equations.

The number of edges in the graphical representation equals the \textit{weight}
of a modular graph function. A translation between graphs at higher weight and
their names in \rcites{Green:2008uj, Green:2013bza} is provided in
\tabref{tab:graphs} in \appref{app:translate}. In terms of modular graph
functions, the $\ap$-expansion of the four-point integral \eqn{Acyc1} reads
\begin{align}
  M_4^\closed(s_{ij}\db\tau) \ &= \ 1 + 2\Dcyc{G2} (s_{12}^2 +s_{12}s_{23} +s_{23}^2) + ( \Dcyc{G3} + 4  \Dcyc{G111}) s_{12}s_{23} s_{13} \notag \\
&\quad+ \frac{1}{6} \Big(\Dcyc{G4} +9 \Dcyc{G2}^2 + 6 
\Dcyc{G1111}\Big)(s_{12}^2 +s_{12}s_{23} +s_{23}^2)^2  \label{Astructure}  \\
&
\quad+\frac{1}{12} 
\Big( \Dcyc{G5}  +48 \Dcyc{G111}\Dcyc{G2} + 12 \Dcyc{G2111} \nnl
  &\qquad\qquad- 12 \Dcyc{G221} + 16 \Dcyc{G311} + 14 \Dcyc{G3} \Dcyc{G2}  - 
24 \Dcyc{GC221}\Big)  \notag \\
& \qquad\quad\qquad\times s_{12}s_{23} s_{13} (s_{12}^2 +s_{12}s_{23} +s_{23}^2)  +\CO(\ap^6)  \,, \notag
\end{align}
where we have used the relations $s_{12}=s_{34}$, $s_{14}=s_{23}$ and
$s_{13}=s_{24}=-s_{12}-s_{23}$ among
four-particle Mandelstam variables.
Since $M_4^\closed(s_{ij}\db\tau)$ is the only integral
contributing to the four-point amplitude, the one-loop contribution to
$D^{2w}R^4$ operators in the effective action follows from integrating
\eqn{Astructure} over the fundamental domain with respect to~$\tau$
\cite{Green:1999pv, Green:2008uj, DHoker:2015gmr}. Closed-string one-loop
amplitudes for $n\geq 5$ points, however, involve a variety of additional
integrals besides $M_n^\closed(s_{ij}\db\tau)$ \cite{Richards:2008jg,
Green:2013bza, Mafra:2016nwr, Basu:2016mmk}. Similarly,
one-loop amplitudes involving massless states of the heterotic string will involve
more general integrals \cite{Lerche:1987qk, Stieberger:2002wk, Dolan:2007eh, Basu:2017nhs}.

The complexity of modular graph functions is correlated with the number of
loops in its graphical representation. We will later on define a notion of depth 
for modular graph functions which relates to the depth of iterated Eisenstein integrals
and which is conjecturally bounded from above by the loop order of the graph.
One-loop graphs give rise to the simplest class of modular graph functions: 
These are non-holomorphic Eisenstein series,
\begin{equation}
  \Dcyc{G2} = \EE{2} \, ,\quad \ \ \Dcyc{G111} = \EE{3} \, ,\quad \ \ 
  \Dcyc{G1111}=\EE{4} \, ,\quad  \ \ \Dcyc{G11111}=\EE{5} \, , \ \ \ldots
\label{Acyc10g}
\end{equation}
which are defined by the lattice sums
\begin{align}
  \EE{k}(\tau) &:=\bigg(\frac{\Im(\tau)}{\pi}\bigg)^k\suml_{(m,n)\neq(0,0)} \frac{1}{|m+\tau n|^{2k}}  \label{nheis} \\
	    &\phantom{:}=
  \nhehat_k(y) - 8y(2k-1)!\suml_{j=0}^{k-1}\binom{2k-2-j}{k-1}\frac{1}{j!}(4y)^{j-k}\,\Re[\cez(2k,\underbrace{0,\ldots,0}_{2k-2-j};q)] \notag
\end{align}
with $y=\pi\Im(\tau)$, Bernoulli numbers $B_{2k}$ and 
\begin{equation}
  \nhehat_k(y) = (-1)^{k-1} \frac{ B_{2k} }{(2k)!} (4y)^k + \frac{ 4 (2k-3)! }{(k-2)! \, (k-1)!} \, \zeta_{2k-1} (4y)^{1-k} \,.
\label{eisen1}
\end{equation}
For generic modular graph functions, a lattice-sum representation generalizing
the first line of \eqn{nheis} can be straightforwardly deduced from the
Fourier-expansion of the Green function \eqn{Acyc2} with respect to $\frac{ \Im z}{\Im
\tau}$ \cite{Green:1999pv},
\begin{equation}
G_{ij}(\tau) = {\Im \tau \over \pi} \suml_{(m,n)\neq(0,0)} \frac{\ee^{2\pi i(n\alpha_{ij} - m \beta_{ij})}}{|m+\tau n|^{2}}   \ , \ \ \ \ \ \
z_{ij} = \alpha_{ij} + \tau \beta_{ij} \ , \ \ \ \ \ \
\alpha_{ij},\beta_{ij} \in \mathbb R \ .
\label{grfct}
\end{equation}
However, the $q$-expansions of modular graph functions beyond $\EE{k}$ have not
been spelt out in the literature before, and we will propose new results in
terms of iterated Eisenstein integrals $\cez$ with $q$-expansion \eqn{qgamma1}
in \secref{sec:modezero}. 

\subsubsection{Laurent polynomials in the zero modes}
\label{ssec:modularZERO}

Modular graph functions associated with a one-particle irreducible graph ${\cal
G}$ admit a double expansion of the form 
\begin{equation}\label{ModGraphExpansion}
\DcycLetter[\CG]= \sum_{m,n=0}^{\infty} c^{{\cal
G}}_{m,n}(y)q^m \bar q^n,
\end{equation}
where the coefficients $c^{\CG}_{m,n}(y)$ are Laurent polynomials in
$y=\pi\Im(\tau)$ of maximum degree equal to the number of edges (or weight) $w$
of $\cal G$ and minimum degree $1-w$ \cite{Zerbini:2015rss}. A variety of
results is available on the polynomial $c^{\CG}_{0,0}(y) =:
\smallDcycLetter[\CG]$ which describes the behavior of the corresponding
modular graph function at the cusp $\tau \rightarrow i\infty$. In abuse of
nomenclature, the polynomial $\smallDcycLetter[\CG]$ will be referred to
as the \textit{zero mode}.  Apart from the zero modes $\nhehat_k(y)$ for the
polygonal graphs in \eqn{eisen1}, the results to be compared with an
open-string setup below read \cite{Green:2008uj, DHoker:2016quv} 
\begin{align}
\smallDcyc{G211} &= \frac{ 2y^4}{14175} + 
\frac{y \zeta_3}{ 45}+ \frac{5 \zeta_5}{12y} - \frac{ \zeta_3^2}{4y^2}  
+ \frac{ 9 \zeta_7}{16y^3} 
\label{laur1}
\\
\smallDcyc{G2111} &= \frac{ 2y^5 }{155925} + \frac{ 2  y^2\zeta_3}{945} - \frac{ \zeta_5}{180} + \frac{ 7 \zeta_7}{16y^2} - \frac{ \zeta_3 \zeta_5}{2y^3} + \frac{ 43 \zeta_9}{64y^4} 
\label{laur2}
\end{align}
at weight four and five as well as
\begin{align}
\smallDcyc{GC222} &=  \frac{38 y^6}{91216125} +  \frac{\zeta_7}{24 y} -  \frac{7 \zeta_9}{16 y^3} + 
    \frac{15 \zeta_5^2}{16 y^4} -  \frac{81 \zeta_{11}}{128 y^5}
    \label{laur3}
\\
\smallDcyc{GC411} &=  \frac{808 y^6}{638512875} +  \frac{ y^3 \zeta_3 }{4725} -  \frac{y \zeta_5}{1890} + 
    \frac{\zeta_7}{720 y} +  \frac{ 23 \zeta_9}{64 y^3} -  \frac{\zeta_5^2 + 
      30 \zeta_3 \zeta_7}{64 y^4} +  \frac{167 \zeta_{11}}{256 y^5}
      \label{laur4}
\\
\smallDcyc{GC321}&=  \frac{43 y^6}{58046625} +  \frac{y \zeta_5}{630} +  \frac{\zeta_7}{144 y} + 
    \frac{7 \zeta_9}{64 y^3} -  \frac{17 \zeta_5^2}{64 y^4} + 
    \frac{99 \zeta_{11}}{256 y^5}
    \label{laur5}
\\
\smallDcyc{GC2211}&=  \frac{103 y^6}{13030875} +  \frac{y^3 \zeta_3}{2025} +  \frac{y \zeta_5}{54}
 -  \frac{\zeta_3^2}{90} -  \frac{\zeta_7}{360 y} +  \frac{5 \zeta_3 \zeta_5}{12 y^2} 
 +  \frac{5 \zeta_9 {-} 48 \zeta_3^3}{288 y^3} +  \frac{14 \zeta_3 \zeta_7 {+} 25 \zeta_5^2}{32 y^4} 
 -  \frac{73 \zeta_{11}}{128 y^5}
 \label{laur6}
\end{align}
at weight six. While the above examples exclusively involve zeta values of
depth\footnote{The depth $r$ of MZVs $\zeta_{n_1,\ldots ,n_r}$ is not a
grading, thus it is often possible that the same MZV has two different
representations where the depth changes; for instance $\zm_3=\zm_{1,2}$.
Here, when we say that MZVs have a certain depth, we mean that they cannot be
written as polynomials in MZVs of lower depth.} one, some of the modular graph
functions at weight $w \geq 7$ were shown to involve single-valued MZVs at
depth three, for instance\footnote{There is a typo in the coefficient of
$y^{-4}$ in the corresponding formula in \rcite{Zerbini:2015rss}. }
\cite{Zerbini:2015rss}
\begin{align}
  \smallDcyc{G511} &=    \frac{ 62 y^7}{10945935} {+} \frac{ 2 y^4 \zeta_3 }{243} {+} 
\frac{ 119  y^2 \zeta_5 }{324} {+}  \frac{ 11 y \zeta_3^2  }{27} {+} \frac{ 21 \zeta_7}{16} 
  {+} \frac{46 \zeta_3 \zeta_5}{3 y} {+} \frac{7115 \zeta_9}{288 y^2} {-} \frac{25 \zeta_3^3}{2 y^2}
   {-} \frac{75 \zeta_5^2}{8 y^3}  \label{laurnew} \\
   &{+} \frac{1245 \zeta_3 \zeta_7}{16 y^3} {-} \frac{ 9 (\zeta_{3,5,3} - \zeta_3 \zeta_{3,5} )}{4y^4}
   {-} \frac{ 315\zeta_3^2 \zeta_5 }{8y^4} {-}\frac{ 9573\zeta_{11} }{128y^4} {+} \frac{ 2475 \zeta_5 \zeta_7}{32 y^5} {+} \frac{ 1125 \zeta_3 \zeta_9}{32 y^5}  {-} \frac{1575 \zeta_{13}}{32 y^6} 
\notag
\end{align}
can be rewritten as
\begin{multline}\label{laurnewsv}
  \smallDcyc{G511} =    \frac{ 62 y^7}{10945935} {+} \frac{ y^4 \zeta^{\rm sv}_3 }{243} {+} 
\frac{ 119  y^2 \zeta^{\rm sv}_5 }{648} {+}  \frac{ 11 y (\zeta^{\rm sv}_3)^2  }{108} {+} \frac{ 21 \zeta^{\rm sv}_7}{32} 
  {+} \frac{23 \zeta^{\rm sv}_3 \zeta^{\rm sv}_5}{6 y} {+} \frac{7115 \zeta^{\rm sv}_9}{576 y^2} {-} \frac{25 (\zeta^{\rm sv}_3)^3}{16 y^2}\\
   {-} \frac{75 (\zeta^{\rm sv}_5)^2}{32 y^3}
   {+} \frac{1245 \zeta^{\rm sv}_3 \zeta^{\rm sv}_7}{64 y^3} {-} \frac{ 9 \zeta^{\rm sv}_{3,5,3}}{8y^4}
   {-} \frac{ 405(\zeta^{\rm sv}_3)^2 \zeta^{\rm sv}_5 }{64y^4} {-}\frac{ 9573\zeta^{\rm sv}_{11} }{256y^4} {+} \frac{ 2475 \zeta^{\rm sv}_5 \zeta^{\rm sv}_7}{128 y^5} {+} \frac{ 1125 \zeta^{\rm sv}_3 \zeta^{\rm sv}_9}{128 y^5}  {-} \frac{1575 \zeta^{\rm sv}_{13}}{64 y^6}
  \ .
\end{multline}
It is conjectured that the coefficients of all Laurent
polynomials in \eqn{ModGraphExpansion} can be written in terms of
single-valued MZVs \cite{Zerbini:2015rss}. Finally, the zero modes in modular
graph functions associated with two-point or two-loop graphs are known in
closed form \cite{DHoker:2017zhq}.\footnote{See also D.~Zagier, \textit{Evaluation of S(m,n)}, appendix to ref.~\cite{Green:2008uj}.}

\subsubsection{Relations among modular graph functions}
\label{ssec:modularA}

Modular graph functions corresponding to different graphs are not independent
objects: they satisfy various relations involving (conjecturally only)
single-valued MZVs, starting with the relation proved by Don Zagier
\cite{ZagierStrings} (see also \cite{DHoker:2015gmr})
\begin{equation}
  0 = \Dcyc{G3} -\Dcyc{G111} - \zeta_3  \,.
    \label{eqn:weightthreerelation}
\end{equation}
At weight four and five, the techniques of \cite{DHoker:2015gmr,
DHoker:2015sve, DHoker:2016mwo} led to 

\begin{align}
\Dcyc{G4}  &=  24 \Dcyc{G211}  -   18 \Dcyc{G1111}  + 3  \Dcyc{G2}^2
\label{eqn:weightfourrelation} \\
  40 \Dcyc{G311} &= 300 \Dcyc{G2111} + 120 \Dcyc{G2}\Dcyc{G111} - 276 \Dcyc{G11111} + 7 \zeta_5  \label{compare10} \\
  \Dcyc{G5} &= 60 \Dcyc{G2111} 
  + 10 \Dcyc{G2}\Dcyc{G3} 
  - 48 \Dcyc{G11111} 
  + 16 \zeta_5 \label{compare11} \\
  10 \Dcyc{G221} &= 20 \Dcyc{G2111} - 4 \Dcyc{G11111} + 3 \zeta_5  \label{compare12} \\
 30 \Dcyc{GC221} &=  12 \Dcyc{G11111} +  \zeta_5  \,,
 \label{compare12a}
\end{align}
and the complete set of weight-six relations displayed in \appref{app:w6rel} has been
identified in \rcite{DHoker:2016quv}.

\subsubsection{Laplace equations among modular graph functions}
\label{ssec:modularB}

Various combinations and powers of modular graph functions are related through
a web of eigenvalue equations for the Laplacian $\Delta :=4 (\Im \tau)^2 \frac{
\partial^2 }{\partial \tau \partial \bar \tau}$. While the non-holomorphic
Eisenstein series \eqn{nheis} associated with one-loop graphs satisfy
\begin{equation}
  (\Delta - k(k{-}1)) \EE{k}  = 0  \,,
    \label{lapeq1}
\end{equation}
the systematics of inhomogeneous Laplace eigenvalue equations at two loops has
been described in \rcite{DHoker:2015gmr}, leading for instance to
\begin{align}
 (\Delta- 2) 
 \Dcyc{G211}  &= 9  \EE{4} - \EE{2}^2
 \label{unique11} \\
 (\Delta-6)
 \Dcyc{G2111}&= \frac{86}{5} \EE{5} - 4 \EE{2} \EE{3} + \frac{ \zeta_5}{10} 
  \label{higher07} 
  \end{align}
as well as
\begin{align}
  (\Delta-2)(4 \Dcyc{GC321} + \Dcyc{GC222}) &= 52 \EE{6} - 4 \EE{3}^2 \notag \\
  (\Delta-12)(6 \Dcyc{GC321} - \Dcyc{GC222}) &= 108 \EE{6} - 36 \EE{3}^2 \label{higher10} \\
  (\Delta-12)(6 \Dcyc{GC411} + \Dcyc{GC222}) &= 120 \EE{6} +  12 \EE{3}^2 - 36 \EE{2} \EE{4}  \,. \notag
\end{align}
Laplace equations for the tetrahedral topology \mpostuse[width=0.5cm,align=c]{tetrahedral} 
at three loops\footnote{See \cite{Basu:2015ayg, Basu:2016xrt} for earlier work
on Laplace equations of specific three-loop examples.} are known from
\rcite{Kleinschmidt:2017ege}; we will report on a new weight-six identity
involving less symmetric topologies in \secref{sec:newlapl}. 

\subsubsection{Cauchy--Riemann equations among modular graph functions}
\label{ssec:modularC}

An essential tool in deriving relations between modular graph functions is the
Cauchy--Riemann derivative
\begin{equation}
\nabla := 2i (\Im \tau)^2 \partial_\tau
\label{CR1}
\end{equation}
with $ \partial_\tau := \frac{ \partial }{\partial \tau}$,
which maps modular forms of weight $(0,w)$ to those of weight $(0,w{-}2)$. For
instance, repeated application of the Cauchy--Riemann derivative \eqref{CR1}
mediates between non-holomorphic and holomorphic Eisenstein series
\cite{DHoker:2016mwo}
\begin{equation}
  \Gamma(k)(\pi \nabla)^k \EE{k} = \Gamma(2k) (\Im ( \tau))^{2k} \GG{2k} \,. 
\label{CR2}
\end{equation}
At higher loop order, the Cauchy--Riemann equations 
\begin{align}
  (\pi \nabla)^3  \Dcyc{G211} &= \frac{9}{10} (\pi \nabla)^3 \EE{4} - 6 \Im(\tau)^4 \GG{4} \pi \nabla \EE{2}
\label{CR3}
\\
(\pi \nabla)^3 \Dcyc{G2111} &= \frac{43}{35} (\pi \nabla)^3 \EE{5} - 2 (\pi \nabla \EE{2}) (\pi \nabla)^2 \EE{3}
- 4 \Im(\tau)^4 \GG{4}  \pi \nabla \EE{3}
\label{CR4}
\end{align}
have been instrumental to prove the weight-four and weight-five relations in
eqs.~(\ref{eqn:weightfourrelation}) to (\ref{compare12a})
\cite{DHoker:2016mwo}. The same method has been applied in
\cite{DHoker:2016quv} to derive the weight-six relations in
\eqn{eqn:weightsixrelations} as well as selected relations at weight seven. 

Holomorphic Eisenstein series appear in both the Cauchy--Riemann derivatives
of modular graph functions and the $\tau$-derivative \eqn{eqn:tauder} of eMZVs. 
In \subsecref{ssec:CRD} below, we will report on a
correspondence between eqs.\ (\ref{CR2}) to (\ref{CR4}) and differential
equations of associated combinations of eMZVs.


\section{An open-string setup for graph functions}
\label{sec:opensetup}

In this section, we will describe an open-string setup mimicking the graphical
organization of the closed-string $\ap$-expansion in \subsecref{ssec:modular}.
Choosing auxiliary abelian open-string states, the permutation symmetry of the
closed-string integration measure in \eqn{Acyc12} can be implemented in an
open-string setup. As a consequence, external abelian states allow to rewrite
the low-energy expansion of open-string integrals without one-particle
reducible graphs. Having done so, the structure of the closed-string amplitude
\eqn{Astructure} equals that of the four-point integral for abelian
open-string states. 

The open-string analogues of the modular graph functions will be referred to as
``$A$-cycle graph functions'' and expressed in terms of the $A$-cycle eMZVs
introduced in \subsecref{ssec:eMZV}. Accordingly, the results of their modular 
$S$-transformation will be referred to as ``$B$-cycle graph functions'', and we
will introduce techniques to express them in terms of the same iterated Eisenstein
integrals as employed for $A$-cycle graph functions. These expressions for $B$-cycle
graph functions will be the starting
point for proposing an analogue of the single-valued projection from
\subsecref{ssec:sv} in the one-loop setup and furnish the left-hand side of the
correspondence in \eqn{tree3141}.

\subsection{Definition of \texorpdfstring{$A$}{$A$}- and \texorpdfstring{$B$}{$B$}-cycle graph functions}
\label{ssec:graphfunctions}
\subsubsection{Review of open-string \texorpdfstring{$\ap$}{$\ap$}-expansions}

The color-ordered one-loop amplitude of four non-abelian open-string states
reads\footnote{Given that the normalization of $\ap$ is
  tailored to the closed-string setup in this work, the expressions for $I_{\rm
  4pt}(1,2,3,4)$ given in \cite{Broedel:2014vla} is recovered from
  \eqn{Acyc111} by rescaling $\alpha' \rightarrow 4 \alpha'$. The definitions
  \eqns{tree03}{Acyc2} of the Mandelstam invariants and the Green function on
  the torus are identical to those of \cite{DHoker:2015gmr, DHoker:2015wxz,
  DHoker:2016mwo, DHoker:2016quv} to match the conventions of the references
  for closed-string integrals and modular graph functions. The normalization of
  $s_{ij}$ and $G_{ij}$ chosen in \cite{Broedel:2014vla, Broedel:2017jdo} can
  be obtained from \eqns{tree03}{Acyc2} by rescaling $s_{ij} \rightarrow -4
  s_{ij}$ and $G_{ij} \rightarrow -G_{ij}$, respectively.} 
\begin{align}
&I_{\rm 4pt}(1,2,3,4\db\tau)  :=   \! \! \!  \! \! \!  \! \! \! 
\int \limits_{0\leq z_2\leq z_3\leq z_4\leq 1}   \! \! \!  \! \! \!  \! \! \! \dd z_2 \ \dd z_3 \ \dd z_4 \
  \exp \left(\frac{1}{2} \sum_{i<j}^4 s_{ij} G_{ij}(\tau) \right)  \,,
\label{Acyc111}
\end{align}
with $z_1=0$. The integration domain corresponds to a single-trace contribution
of the non-abelian gauge-group generators\footnote{The contributions from
  cylinder- and M\"obius-strip diagrams to planar one-loop amplitudes are
  obtained by integrating \eqref{Acyc111} over $\tau \in i \mathbb R_+$ and
$\tau \in \tfrac{1}{2}+i \mathbb R_+$, respectively \cite{Green:1984ed}.}.
The open-string Green function $G_{ij}$ can be obtained from the closed-string
version in \eqn{Acyc2} by restricting to real arguments. Comparing with the
definition of elliptic iterated integrals in \eqn{defgam} and the form of the
integration kernel $f^{(1)}$, we find
\begin{equation}
  G_{ij}(\tau)=-2\GLarg{1}{0}{z_{ij}\db\tau} + k(\tau)= -2\GLarg{1}{z_j}{z_i\db\tau}-2\GLarg{1}{0}{z_j\db\tau}+ k(\tau) \,.
\label{gfopen}
\end{equation}
The iterated elliptic integrals $\Gamma$ in \eqn{gfopen} need regularization,
see e.g.~section 4.2.1 of \rcite{Broedel:2017jdo}, which leads to the
scheme-dependent quantity $k(\tau)$. The latter, however, does not depend on
$z_i,z_j$ and thus cancels out from \eqn{Acyc111} after using momentum
conservation $\sum_{i<j} s_{ij}=0$. We will suppress the dependence on $\tau$
henceforth.

The representation \eqn{gfopen} of the Green function has been used to
algorithmically perform the $\ap$-expansion of \eqn{Acyc111} in the framework
of eMZVs, leading to \cite{Broedel:2014vla}
\begin{align}
&I_{\rm 4pt}(1,2,3,4)  = \frac{1}{6} - 2 s_{13}  \omm_A(0,1,0,0)+  2  \omm_A(0,1,1,0,0)\,  \big( s_{12}^2  + s_{23}^2 \big) - 2  \omm_A(0,1,0,1,0) \,s_{12}s_{23}  \notag \\
&  \ \ \ \ \ \ \ \ \ \ \ \ \ \ \ \ \ \ \ \ \, - \, \beta_5 \, (s_{12}^3+2 s_{12}^2 s_{23} + 2s_{12} s_{23}^2+s_{23}^3) \, - \, \beta_{2,3} \, s_{12} s_{23}(s_{12}+s_{23}) \, + \, {\cal O}(\ap^4) 
\label{Acyc6}
\end{align}
with
\begin{align}
\beta_5 &=  \frac{4}{3} \, \big[  \omm_A(0,0,1,0,0,2)+ \omm_A(0,1,1,0,1,0) -  \omm_A(2,0,1,0,0,0) - \zeta_2  \omm_A(0,1,0,0) \big]  \\
\beta_{2,3} &=\frac{\zeta_3}{12}+\frac{8\,\zeta_2}{3} \omm_A(0,1,0,0)-\frac{5}{18} \omm_A(0,3,0,0)
\,.
\end{align}
As can be seen from the non-vanishing contribution at linear order, a single
Green function does {\it not} integrate to zero. This is true in general for
the non-abelian situation: one cannot find a constant $c(\tau)$ such that both
$G_{12}(\tau)+c(\tau)$ and the cyclically inequivalent $G_{13}(\tau)+c(\tau)$
integrate to zero simultaneously within \eqn{Acyc111}. Hence, in presence of
non-abelian open-string states, there is no analogue of the property
\eqn{Acyc3} which eliminates one-particle reducible graphs in the expansion.

\subsubsection{Open-string \texorpdfstring{$\ap$}{$\ap$}-expansion
for abelian states}

Switching from non-abelian to abelian open-string states amounts to
democratically combining all different possible integration domains in
\eqn{Acyc111} and to independently integrating each $z_j$ for $j=2,3,4$ 
over the unit interval. Hence, we will be interested in
symmetrized open-string integrals
\begin{equation}
M_n^{\rm open}(s_{ij} ) :=  \int \dd \mu^{\rm open}_n \ \exp \left( \sum_{i<j}^n s_{ij} \PP_{ij}  \right)
\label{Acyc4}
\end{equation}
with $z_1=0$ and an integration measure analogous to \eqn{Acyc12}:
\begin{equation}
  \int \dd \mu_n^{\rm open}  =  \int^1_0 \dd z_2 \int^1_0 \dd z_3\ldots \int^1_0 \dd z_n \,.
\label{Acyc12op}
\end{equation}
Momentum conservation has been used to trade the Green function \eqn{gfopen}
for\footnote{Note that the right-hand side of \eqn{Acyc5} does not match the
definition of $P_{ij}$ in \rcite{Broedel:2017jdo}:  The propagator of the reference does not satisfy
\eqn{Acyc7a}, as it does not include the term $\omm_A(1,0)$ of \eqn{Acyc5}.}
\begin{equation}
\PP_{1j}    =    \omm_A(1,0) - \GLarg{1}{0}{z_j}
\,, \ \ \ \ \ \ 
\PP_{ij}    = 
\omm_A(1,0) -  \GLarg{1}{z_j}{z_i} - \GLarg{1}{0}{z_j}\,,
\label{Acyc5}
\end{equation}
with $z_1=0$ and $i,j\neq 1$ (we have suppressed the dependence on $\tau$ from
the notation). Note that one can also swap the roles of $z_i$ and $z_j$ in the
rightmost expression since $P_{ij} = P_{ji}$. In analogy to the situation for
the quantity $k(\tau)$ in \eqn{gfopen}, the addition of $\omm_A(1,0)$ in
\eqn{Acyc5} does not contribute to the open-string integral \eqn{Acyc111} after
taking momentum conservation into account. However, including $\omm_A(1,0)$
into the propagator \eqn{Acyc5} ensures that an analogue of the crucial
identity \eqn{Acyc3} from the closed-string setup holds 
\begin{equation}
\int^1_0 \dd z_i \ \PP_{ij}  \ = \ 0 \,,
\label{Acyc7a}
\end{equation} 
as can be checked using the definition \eqn{defgam} of elliptic iterated
integrals. Then, the $\ap$-expansion of the four-point integral \eqn{Acyc4} for abelian
open-string states will be organized in terms of one-particle irreducible
graphs: each integration variable in \eqn{Acyc12op} is represented by a
vertex, and each propagator $P_{ij}$ in \eqn{Acyc5} between vertices $i$ and
$j$ is visualized by an undirected edge
\begin{equation} 
  P_{ij} \,  = \,\,\mpostuse[align=c]{G1label} \,.
\label{}
\end{equation}
In these conventions, the open-string analogue of \eqn{Astructure} reads
\begin{align}
  M_4^{\rm open}(s_{ij} )  &= \ 1 + 2\Acyc{G2} (s_{12}^2 +s_{12}s_{23} +s_{23}^2) + ( \Acyc{G3} + 4  \Acyc{G111}) s_{12}s_{23} s_{13} \notag \\
&\quad+ \frac{1}{6} \Big(\Acyc{G4} +9 \Acyc{G2}^2 + 6 
\Acyc{G1111}\Big)(s_{12}^2 +s_{12}s_{23} +s_{23}^2)^2 \notag  \\
&
\quad+\frac{1}{12} 
\Big( \Acyc{G5}  +48 \Acyc{G111}\Acyc{G2} + 12 \Acyc{G2111} \label{AAstructure} \\
  &\qquad\qquad- 12 \Acyc{G221} + 16 \Acyc{G311} + 14 \Acyc{G3} \Acyc{G2}  - 
24 \Acyc{GC221}\Big)  \notag \\
& \qquad\quad\qquad\times s_{12}s_{23} s_{13} (s_{12}^2 +s_{12}s_{23} +s_{23}^2)  +\CO(\ap^6)  \,, \notag
\end{align}
where the $A$-cycle graph function $\text{\textbf{A}}[\CG]$ associated
with a graph $\CG$ is defined in analogy with the corresponding modular
graph function $\text{\textbf{D}}[\CG]$
\begin{equation}
\text{\textbf{A}}[\CG]  :=  \text{\textbf{D}}[\CG] \, \Big|^{\dd \mu_n^{\rm closed} \ \rightarrow \ \dd \mu_n^{\rm open}}_{G_{ij} \ \rightarrow \ P_{ij}} \,,
\label{defAcyc}
\end{equation} 
for instance
\begin{align}
  \Acyc{G2}&= \int \dd \mu_2^\open \ P_{12}^2 \,,\quad \ \ 
\Acyc{GC321} = \int  \dd \mu_5^\open\, P_{13}P_{34}P_{42} P_{15} P_{52} P_{12} \, . \notag 
\end{align}
Again, the number of edges in the graphical representation equals the
\textit{weight} of an $A$-cycle graph function.

Finally, symmetrizing over the respective integration domains, the four-point
integral in the abelian case coincides with the symmetrization of \eqn{Acyc6},
\begin{equation}
  M_4^{\rm open}(s_{ij})  = \sum_{\rho \in S_3} I_{\rm 4pt}(1,\rho(2,3,4)) \,.
\label{symme}
\end{equation} 
In particular, up to the orders where $I_{\rm 4pt}(1,2,3,4)$ is available,
\eqn{symme} has been used as a consistency check for the explicit results for
the $A$-cycle graph functions in \eqn{AAstructure} to be obtained in the next
section.

Although the $n$-point amplitude of the open superstring involves many
integrals beyond \eqn{Acyc4} \cite{Tsuchiya:1988va, Stieberger:2002wk,
Broedel:2014vla, Mafra:2016nwr, Mafra:2017ioj}, we still want to study
$A$-cycle graph function with $n \geq 5$ vertices for the sake of their
parallels with modular graph functions.


\subsubsection{\texorpdfstring{$B$}{$B$}-cycle graph functions}

The open-string integral \eqn{Acyc111} and the measure \eqn{Acyc12op} are
expressed in a parametrization of the cylinder worldsheet, where one of the
boundary components is the $A$-cycle. By a modular transformation, this setup
is related to a parametrization of the boundary component through the path from
$0$ to $\tau$, i.e.~the $B$-cycle (cf.~\figref{figureone} in
\secref{ssec:eMZV}). In order to compare open-string quantities with modular
graph functions below, we will study the image of $A$-cycle graph functions
under the $S$-transformation $\tau \rightarrow -\frac{1}{\tau}$ (cf.~below
\eqn{eqn:defommB}),
\begin{equation}
 \text{\textbf{B}}[\CG] :=  \text{\textbf{A}}[\CG]  \, \big|_{ \tau \rightarrow -\frac{1}{\tau} }\,,
\label{bgraph}
\end{equation} 
which will be referred to as $B$-cycle graph functions, and can be expressed in
terms of $B$-cycle eMZVs by \eqn{eqn:fromAtoB}. Techniques for their
systematic evaluation in terms of $A$-cycle quantities $\cez(\ldots; \tau)$
with known $q$-expansion \eqn{qgamma1} will be discussed in \secref{sec:evalB}. The main
motivation to do this comes from the fact that the asymptotic expansion at the
cusp of $B$-cycle eMZVs \eqref{B-cycleAsymptExp} looks more suitable to be
compared with the asymptotic expansion of modular graph functions
(\ref{ModGraphExpansion}) than the simple Fourier expansion of their $A$-cycle
counterparts.


\subsection{Evaluating \texorpdfstring{$A$}{$A$}-cycle graph functions}
\label{sec:eval}

The representation of the propagator in \eqn{Acyc5} guarantees that the
low-energy expansion of open-string integrals \eqn{Acyc4} is expressible in
terms of elliptic iterated integrals. As will be argued below, there is no
bottleneck in algorithmically computing $A$-cycle graph functions of arbitrary
complexity by means of the techniques developed in \rcites{Broedel:2014vla,
Broedel:2015hia}.


\subsubsection{\texorpdfstring{$A$}{$A$}-cycle graph functions at weight two}

The simplest non-trivial $A$-cycle graph function at the second order of
\eqn{AAstructure} can be computed using the definition \eqn{defgam} of
elliptic iterated integrals,
\begin{align}
\Acyc{G2}  & =  \int^1_0 \dd z_2 \ \Big\{ \omm_A(1,0)^2
- 2 \omm_A(1,0) \Gamma\left( \begin{smallmatrix}
1 \\0
\end{smallmatrix} ;  z_2\right) +  \Gamma\left( \begin{smallmatrix}
1 \\0 
\end{smallmatrix} ;  z_2\right)^2 
\Big\}  \label{Acyc8}\\
& =  2\omm_A(1,1,0) -2 \omm_A(1,0)^2+ \omm_A(1,0)^2  = \omm_A(2,0,0) + \frac{ 5 \zm_2}{6}\,. \notag 
\end{align}
Here and below we have been using relations between eMZVs like $2\omm_A(1,1,0)
= \frac{5\zm_2}{6} + \omm_A(1,0)^2 + \omm_A(2,0,0)$, which can be found on
the website \cite{WWWe} along with various generalizations up to and including
length six.  In \eqn{Acyc8} as well as in all computations of $A$-cycle graph
functions below, the term $\omm_A(1,0)$ in the propagator \eqn{Acyc5} avoids
the appearance of divergent eMZVs.


\subsubsection{\texorpdfstring{$A$}{$A$}-cycle graph functions at weight three}

The $A$-cycle graph functions at the third order of \eqn{AAstructure} can be
computed via
\begin{align}
\Acyc{G3} &= \int \dd \mu_2^\open \ P_{12}^3 \notag \\
&= \int^1_0 \dd z_2 \ 
\Big\{ \omm_A(1,0)^3
- 3 \omm_A(1,0)^2 \Gamma\left( \begin{smallmatrix}
1 \\0
\end{smallmatrix} ;  z_2\right)
+ 3 \omm_A(1,0) \Gamma\left( \begin{smallmatrix}
1 \\0
\end{smallmatrix} ;  z_2\right)^2
- \Gamma\left( \begin{smallmatrix}
1 \\0
\end{smallmatrix} ;  z_2\right)^3
\Big\}
  \notag \\
&= -6 \omm_A(1,1,1,0) + 6 \omm_A(1,1,0) \omm_A(1,0) - 2 \omm_A(1,0)^3\notag \\
&
=  \frac{ \zeta_3}{2} + 8 \zm_2 \omm_A(0, 1, 0, 0) - \frac{1}{3} \omm_A(0, 3, 0, 0) \label{Acyc13} \\
\Acyc{G111}&= \ \int  \dd \mu_3^{\rm open} \ \PP_{12}\PP_{13}\PP_{23}  \notag \\
&=  \omm_A(1,0)^3 - 2 \int^1_0 \dd z_3 \int^{z_3}_0 \dd z_2 \
\Gamma\left( \begin{smallmatrix}
1 \\0
\end{smallmatrix} ;  z_2\right)
\Gamma\left( \begin{smallmatrix}
1 \\0
\end{smallmatrix} ;  z_3\right)
\,
\Big\{ \Gamma\left( \begin{smallmatrix}
1 \\0
\end{smallmatrix} ;  z_3\right)
+\Gamma\left( \begin{smallmatrix}
1 \\z_3
\end{smallmatrix} ;  z_2\right)
\Big\}
\notag \\
&=   \omm_A(1,0)^3 + 2\int^1_0 \dd z_3 \ \Gamma\left( \begin{smallmatrix}
1 \\0
\end{smallmatrix} ;  z_3\right) \GLarg{1 &0 &1}{z_3 &0 &0 }{z_3} \notag \\
&= 2 \zm_2 \omm_A(0, 1, 0, 0) - \frac{1}{3} \omm_A(0, 3, 0, 0)  \,.\label{Acyc13a}
\end{align}
In \eqn{Acyc13}, the relevant eMZV relation is 
\begin{align}
\omm_A({0, 1, 1, 1}) &= \frac{\zeta_3}{12} - \frac{ \zeta_2 }{4}  \omm_A(0,1) + \frac{1}{6}\omm_A(0,1)^3
+ \frac{ 1 }{36} \omm_A(0,3) \\
& \! \!  \! \!  \! \!  \! \!  \! \! + \frac{ 1 }{2}  \omm_A(0,1) \omm_A(0,0,2)
+4 \zm_2 \omm_A(0,0,0,1) - \frac{1}{6} \omm_A(0,0,0,3) \,,  \notag 
\end{align}
and the last step of \eqn{Acyc13a} involves the identity (\ref{appG2}) for
$\GLarg{1 &0 &1}{z_3 &0 &0 }{z_3}$ along with the eMZV relations from appendix
I.2 of \rcite{Broedel:2017jdo}. Moreover, in \eqns{Acyc13}{Acyc13a} we have
replaced the integration domains according to $\int^1_0 \dd z_2 \int^1_0 \dd
z_3 \rightarrow 2  \int^1_0 \dd z_3  \int^{z_3}_0 \dd z_2$, which is valid along
with any monomial $\PP_{12}^m \PP_{13}^n \PP_{23}^q$ due to the symmetry
$\PP_{ij}=\PP_{ji}$ of the propagator, i.e.~for any three-vertex diagram.


\subsubsection{Computing \texorpdfstring{$A$}{$A$}-cycle graph
functions at higher weight}

$A$-cycle graph functions with higher numbers of vertices $n$ can be
algorithmically computed by iterating the manipulations in \eqn{Acyc13a}.
Among other things, the recursive techniques of \cite{Broedel:2014vla} to
eliminate the appearance of the argument $z$ in the second line of the elliptic 
iterated integral $\GLarg{n_1 &n_2 &\ldots &n_r}{z &0 &\ldots &0 }{z}$ -- see
e.g.\ \eqn{appG2} -- play a key role. As will be explained in the following, $A$-cycle
graph functions with an arbitrary number of vertices or edges can always be
expressed in terms of eMZVs. 

In order to connect with the definition \eqref{defgam} of elliptic iterated
integrals, the integration region $[0,1]^{n-1}$ of the measure \eqn{Acyc12op}
has to be decomposed into simplicial cells defined by $0\leq z_{\rho(2)}\leq
z_{\rho(3)}\leq\ldots\leq z_{\rho(n)}\leq 1$ with $\rho \in S_{n-1}$. Using the
symmetry $P_{ij}=P_{ji}$ of the propagator, this ordering is equivalent to its
reversal $0\leq z_{\rho(n)}\leq \ldots\leq z_{\rho(3)}\leq z_{\rho(2)}\leq 1$,
that is, only $\frac{1}{2}(n{-}1)!$ inequivalent cells need to be considered.
Different cells benefit from different representations of the propagators,
e.g.\ in situations with $z_2<z_3$, it is preferable to use the expression 
\begin{equation}
\PP_{23}    = \omm_A(1,0)-  \Gamma\left( \begin{smallmatrix}
1 \\z_3
\end{smallmatrix} ;  z_{2}\right) - \Gamma\left( \begin{smallmatrix}
1 \\0
\end{smallmatrix} ;  z_{3}\right)
\ \ \ 
\textrm{rather than}
\ \ \ \PP_{23}    =  \omm_A(1,0)- \Gamma\left( \begin{smallmatrix}
1 \\z_2
\end{smallmatrix} ;  z_{3}\right) - \Gamma\left( \begin{smallmatrix}
1 \\0
\end{smallmatrix} ;  z_{2}\right)
\end{equation}
as done in \eqn{Acyc13a}. 

Compact expressions for $A$-cycle graph functions are tied to expressing the
eMZVs in terms of a basis over $\mathbb Q[(2\pi i)^{\pm 1}]$-combinations of MZVs.
For certain ranges of their length and weight, an exhaustive list of such
relations among eMZVs is available for download \cite{WWWe}, but already for
$A$-cycle graph functions at weight four, some of the intermediate steps exceed
the scope of this website.  In deriving the subsequent results on $A$-cycle
graph functions of weight $w\leq 6$, we have expressed the eMZVs in terms of
iterated Eisenstein integrals \eqn{eqn:iteis0} to automatically attain the
desired basis decomposition. Using this method, the divergent eMZV
$\omm_A(1,0)$ could be shown to drop out in all cases considered, which is a
strong consistency check for our calculational setup.


\subsubsection{\texorpdfstring{$A$}{$A$}-cycle graph functions at weight four and beyond}

The strategy outlined in the previous section gives rise to the following
expressions for the three $A$-cycle graph functions at weight four:
\begin{align}
\Acyc{G4}  &=    15 \omm_A({0, 0, 2})^2  - 30 \omm_A({0, 0, 0, 0, 4}) + 
3 \omm_A({0, 0, 4}) - 24 \omm_A({0, 0, 0, 2, 2}) \notag \\
& \ \ 
  - 48 \zm_2 \omm_A({0, 0, 0, 0, 2})
+13 \zm_2 \omm_A({0, 0, 2}) +\frac{343 \zm_4}{24} 
 \label{eis7}
 \\
\Acyc{G211} &=  \frac{1}{2} \omm_A({0, 0, 2})^2 - \frac{1}{2} \omm_A({0, 0, 0, 0, 4}) - 
 \omm_A({0, 0, 0, 2, 2})  \notag \\
 & \ \ + \frac{7 \zm_2}{3}   \omm_A({0, 0, 2}) - 
 14 \zm_2 \omm_A({0, 0, 0, 0, 2}) +  \frac{301 \zm_4}{180} 
 \label{eis8}
\\
\Acyc{G1111} &=   \omm_A(0,0,0,0,4) - \frac{1}{6} \omm_A(0,0,4) +  
 \frac{4 \zm_{2}}{3}  \omm_A( {0, 0, 2})   - 8 \zm_2 \omm_A({0, 0, 0, 0, 2})  + \frac{311 \zm_4}{360}  \,.\label{eis8a}
 \end{align}
At weight five there are six $A$-cycle graph functions, for example
\begin{align}
\Acyc{G2111}&=  \frac{1}{90} \omm_A(0, 5) + \frac{2}{3} \omm_A(0, 0, 0, 5) - \frac{1}{3} \omm_A(0, 0, 2, 3) 
+ 2 \omm_A(0, 3) \omm_A(0, 0, 0, 0, 2)  \notag \\
 &- 6 \omm_A(0, 0, 0, 0, 0, 5) + 
 2 \omm_A(0, 0, 0, 0, 1, 4) - \frac{\zm_2}{3}  \omm_A(0, 3)  + 
 \frac{8 \zm_2}{9} \omm_A(0, 0, 3, 0) \notag \\
 & + 
 24  \zm_2 \omm_A(0, 0, 0, 0, 0, 3) - 
 16 \zm_2 \omm_A(0, 0, 0, 0, 1, 2)  + \frac{2}{3} \zm_2 \zm_3 + 
 7 \zm_4 \omm_A(0, 0, 1, 0)  \notag \\
 & - 52 \zm_4 \omm_A(0, 0, 0, 1, 0, 0) 
  \label{w5D}
 \\
\Acyc{G11111} &= -\frac{7}{360} \omm_A({0, 5}) + \frac{1}{6} \omm_A({0, 0, 0, 5}) - \omm_A({0, 0, 0, 0, 0, 5})
 -  \frac{\zm_2}{12}  \omm_A({0, 3}) + \frac{5 \zm_2}{9}  \omm_A({0, 0, 3, 0})     \notag \\
 & +  10 \zm_2 \omm_A({0, 0, 0, 0, 0, 3})
- \frac{\zm_4}{2}  \omm_A({0, 0, 1, 0})  - 9 \zm_4\omm_A({0, 0, 0, 1, 0, 0})  \,,
  \label{w5F}
 \end{align}
and expressions of comparable complexity for
$\Acyc{G5},\Acyc{G311},\Acyc{G221}$ and $\Acyc{GC221}$ are displayed in
\appref{app:higher5}. Analogous results at weight six are available from the
authors. 


\subsection{Evaluating \texorpdfstring{$B$}{$B$}-cycle graph functions}
\label{sec:evalB}
 
In this section, we compute modular transformations of $A$-cycle eMZVs. For
this purpose it will be convenient to represent $A$-cycle graph functions in
terms of iterated Eisenstein integrals \eqn{eqn:iteis0}
\begin{align}
\Acyc{G2} &= - 6 \cez(4,0) + \frac{1}{2} \zm_2
\notag \\
\Acyc{G3} &= \frac{3}{2} \zm_3 -6 \cez(4,0,0) - 60 \cez(6,0,0) \notag \\
\Acyc{G111} &= \frac{1}{4} \zm_3 -\frac{3}{2} \cez(4,0,0) - 60 \cez(6,0,0) \label{AcycEE}\\
\Acyc{G211}&=  - 36 \cez(4,4,0,0) - 756 \cez(8,0,0,0) -70 \cez(6,0,0,0) - \frac{1}{10}  \cez(4,0,0,0)+ \frac{3}{8} \zm_4 \notag \\
\Acyc{G1111} &= - 840 \cez(8, 0, 0, 0) -40\cez(6, 0, 0, 0) +\frac{ 1 }{ 8} \zm_4 \ , \notag
\end{align}
and we will now present two methods to compute their $S$-transformation. Both
of these methods leave certain additive constants built from MZVs undetermined.
These constants can be either determined numerically or by a method of Enriquez
\cite{Enriquez:Emzv}, which allows to infer constant terms of $B$-cycle eMZVs
from the Drinfeld associator, see \appref{app:constantB} for more details.

In subsections \ref{sssec:Brown} to \ref{ssec:Bcyc3}, the method of obtaining
$B$-cycle eMZVs from $A$-cycle eMZVs as developed by Brown is
explained. An alternative method using differential equations is provided in
\subsecref{ssec:Bdiff}.


\subsubsection{Conversion to Brown's iterated Eisenstein integrals}
\label{sssec:Brown}

In this subsection we want to briefly recall the theory of iterated integrals
of Eisenstein series, developed by Brown in \rcite{Brown:mmv}, and
explain how one can use it to get the $q$-expansion of $B$-cycle eMZVs. The key
idea is to express the iterated Eisenstein integrals appearing in $A$-cycle
graph functions in terms of the iterated integrals
\begin{equation}
\GGG{j_1 &j_2 &\ldots &j_r}{k_1 &k_2 &\ldots &k_r}{\tau} :=
\int \limits_{\tau}^{i\infty} \dd \tau_r \ \tau_r^{j_r} \GG{k_r}(\tau_r) \int \limits_{\tau_r}^{i\infty} \dd \tau_{r-1} \ \tau_{r-1}^{j_{r-1}} \GG{k_{r-1}}(\tau_{r-1}) \, \ldots \,
\int \limits_{\tau_{2}}^{i\infty} \dd \tau_1 \  \tau_1^{j_1} \GG{k_1}(\tau_1)\, ,
\label{modular1}
\end{equation}
which already appeared in \eqn{prevBrown}, and are regularized as explained in
\subsecref{SectionItEisInt}. The modular properties of the functions $\CG$ are
known from \rcite{Brown:mmv} (for a certain range of the powers $j_i$'s) and
will be discussed in the next subsection. 

The translation between the expressions \eqn{eqn:iteis} for iterated Eisenstein
integrals $\ce$ and \eqn{modular1} can be conveniently extracted from the
respective generating series
\begin{align}
\ZE_{\underline{k}}(Y_0,Y_1,\ldots ,Y_r;\tau)&:= 
\sum_{p_0,p_1,\ldots ,p_r\geq 0}\frac{1}{(2\pi i)^{2p_0}} \bigg[ \prod_{i=1}^r(2\pi i)^{k_i-2p_i-1} \bigg]  \notag \\
&\times \ce(0^{p_0},k_1,0^{p_1},\ldots ,k_r,0^{p_r};\tau)Y_0^{p_0}Y_1^{p_1}\cdots Y_r^{p_r} \label{modular41} \\
\mathbb{G}_{\underline{k}}(T_1,T_2,\ldots ,T_r;\tau)&:=\sum_{p_1,\ldots ,p_r\geq 0} \bigg[ \prod_{i=1}^r\frac{1}{p_i!}\bigg(\frac{T_i}{2\pi i}\bigg)^{p_i} \bigg]
\GGG{p_1 &p_2 &\ldots &p_r}{k_1 &k_2 &\ldots &k_r}{\tau}  \notag 
\end{align}
with formal variables $Y_i$ and $T_i$. Here and in later places, we are using multi-index
notation $\underline{k}:=(k_1,k_2,\ldots, k_r)$, i.e.\ eqns.~\eqref{modular41} define two generating
series for any fixed $r$-tuple $\underline{k}$. As will be shown in
\appref{app:GvsE}, the series in eqns.~\eqref{modular41} are related via
\begin{align}
\ZE_{\underline{k}}(Y_0,Y_1,\ldots ,Y_r;\tau)&=
\exp\bigg(\frac{\tau Y_r}{2\pi i}\bigg) \, \ZG_{\underline{k}}(Y_0-Y_1,Y_1-Y_2,\ldots ,Y_{r-1}-Y_r;\tau)
\label{modular42} \\
\mathbb{G}_{\underline{k}}(T_1,T_2,\ldots ,T_r;\tau)&=
\exp\bigg({-}\frac{\tau U}{2\pi i}\bigg) \, \ZE_{\underline{k}}(T_1{+}T_2{+}\cdots {+}T_r{+}U,T_2{+}\cdots {+}T_r{+}U,\ldots ,T_r{+}U,U;\tau) \,,
\notag
\end{align}
where the dependence of the right-hand side on the formal variable $U$ drops
thanks to shuffle relations.  By isolating the coefficients of suitable
monomials in the formal variables, \eqn{modular42} translates into the
following relations at depth one and two,
\begin{align}
\ce(0^{p_0},k_1,0^{p_1};\tau)&= (2\pi i)^{p_0+p_1-k_1+1}
\sum_{\alpha_1+\beta_1=p_1}\frac{(-1)^{\alpha_1}\tau^{\beta_1}}{p_0!\alpha_1!\beta_1!}
\GGG{p_0+\alpha_1}{k_1}{\tau} \,,
\label{modular43A}
\\
\ce(0^{p_0},k_1,0^{p_1},k_2,0^{p_2};\tau)&= (2\pi i)^{p_0+p_1+p_2-k_1-k_2+2}
\sum_{\substack{\alpha_1+\beta_1=p_1\\ 
\alpha_2+\beta_2=p_2}}\frac{(-1)^{\alpha_1+\alpha_2}\tau^{\beta_2}}{p_0!\alpha_1!\beta_1!\alpha_2!\beta_2!}
\GGG{p_0+\alpha_1 &\beta_1+\alpha_2}{k_1 &k_2}{\tau} \,, \notag
\end{align}
and conversely
\begin{align}
\GGG{p_1 }{k_1 }{\tau} &= (2\pi i)^{k_1-p_1-1}p_1!\ce(0^{p_1},k_1;\tau)
\label{modular44} \\
\GGG{p_1 &p_2}{k_1 &k_2}{\tau}
&=(2\pi i)^{k_1+k_2-p_1-p_2-2}\sum_{a+b=p_2}\frac{(p_1{+}a)!p_2!}{a!}\ce(0^{p_1+a},k_1,0^{b},k_2;\tau) \,.
\notag
\end{align}
%


\subsubsection{Modular transformations of Brown's iterated Eisenstein integrals}

The modular transformation of Brown's iterated Eisenstein integrals \eqn{modular1}
can be compactly encoded in another generating function 
\begin{equation}
I^E(\tau,\infty):=1+\int \limits_{\tau}^{i\infty} \Theta^E(X_1,Y_1,\tau_1)\, +\int \limits_{\tau}^{i\infty} \Theta^E(X_2,Y_2,\tau_2)\int \limits_{\tau_1}^{i\infty}\Theta^E(X_1,Y_1,\tau_1) \,+\ldots \,,
\label{modular2}
\end{equation}
where Eisenstein series are combined with non-commutative formal
variables\footnote{Brown developed the theory for the full space of modular
  forms. Here, we specialize his construction to iterated integrals of
  Eisenstein series only, so we keep his original notation, adding the
superscript $E$ which stands for Eisenstein.  Moreover, we chose a different
normalization convention for Eisenstein series.} $\texttt{g}_k$ 
\begin{equation}
\Theta^E(X,Y,\tau) =\dd \tau \sum_{k\geq 4}\GG{k}(\tau) \, (X-\tau Y)^{k-2} \, \texttt{g}_k \,.
\label{modular3}
\end{equation}
As a special case of a lemma proved by Brown in \rcite{Brown:mmv}, there exists
a series $\CC^E_S$ in infinitely many non-commutative variables $\texttt{g}_k$
and infinitely many pairs of commutative variables $(X_i,Y_i)$ such
that\footnote{In \cite{Brown:mmv}, the position of the factors on the right-hand 
side is reversed because of our opposite convention for iterated
integrals.}
\begin{equation}\label{modular43B}
I^E(\tau,\infty)=\CC^E_S I^E \Big(-\tfrac{1}{\tau},\infty \Big) |_S \,,
\  \ \ \ \ \ 
I^E\Big(-\tfrac{1}{\tau},\infty \Big)=\big(\CC^E_S\big)^{-1}|_S\,I^E(\tau,\infty)|_S \, ,
\end{equation}
where $|_S$ acts on a function $F(X_i,Y_i)$ of the commutative variables $X_i,Y_i$ according to
\begin{equation}\label{defSact}
F(X_i,Y_i) |_S = F(-Y_i,X_i) \,.
\end{equation}
The series $\CC^E_S$ does not depend on $\tau$, and its coefficients are called
\emph{multiple modular values} (of Eisenstein series). In all cases relevant to
the computation of $B$-cycle graph functions at weight $w\leq 7$, these
coefficients are $\ZQ[2 \pi i]$-linear combinations of MZVs of known
transcendentality whose  composition can be obtained either numerically, using
the fact that (by \eqn{modular43B})
\begin{equation}
\CC^E_S=I^E(i,\infty)\big(I^E(i,\infty)|_S\big)^{-1},
\end{equation} 
or by matching with the method of Enriquez reviewed in \appref{app:constantB}. 

The desired modular transformations of iterated Eisenstein integrals can be
extracted from the series in \eqn{modular43B}: To isolate the coefficients of
any non-commutative word $\texttt{g}_{k_1}\texttt{g}_{k_2}\cdots
\texttt{g}_{k_r}$ in the above generating series $I^E(\tau,\infty)$ and
$\CC^E_S$, we will write $I^E(\tau,\infty)(k_1,k_2,\ldots ,k_r)$ and
$\CC^E_S(k_1,k_2,\ldots ,k_r)$, respectively. In terms of Brown's iterated
Eisenstein integrals \eqn{modular1}, we find
\begin{align}
&I^E(\tau,\infty)(k_1,k_2,\ldots ,k_r)= \sum_{j_1=0}^{k_1-2} \ \sum_{j_2=0}^{k_2-2}\cdots \sum_{j_r=0}^{k_r-2}  \bigg[
\prod_{i=1}^r (-1)^{j_i} \binom{k_i-2}{j_i} 
\bigg] \label{CoeffOfI} \\
& \ \  \times 
\GGG{j_1 &j_2 &\ldots &j_r}{k_1 &k_2 &\ldots &k_r}{\tau}
X_1^{k_1-2-j_1}X_2^{k_2-2-j_2}\cdots X_r^{k_r-2-j_r}Y_1^{j_1}Y_2^{j_2}\cdots Y_r^{j_r} \,. \notag
\end{align}
In the case of a single integration, one gets abelian cocycles $\CC^E_S(k)$,
also called period polynomials, very well known after the work of Eichler,
Shimura and Manin in the case of cusp forms, and worked out for Eisenstein
series in \rcites{ZagierF,Haberland}. In particular, it was proven that
\begin{equation}\label{CocycleEis}
\CC^E_S(2k)=\frac{2\pi i}{2k{-}1}\Big(\zeta_{2k-1}(Y^{2k-2}-X^{2k-2})-(2\pi i)^{2k-1}\sum_{i=1}^{k-1}\frac{B_{2i}B_{2k-2i}}{(2i)!(2k{-}2i)!}X^{2i-1}Y^{2k-2i-1}\Big)\, .
\end{equation}


\subsubsection{\texorpdfstring{$B$}{$B$}-cycle eMZVs from Brown's
iterated Eisenstein integrals}
\label{ssec:Bcyc3}

The computation of $B$-cycle eMZVs from \eqn{modular43B} follows a simple idea
which has already been used in \rcite{ZerbiniThesis} at depth one: once the
underlying $\ce(\underline{k};-\frac{1}{\tau})$ are related to the coefficients
\eqn{modular1} of the series $I^E(\tau,\infty)$, one can use Brown's result.
In particular, by inserting \eqn{CoeffOfI} into the special cases of  
\begin{align}
I^E\Big(-\tfrac{1}{\tau},\infty\Big)(k_1)&= I^E(\tau,\infty)(k_1)|_S -\CC^E_S(k_1)|_S \label{modular46} 
\\
I^E\Big(-\tfrac{1}{\tau},\infty\Big)(k_1,k_2)&= I^E(\tau,\infty)(k_1,k_2)|_S-\CC^E_S(k_1)|_SI^E(\tau,\infty)(k_2)|_S \notag \\
& \ \ \ \ \  +\CC^E_S(k_1)|_S\CC^E_S(k_2)|_S-\CC^E_S(k_1,k_2)|_S  \label{modular47} 
\end{align}
of \eqn{modular43B}, we arrive at
\begin{align}
\GGG{j_1}{k_1}{-\tfrac{1}{\tau}} &= (-1)^{j_1}\GGG{k_1-2-j_1}{k_1}{\tau}-\binomial{k_1-2}{j_1}^{-1}c_{k_1-2-j_1}(k_1),
\label{modular48}\\
\GGG{j_1 &j_2}{k_1 &k_2}{-\tfrac{1}{\tau}} &= (-1)^{j_1+j_2}\GGG{k_1-2-j_1 &k_2-2-j_2}{k_1 &k_2}{\tau}-(-1)^{j_2}\binomial{k_1{-}2}{j_1}^{-1}c_{k_1-2-j_1}(k_1) \, \GGG{k_2-2-j_2}{k_2}{\tau} \notag \\
&\! \! \! \! \! \! \! \! \! + \binomial{k_1{-}2}{j_1}^{-1}\binomial{k_2{-}2}{j_2}^{-1}\big(c_{k_1-2-j_1}(k_1)c_{k_2-2-j_2}(k_2)-c_{k_1-2-j_1,k_2-2-j_2}(k_1,k_2)\big) \, ,
\label{modular49}
\end{align}
where the quantities $c_{\ldots}(k_1,\ldots)$ are defined by the expansion
\begin{equation}
\CC^E_S(k_1,\ldots ,k_r)=\sum_{j_1=0}^{k_1-2}\cdots \sum_{j_r=0}^{k_r-2} c_{j_1,\ldots ,j_r}(k_1,\ldots ,k_r)X_1^{k_1-2-j_1}\cdots X_r^{k_r-2-j_r}Y_1^{j_1}\cdots Y_r^{j_r}.
\end{equation}
One must be warned that not all $\ce(\underline{k};-\frac{1}{\tau})$ can be
computed in this way: if $\underline{k}$ contains too many zeros, \eqn{modular42}
gives rise to $\GGG{j_1 &j_2 &\ldots &j_r}{k_1 &k_2 &\ldots &k_r}{\tau}$ with
$j_i \notin \{0,1,\ldots,k_i{-}2\}$ which are excluded from the building block
\eqn{modular3} of Brown's series \eqn{modular2}. However, this
method always applies to the special linear combinations of
$\ce(\underline{k};-\frac{1}{\tau})$ given by eMZVs and therefore selected
by a certain derivation algebra \cite{KZB, Hain, Broedel:2015hia}: This is a 
consequence of Proposition 6.3 of ref. \cite{Brown:2017qwo2}, and in fact,
the linear combinations of $\ce(\underline{k};-\frac{1}{\tau})$ descending from 
eMZVs are contained in a proper subset of the iterated integrals \eqn{prevBrown}.
Putting all of this together, one obtains a closed formula at depth one for
$p_0+p_1\leq k_1-2$
\begin{align}
\ce(0^{p_0},k_1,0^{p_1};-\tfrac{1}{\tau} ) &= (-1)^{p_1}(2\pi i)^{p_0+p_1+1-k_1}\sum_{\alpha+\beta=p_1}\frac{(k_1-2-p_0-\alpha)!}{\alpha!\beta!\tau^\beta} \\
&\times \Big((-2\pi i)^{p_0+\alpha}\ce(0^{k_1-2-p_0-\alpha},k_1;\tau)-\frac{(p_0+\alpha)!}{(k_1-2)!}c_{k_1-2-p_0-\alpha}(k_1)\Big) \notag \, ,
\label{modular91}
\end{align}
and higher-depth expressions such as\footnote{We do not have a closed
formula like \eqn{CocycleEis} for multiple modular values at depth $\geq 2$, so
for the purposes of this paper, we contented ourselves to guessing their
representations as MZVs based on five hundred digits numerical 
approximations. In all cases up to weight six, these representations
have been confirmed through the analytic method of \appref{app:constantB}.}
\begin{align}
(2\pi i)^6 &\ce(6,4;-\tfrac{1}{\tau}) = \frac{\zeta_{3,5}}{75} + \frac{\zeta_3\zeta_5}{15} - \frac{503\zeta_8}{10800} -  \frac{2\zeta_5}{5}\ce(0,0,4;\tau)  \\
&+48\big(\ce(0,0,0,0,6,0,0,4;\tau) + 5\ce(0,0,0,0,0,6,0,4;\tau) + 15\ce(0,0,0,0,0,0,6,4;\tau)\big), \notag
\end{align}
as well as (setting $T=\pi i\tau$) 
\begin{align}
&\ce(4, 4, 0, 0;-\tfrac{1}{\tau})
 = \ce(0,0,4,4;\tau)-\frac{\zeta_3\ce(4;\tau)}{6}+\frac{209\pi^4}{11664000} \notag \\
&-\frac{1}{T}\Big(\ce(0,0,4,0,4;\tau)+3\ce(0,0,0,4,4;\tau)-\frac{\zeta_3\ce(0,4;\tau)}{6}+\frac{\zeta_2\zeta_3}{360}-\frac{5\zeta_5}{432}\Big)  \label{modular91a} \\
&+\frac{1}{2T^2}\Big( \frac{1}{2}\ce(0,0,4;\tau)^2 -\frac{\zeta_3\ce(0,0,4;\tau)}{6}+\frac{\zeta_3^2}{72}\Big)\, , \notag\\[12pt]
&\ce(4, 0, 4, 0, 0;-\tfrac{1}{\tau}) + 3 \ce(4, 4, 0, 0, 0;-\tfrac{1}{\tau})=-\frac{\pi^4}{108}\ce(4;\tau)-2\pi^2\ce(0,4,4;\tau)-\frac{\zeta_2\zeta_3}{360}+\frac{5\zeta_5}{432}\notag \\
&-\frac{\pi^2}{T}\Big(\frac{\zeta_3}{3}\ce(4;\tau)-\frac{\zeta_2}{18}\ce(0,4;\tau)-2\ce(0,4,0,4;\tau)-6\ce(0,0,4,4;\tau)
 -\frac{ 167 \zeta_4}{32400}\Big)\notag \\
&-\frac{\pi^2}{T^2}\Big(-\frac{\zeta_3}{3}\ce(0,4;\tau)+\frac{\zeta_2}{36}\ce(0,0,4;\tau)-\frac{\zeta_2\zeta_3}{540}-\frac{5\zeta_5}{432}+\ce(0,4,0,0,4;\tau)  \label{modular92} \\
& \ \ \ \ +4\ce(0,0,4,0,4;\tau)+9\ce(0,0,0,4,4;\tau)\Big)
+ \frac{\pi^2}{T^3}\Big( \frac{1}{2} \ce(0,0,4;\tau)^2 - \frac{\zeta_3}{6}\ce(0,0,4;\tau)
+\frac{\zeta_3^2}{72}\Big)\notag
\end{align}
and the modular transformations given in \appref{moremod}.
In all examples of $A$-cycle graph functions tested so far we indeed landed on iterated
integrals of the kind \eqn{modular1} with $j_i\leq k_i-2 $, whose
$S$-transform can therefore be computed as explained above.
Note that the relative factor of 3 on the left-hand side of \eqn{modular92}
is crucial to obey this criterion.

In order to determine the $q$-expansion of $B$-cycle graph functions, the
iterated Eisenstein integrals on the right-hand side of \eqn{modular91} and the
above depth-two examples need to be cast into the form $\cez(k,\ldots)$ with
$k\neq 0$ such that \eqn{qgamma1} becomes applicable. This can always be
achieved by first applying shuffle relations such as $\ce(0,4;\tau) =
\ce(0;\tau)\ce(4;\tau) {-} \ce(4,0;\tau) $ and $\ce(0,0,4;\tau) =
\ce(4,0,0;\tau)  - \ce(0) \ce(4,0;\tau) + \ce(0,0;\tau) \ce(4;\tau)$ to attain
the form $\ce(k,\ldots)$ with $k\neq 0$. Then, the conversion between
$\ce(\ldots)$ and $\cez(\ldots)$ follows from the definitions
\eqns{eqn:iteis}{eqn:iteis0} of the respective iterated Eisenstein integrals,
along with
\begin{equation}
\ce(\underbrace{0,0,\ldots,0}_n;\tau) = \cez(\underbrace{0,0,\ldots,0}_n;\tau) = \frac{1}{n!} (2\pi i \tau)^n \,,
\label{onlyzero}
\end{equation}
for instance $\ce(4,0;\tau) = \cez(4,0;\tau) +\frac{\pi^2 \tau^2}{360}$ and
$\ce(4,0,0;\tau) = \cez(4,0,0;\tau) + \frac{i\pi^3 \tau^3}{540}$.  At depth
larger than one, this might introduce further instances of $\cez(0,\ldots)$
with zero in the first entry which call for additional shuffle manipulations.
This can be illustrated through the following example at depth two
\begin{align}
\ce(4, 4, 0, 0)  &= \cez(4, 4, 0, 0)   - \frac{2 \zeta_4}{(2\pi i)^4} \big[ \cez(4, 0, 0, 0)+\cez(0, 4, 0, 0) \big] + 
\Big( \frac{2 \zeta_4}{(2\pi i)^4}  \Big)^2 \cez(0, 0, 0, 0) \notag \\
&=\cez(4, 4, 0, 0)  +\frac{1}{360} \cez(4, 0, 0, 0) 
-\frac{i \pi \tau}{360}  \cez(4, 0, 0)
+ \frac{\pi^4 \tau^4}{777600} \ ,
\label{thisex}
\end{align}
where we have inserted $\cez(0, 4, 0, 0)=\cez(0)\cez(4, 0, 0) - 3\cez(4, 0, 0,
0)$ in passing to the second line.  A formula for the most general case can be
found in \appref{app:gamma0}.

\subsubsection{\texorpdfstring{$B$}{$B$}-cycle eMZVs from
differential equations}
\label{ssec:Bdiff}

As an alternative and recursive method to determine modular transformations of
iterated Eisenstein integrals \eqns{eqn:iteis}{eqn:iteis0}, one can take
advantage of the differential equation
\begin{align}
\tau^2 2\pi i \partial_\tau \ce(\underline{m},2k; -\tfrac{1}{\tau}) &= -   (2\pi i)^{2-2k}  \tau^{2k} \GG{2k}(\tau)  \ce(\underline{m}; -\tfrac{1}{\tau})  
\label{diffmod1}
\\
\tau^2 2\pi i \partial_\tau \cez(\underline{m},2k; -\tfrac{1}{\tau}) &= - (2\pi i)^{2-2k} \big[   \tau^{2k} \GG{2k}^0(\tau) + 2 \zm_{2k}(\tau^{2k}-1)
\big]  \cez(\underline{m};-\tfrac{1}{\tau})  
\label{diffmod2}
\end{align} 
for $k\neq 0$ as well as 
\begin{align}
\tau^2 2\pi i \partial_\tau \ce(\underline{m},0;-\tfrac{1}{\tau}) =    (2\pi i)^{2}   \ce(\underline{m};-\tfrac{1}{\tau})  
\,, \ \ \ \ 
\tau^2 2\pi i \partial_\tau \cez(\underline{m},0;-\tfrac{1}{\tau}) =  (2\pi i)^{2}   \cez(\underline{m};-\tfrac{1}{\tau})   \,,
\label{diffmod3}
\end{align} 
resulting from their recursive definition. With this method, the expression for
$\ce(n,0,0,\ldots,0;-\tfrac{1}{\tau} ) $ in \eqn{modular91} with $j{-}1$
successive zeros follows from integrating \eqns{diffmod1}{diffmod3} $j$ times,
and the multiple modular values \eqn{CocycleEis} arise as the integration
constants of the respective $j$ steps. So the modular transformation is
performed separately on each integration kernel in the iterated Eisenstein integrals. At higher
depth, these integration constants can be obtained numerically or by matching
with Enriquez's method reviewed in \appref{app:constantB}. In all cases we have
checked the approach of this subsection matches the results obtained from
Brown's theory.


\subsubsection{Examples of \texorpdfstring{$B$}{$B$}-cycle graph functions}
\label{sec:evalBEX}

In applying the modular transformation \eqn{modular91} at depth one to the
$A$-cycle graph functions in \eqn{AcycEE}, we have to take the offsets between
the $\ce(\ldots)$ and $\cez(\ldots)$ into account.  From the discussion around
\eqn{onlyzero}, we have
\begin{align}
\cez(4,0;-\tfrac{1}{\tau}) &= \ce(4,0;-\tfrac{1}{\tau}) - \frac{\pi^2}{360\tau^2} 
= \ce(4,0;\tau) + \frac{i}{\pi \tau}  \ce(4,0,0;\tau) - \frac{i \zeta_3}{6\pi \tau} + \frac{\pi^2}{216}   - \frac{\pi^2}{360\tau^2}  \notag \\
&=  \cez(4,0;\tau) - \frac{1}{T}  \cez(4,0,0;\tau) - \frac{ T^2}{1080} + \frac{\zeta_2}{36} + \frac{ \zeta_3}{6T}+ \frac{\zeta_4}{4T^2}
\label{simex1}
\end{align}
with $T= \pi i\tau$, and by similar manipulations,
\begin{equation}
\cez(4,0,0;-\tfrac{1}{\tau})  = -\frac{\pi^2}{T^2} \cez(4,0,0;\tau) +  \frac{ T \zeta_2}{90}   + \frac{ \zeta_3}{6} 
+ \frac{ 5 \zeta_4}{6 T} + \frac{ \zeta_2 \zeta_3}{  T^2}  + \frac{7 \zeta_6}{ 4 T^3} \,.
\label{simex2}
\end{equation}
Following the same strategy at higher weight, one obtains the following 
expressions for $B$-cycle graph functions:
\begin{align}
\Bcyc{G2} &=
\frac{ T^2}{180}  + \frac{\zeta_2}{3} - \frac{ \zeta_3}{T} - \frac{3 \zeta_4}{2 T^2}  - 6 \cez(4, 0) + \frac{6 \cez(4, 0, 0)}{T} \notag \\
\Bcyc{G111} &= -\frac{T^3}{3780}  - \frac{T \zeta_2}{60}- \frac{ \zeta_4}{T}   - \frac{3 \zeta_2 \zeta_3}{2 T^2} 
 + \frac{3 \zeta_5}{2 T^2} - \frac{\zeta_6}{8 T^3} \notag \\
 &- 60 \cez(6, 0, 0) + \frac{180 \cez(6, 0, 0, 0)}{T} - \frac{
 180 \cez(6, 0, 0, 0, 0)}{T^2} + \frac{ 9 \zeta_2 \cez(4, 0, 0) }{T^2} 
\notag \\
\Bcyc{G1111} &= \frac{T^4}{75600} 
+\frac{T^2 \zeta_2}{945}   + \frac{13 \zeta_4}{180} 
+ \frac{4 \zeta_6}{3 T^2}
+ \frac{ 4 \zeta_2 \zeta_5}{T^3} 
 - \frac{5 \zeta_7}{ 2 T^3} + \frac{95 \zeta_8}{24 T^4}
 \label{bcyc1} \\
 &- 840 \cez(8, 0, 0, 0) + 
\frac{ 5040 \cez(8, 0, 0, 0, 0)}{T} 
- \frac{ 12600 \cez(8, 0, 0, 0, 0, 0)}{T^2} 
\notag \\
&+ \frac{ 12600 \cez(8, 0, 0, 0, 0, 0, 0)}{T^3}  + \frac{ 240  \zeta_2 \cez(6, 0, 0, 0)}{T^2} 
- \frac{ 480 \zeta_2 \cez(6, 0, 0, 0, 0) }{T^3} \,.
\notag
 \end{align}
We have rewritten the integrals $\ce$ following from the above modular
transformations in terms of $\cez$ to make the $q$-expansion of the $B$-cycle
graph functions accessible from \eqn{qgamma1}.  Moreover, this highlights the
property of $B$-cycle eMZVs that coefficients of $q^{n}$ are Laurent
polynomials in $\tau$. The change of variables from $\pi i\tau$ to $T$ absorbs
all extra powers of $\pi i$ and yields $\mathbb Q$-linear combinations of MZVs
as Laurent coefficients, as remarked in \eqn{Texpansion}. Hence, these
Laurent polynomials can be thought of as the open-string antecedents of the
zero modes $\smallDcycLetter[\CG]$ of modular graph functions discussed in
\secref{ssec:modularZERO}.  Accordingly, we will denote the coefficient of
$q^0$ in the $B$-cycle graph function $\text{\textbf{B}}[\CG] $ by
$\text{\textbf{b}}[\CG] $, e.g.~one finds
\begin{equation}
\smallBcyc{G2}=
\frac{ T^2}{180}  + \frac{\zeta_2}{3} - \frac{ \zeta_3}{T} - \frac{3 \zeta_4}{2 T^2}
 \,, \ \ \ \ \ \ 
\smallBcyc{G111}=
-\frac{T^3}{3780}  - \frac{T \zeta_2}{60}- \frac{ \zeta_4}{T}   - \frac{3 \zeta_2 \zeta_3}{2 T^2} 
 + \frac{3 \zeta_5}{2 T^2} - \frac{\zeta_6}{8 T^3} \, ,
\end{equation}
and a method to determine such $\text{\textbf{b}}[\CG] $ from the Drinfeld
associator is presented in \appref{app:constantB}. This method goes back to
Enriquez \cite{Enriquez:Emzv}, where a generating series for the constant terms
of $A$-cycle and $B$-cycle eMZVs is given, and a procedure to extract the
constant terms of individual $A$-cycle eMZVs is explained in section 2.3 of
\rcite{Broedel:2015hia}.

At depth two, the modular transformation \eqn{modular91a} of $\ce(4,4,0,0)$
leads to
 \begin{align}
&\Bcyc{G211}  - \tfrac{9}{10}\Bcyc{G1111}  = - \frac{T^4}{324000} 
+\frac{ 17 T^2 \zeta_2}{18900}   - \frac{T \zeta_3}{180}  + \frac{ 253 \zeta_4}{1800}
- \frac{ 5 \zeta_5}{12 T}  +  \frac{49 \zeta_6}{80 T^2}    \notag \\
& - \frac{ \zeta_3^2}{4 T^2}
-  \frac{3 \zeta_3 \zeta_4}{2 T^3}  
  + \frac{17 \zeta_5 \zeta_2}{5 T^3}  + \frac{ 277 \zeta_8}{48 T^4} 
 +\Big(\frac{T}{30}  + \frac{3   \zeta_3}{T^2} +   \frac{ 9\zeta_4 }{ T^3}    \Big) \cez(4, 0, 0)
  - \frac{9 \cez(4, 0, 0)^2}{T^2}  \notag \\
  &  + \frac{36}{T}  \big[  \cez(4, 0, 4, 0, 0) + 3\cez(4, 4, 0, 0, 0) + \tfrac{1}{360}
\cez(4, 0, 0, 0, 0) \big] \label{bcyc3} \\
   & -  36\big[ \cez(4, 4, 0, 0) + \tfrac{1}{360} \cez(4, 0, 0, 0)   \big]
 + \frac{204 \zeta_2 \cez(6, 0, 0, 0)}{T^2}- \frac{ 408 \zeta_2 \cez(6, 0, 0, 0, 0)}{T^3}   \,, \notag
\end{align}
where the specific linear combination of $B$-cycle graph functions will be
motivated in \secref{sec:modezero}.  Note that the combination $\cez(4, 4, 0,
0) + \tfrac{1}{360} \cez(4, 0, 0, 0) $ in the last line can be recombined to
$\ce(4, 4, 0, 0)$ according to \eqn{thisex}, and a similar statement applies to
the length-five combination in the third line of \eqn{bcyc3}.

By the modular transformation \eqn{modular98}, the $A$-cycle graph functions
\eqns{w5D}{w5F} at weight five are mapped to the $B$-cycle graph function
 \begin{align}
 &\Bcyc{G2111}  - \tfrac{43}{35}\Bcyc{G11111}  = \frac{ T^5}{2381400}  + \frac{T^2 \zeta_3}{3780}  - \frac{\zeta_5}{360}
  + \frac{7 \zeta_7}{8 T^2} + \frac{\zeta_3 \zeta_5}{T^3}   
- \Big( \frac{T^2}{630}  + \frac{ 6  \zeta_5}{T^3} \Big) \cez(4, 0, 0)  \notag \\
&  + \Big( \frac{ 2T}{3}   + \frac{ 60   \zeta_3}{T^2} \Big) \cez(6, 0, 0, 0)
 - \Big(  4     + \frac{ 120 \zeta_3}{T^3} \Big) \cez(6, 0, 0, 0, 0) \notag \\
&  - \frac{ 360 \cez(4, 0, 0) \cez(6, 0, 0, 0)}{T^2}  + \frac{
 720 \cez(4, 0, 0) \cez(6, 0, 0, 0, 0)}{T^3} \notag \\
& - 720 \cez(4, 6, 0, 0, 0) + \frac{1}{42} \cez(4, 0, 0, 0, 0)    - 
 240 \cez(6, 0, 4, 0, 0) - 720 \cez(6, 4, 0, 0, 0) \notag \\
 &  + \frac{720 \cez(4, 0, 6, 0, 0, 0)}{T} 
 + \frac{ 4320 \cez(4, 6, 0, 0, 0, 0)}{T}
 - \frac{ \cez(4, 0, 0, 0, 0, 0)}{14 T} \label{bcyc5} \\
 &+ \frac{ 720 \cez(6, 0, 0, 4, 0, 0)}{T}
  + \frac{ 2160 \cez(6, 0, 4, 0, 0, 0)}{T}
   + \frac{ 4320 \cez(6, 4, 0, 0, 0, 0)}{T}  \notag \\
 &  + \frac{10 \cez(6, 0, 0, 0, 0, 0)}{T}- \frac{1440 \cez(4, 0, 6, 0, 0, 0, 0)}{T^2}
  - \frac{ 7200 \cez(4, 6, 0, 0, 0, 0, 0)}{T^2}  \notag \\
 & + \frac{\cez(4, 0, 0, 0, 0, 0, 0)}{ 14 T^2}- \frac{ 10 \cez(6, 0, 0, 0, 0, 0, 0)}{T^2}
  - \frac{ 720 \cez(6, 0, 0, 0, 4, 0, 0)}{T^2} \notag \\
  & - \frac{ 2160 \cez(6, 0, 0, 4, 0, 0, 0)}{T^2} - \frac{ 4320 \cez(6, 0, 4, 0, 0, 0, 0)}{T^2}
   - \frac{ 7200 \cez(6, 4, 0, 0, 0, 0, 0)}{T^2}  \ {\rm mod} \ \zeta_2   \, . \notag
 \end{align}
For reasons to be explained in \subsecref{sec:funrules} below, we have
suppressed terms of the form $\pi^{2k} \cez(\ldots) T^m$ with $k\geq 1$ and
$m\in\mathbb Z$ and refer to their omission by ${\rm mod} \ \zeta_2$. The modular
transformations in \appref{moremod} lead to similar expressions for $B$-cycle
graph functions at weight six which are available from the authors upon
request. We have also determined numerically a Laurent polynomial at weight
seven\footnote{Here we have been employing results from the multiple zeta value
data mine \cite{Blumlein:2009cf}.}
\begin{align}
\smallBcyc{G511}
 &= 
 -\frac{31 T^7}{700539840 }
 -  \frac{ 5251 T^5 \zeta_2}{233513280}
 +  \frac{T^4 \zeta_3}{3888 }
 -  \frac{7405 T^3 \zeta_4}{598752}
  +  \frac{119 T^2 \zeta_5}{2592}
  +   \frac{31 T^2 \zeta_2 \zeta_3 }{864}  \notag \\
 & - \frac{ 11 T \zeta_3^2  }{ 216 } - \frac{15527 T \zeta_6}{10368}
+ \frac{21 \zeta_7}{32} +  \frac{67 \zeta_2 \zeta_5}{27}   + \frac{167 \zeta_3 \zeta_4}{48} 
 - \frac{ 23 \zeta_3 \zeta_5}{3 T}
 - \frac{80017 \zeta_8}{ 1296 T}
 + \frac{3 \zeta_2 \zeta_3^2}{T}  \notag \\
 &- \frac{25 \zeta_3^3}{4 T^2} + \frac{7115 \zeta_9}{144 T^2}
+ \frac{21 \zeta_2 \zeta_7}{T^2} + \frac{35 \zeta_4 \zeta_5}{6 T^2} - \frac{6613 \zeta_3 \zeta_6}{288 T^2}
 + \frac{75 \zeta_5^2}{4 T^3} - \frac{1245 \zeta_3 \zeta_7}{8 T^3}
  - \frac{48 \zeta_{3, 5} \zeta_2}{T^3}   \notag \\
  &+ \frac{ 443 \zeta_2 \zeta_3 \zeta_5}{T^3}
  - \frac{275 \zeta_3^2 \zeta_4}{8 T^3} + \frac{941869 \zeta_{10}}{ 5760 T^3}
   - \frac{ 9573 \zeta_{11}}{ 16T^4} - \frac{ 18 \zeta_{3,5,3} }{T^4} 
   - \frac{405 \zeta_3^2 \zeta_5}{  4T^4} 
   + \frac{195 \zeta_2 \zeta_3^3}{ 2T^4} \notag \\
    &+ \frac{27745 \zeta_5 \zeta_6}{ 48T^4}
   - \frac{ 3795 \zeta_4 \zeta_7}{ 16T^4}   + \frac{17731 \zeta_3 \zeta_8}{ 16T^4} + \frac{15875 \zeta_2 \zeta_9}{12 T^4}
  - \frac{2475 \zeta_5 \zeta_7}{4 T^5} - \frac{1125 \zeta_3 \zeta_9}{ 4 T^5} \notag \\
  &   + \frac{90 \zeta_{3, 5} \zeta_4}{T^5}      + \frac{450 \zeta_{3, 7} \zeta_2}{7 T^5}
      - \frac{ 165 \zeta_3 \zeta_4 \zeta_5}{2 T^5}  + \frac{3375 \zeta_2 \zeta_5^2}{7 T^5}
         + \frac{ 3335 \zeta_3^2 \zeta_6}{4 T^5}  + \frac{3960 \zeta_2 \zeta_3 \zeta_7}{7 T^5}  \notag \\
&    + \frac{ 93091945 \zeta_{12}}{11056 T^5}
- \frac{1575 \zeta_{13}}{T^6}   + \frac{13275 \zeta_2 \zeta_{11}}{4 T^6}  + \frac{7425 \zeta_4 \zeta_9}{8 T^6}
  + \frac{129465 \zeta_6 \zeta_7}{16 T^6}    + \frac{ 233525 \zeta_5 \zeta_8}{48 T^6}  \notag \\
  & + \frac{160053 \zeta_3 \zeta_{10}}{64 T^6} 
+ \frac{15301285 \zeta_{14}}{768 T^7} \label{b511}
\end{align}
comprising the depth-three MZV $\zeta_{3,5,3}$ along with $T^{-4}$ which 
will be argued to harmonize with the Laurent polynomial \eqn{laurnew}
of the corresponding modular graph function.


\section{Open versus closed strings}
\label{sec:openVSclosed}
\begin{figure}[h]
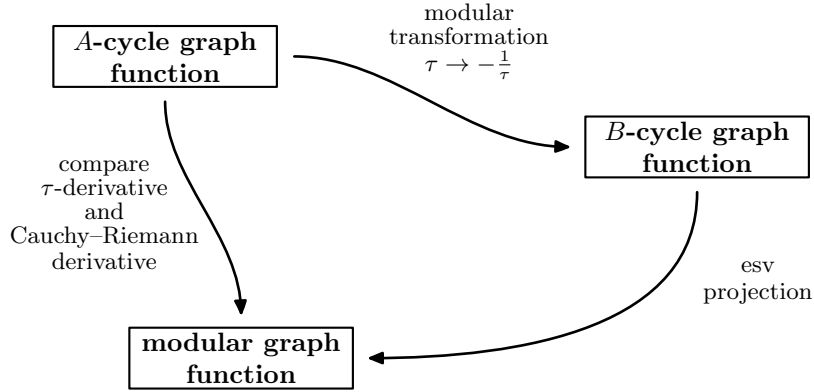

  \begin{center}
 \mpostuse{overview}\,.
 \end{center}
 \caption{Two paths for the calculation of modular graph functions}
 \label{fig:result}
\end{figure}

In this section, we are going to establish and discuss the relation and
connection between open-string graph functions and modular graph functions. The
reason and origin for our investigations is a stunning similarity of the
relations satisfied by open-string graph functions and their corresponding
modular graph functions: in \subsecref{sec:ADrels} we are going to spell out
commonalities and differences in order to establish a clear starting point.
Given this similarity, it is an obvious question, whether modular graph
functions can be eventually calculated from their open-string analogues.
Anticipating the main result of this article, the answer is indeed positive: we
can obtain modular graph functions from $A$-cycle graph functions performing
the operations noted at the arrows in \figref{fig:result}.
The two different paths which can be taken in order to obtain modular
graph functions from $A$-cycle graph functions are as follows: 
\begin{itemize}
  \item the first path starts from $A$-cycle graph functions and employs the
    similarity between the $\tau$-derivative on the $A$-cycle graph
    functions and the Cauchy--Riemann derivative acting on modular graph
    functions. Using the appropriate derivatives multiple times on both sides
    of the correspondence allows to successively infer the elements of the
    modular graph functions from their $A$-cycle graph analogues. This method
    is described in \subsecref{sec:modezero}.
  \item for the second path one converts $A$-cycle graph functions into
    $B$-cycle ones. Following this step, the projection \emph{esv} is
    applied, which we conjecture to be an elliptic analogue of the
    single-valued projection \emph{sv} mentioned in \eqn{tree07a}. While the
    conversion from $A$- to $B$-cycle graph functions using a modular
    transformation has been described in \subsecref{sec:evalB}, the map esv
    will be described and discussed in \subsecref{sec:funrules}. 
\end{itemize}
Both methods yield the same results, which are simultaneously in agreement with
all expressions for modular graph functions calculated before~\cite{Green:1999pv, 
Green:2008uj, DHoker:2015gmr}.


\subsection{Comparing relations among
\texorpdfstring{$A$}{$A$}-cycle graph functions with relations
among modular graph functions}
\label{sec:ADrels}

Given the common graphical representation of $A$-cycle graph functions and
modular graph functions it is tempting to investigate, whether the known
relations for modular graph functions reviewed in \subsecref{ssec:modularA}
have an echo for $A$-cycle graph functions. 

The simplest relation among modular graph
functions, $\Dcyc{G3}-\Dcyc{G111} = \zeta_3$ at weight three, translates into  
\begin{align}
\Acyc{G3} - \Acyc{G111}&=  \frac{1}{2} \zm_3 + 6 \zm_2 \omm_A(0,1,0,0) \,,
\label{nontriv}
\end{align}
and the weight-four relation \eqn{eqn:weightfourrelation} leads to
\begin{align}
\Acyc{G4} - 24 \Acyc{G211} + 18 \Acyc{G1111} - 3 \Acyc{G2}^2  
&=   144 \zm_2  \omm_A(0,0,0,0,2) - 24 \zm_2 \omm_A(0,0,2)  - \frac{31}{2} \zeta_4 \,,
 \label{noA1111}
 \end{align}
where the right-hand sides have been obtained by simply plugging in our results
from \subsecref{sec:eval}.

The right-hand sides of the corresponding equations for modular graph
functions, \eqns{eqn:weightthreerelation}{eqn:weightfourrelation} read
$\zeta_3$ and $0$, respectively.  This is rather suggestive: a relation between
$A$-cycle graph functions might imply a valid relation between modular graph
functions by formally replacing 
\begin{equation}
\omm_A(\underline{m}) \, \zeta_{\underline{n}} 
  \rightarrow \omm_A(\underline{m})\,  \zeta^{\rm sv}_{\underline{n}}  \,,
\label{adhoc}
\end{equation} 
where the single-valued projection of MZVs has been discussed in
\eqns{tree07a}{tree07}, and we remind the reader of the multi-index notation
$\underline{n} = (n_1,n_2,\ldots,n_r)$. The ad-hoc prescription \eqn{adhoc} has
the desired effect of replacing $\frac{1}{2}\zeta_3$ with $\zeta_3$ on the
right-hand side of \eqn{nontriv}\footnote{When $A$-cycle eMZVs are expressed in terms of
  iterated Eisenstein integrals $\ce$, the prescription in \eqn{adhoc} might
  seem to be tension with the intuition from the single-valued projection of
  MZVs.  For instance, $\zm_2 \omm_A(0,1,0,0\db\tau)$ can be represented as
  $\frac{1}{8} \zeta_3 -\frac{ 3}{4} \cez(4,0,0;q)$. Demanding consistency
  after application of \eqn{adhoc} to both expressions would yield a constraint
  for a replacement of $\cez(4,0,0;q)$. As will become clear in
  \subsecref{sec:funrules}, the formulation of \eqn{adhoc} in terms of
  $\omega_A$ is very natural after converting the $A$-cycle eMZVs $\omm_A$ to
$B$-cycle eMZVs $\omm_B$ by a modular transformation. }.  Similarly, $\zeta_4$
is mapped to zero on the right-hand side of \eqn{noA1111}, and all instances of
$\zeta_2 \omm_A(\underline{m})$ are suppressed. For brevity of notation below,
let $X$ be the rational vector space generated by products of classical and
elliptic MZVs vanishing after applying \eqn{adhoc}. That is
\begin{equation}\label{DefX}
  X :=\langle \zm_2\omega_A(\underline{m}),\zm_4\omega_A(\underline{m}),\ldots,(2\zm_{3,5}+5\zm_3\zm_5)\omega_A(\underline{m}),\ldots\rangle_\mathbb{Q} \,.
\end{equation}
In the equations below we will write ``mod $X$'', which means that we are not
writing terms from the space $X$.
At weight five, the expressions in eqs.~\eqref{w5D}, \eqref{w5F} and
\appref{app:higher5} for $A$-cycle graph functions lead to the relations
\begin{align}
40  \Acyc{G311}&= 300 \Acyc{G2111} + 120  \Acyc{G2}  \Acyc{G111} - 276  \Acyc{G11111} +\frac{ 7}{2} \zeta_5  \ \ \  {\rm mod} \ X  \label{newcompare10} \\
 \Acyc{G5} &= 60 \Acyc{G2111}  + 10  \Acyc{G2}  \Acyc{G3} - 48 \Acyc{G11111} 
  + 8 \zeta_5  \ \ \  {\rm mod} \ X  \label{newcompare11} \\
10  \Acyc{G221}&= 20 \Acyc{G2111}  - 4 \Acyc{G11111} + \frac{3}{2} \zeta_5   \ \ \  {\rm mod} \ X   \label{newcompare12} \\
 30  \Acyc{GC221}&= 12 \Acyc{G11111} + \frac{ \zeta_5}{2}   \ \ \  {\rm mod} \ X  \, ,
 \label{newcompare12a}
\end{align}
which -- upon employing \eqn{adhoc} -- yield relations (\ref{compare10}) to
(\ref{compare12a}) among modular graph functions. The validity of the
connection between $A$-cycle graph functions and modular graph functions
described above has been also checked for relations between $A$-cycle graph
functions of weight six, see \appref{app:w6rel} -- applying \eqn{adhoc}
reproduces the relations \eqn{eqn:weightsixrelations} among modular graph
functions from open-string input\footnote{More precisely, we have been
  calculating only 12 out of the 13 $A$-cycle graph functions at weight six,
  since $\Acyc{G111111}$ is beyond the reach of our current computer
  implementation.  Instead, we have inferred a conjectural expression for
$\Acyc{G111111} \ {\rm mod} \ X$ from one of the relations in
\eqn{eqn:weightsixrelations}. Hence, only seven out of the eight relations in
\appref{app:w6rel} could be used as a check.}. 

Note that the prescription in \eqn{adhoc} is ill-defined as it does depend on
the particular representations of eMZVs. In particular, there exist many
relations among eMZVs and classical MZVs: For instance, the combination
$2\omm_A(0,2,2)+ \omm_A(0,0,4) = 3 \zeta_4$ should in principle be annihilated
by applying \eqn{adhoc}, but the definition in \eqn{adhoc} leaves both terms
$\omm_A(0,2,2)$ and $\omm_A(0,0,4)$ inert. However, this does not affect the
statement of the following conjecture: Given a polynomial $\CP$ in $A$-cycle
graph functions and MZVs such that 
\begin{equation}
{\cal P}( \text{\textbf{A}}[\CG],\zeta_{\underline{n}}) = 0 \ {\rm mod} \ X \,,
\label{adhoc1}
\end{equation} 
with some graphs $\CG$, one can replace
$\text{\textbf{A}}[\CG] \rightarrow \text{\textbf{D}}[\CG]$ and
$\zeta_{\underline{n}} \rightarrow  \zeta^{\rm sv}_{\underline{n}}$ in that
polynomial to obtain a relation between modular graph functions
\begin{equation}
{\cal P}( \text{\textbf{D}}[\CG],\zeta^{\rm sv}_{\underline{n}}) = 0 \,.
\label{adhoc2}
\end{equation} 
While this alone is a beautiful result, we would like to turn it into a
formalism to actually \textit{compute} modular graph functions. The next two
subsections are dedicated to the description of the two possible methods
outlined in \figref{fig:result}.

We emphasize that \eqns{adhoc1}{adhoc2} only apply to modular graph functions
and $A$-cycle graph functions that are defined by integrating monomials in
Green functions (involving any number of punctures).  In superstring amplitudes
involving five or more legs and all massless heterotic-string amplitudes,
however, one may encounter more general integrals which will require extensions
of the correspondence in \eqns{adhoc1}{adhoc2} between open- and closed-string
expressions. The same caveat applies to the rest of this section.


\subsection{Modular graph functions from
\texorpdfstring{$A$}{$A$}-cycle graph functions}
\label{sec:modezero}

Given that relations among $A$-cycle graph functions can be mapped to those of
modular graph functions, a natural follow-up question concerns a mapping
between the respective functions of $\tau$ themselves: eMZVs on the open-string
side and, as will be shown below, real parts of iterated Eisenstein integrals
on the closed-string side.  For this purpose, we compare first-order
differential operators, namely $ \partial_\tau := \frac{ \partial }{\partial
\tau}$ acting on $A$-cycle graph functions and the Cauchy--Riemann derivative
$\nabla$ defined in \eqn{CR1} acting on modular graph functions.


\subsubsection{\texorpdfstring{$\boldsymbol{\tau}$}{$\tau$}-derivatives versus
Cauchy--Riemann equations}
\label{ssec:CRD}

From the representation of $A$-cycle graph functions in terms of eMZVs, their
$\tau$-derivatives can be conveniently computed using \eqn{eqn:tauder}. For
instance, the expressions \eqns{Acyc8}{Acyc13a} straightforwardly imply that
\begin{align}
2\pi i \partial_\tau \Acyc{G2} &= -2 \omm_A(0,3)  \notag \\
(2\pi i \partial_\tau)^2 \Acyc{G2} &= 6 \GG{4}^0   \label{Cauchy2}
\end{align}
as well as
\begin{align}
2\pi i \partial_\tau \Acyc{G111} &= 3  \omm_A(0,0,4)  -6 \zeta_2  \omm_A(0,0,2) - 4 \zeta_4  \notag \\
(2\pi i \partial_\tau)^2 \Acyc{G111}&= -12 \omm_A(0,5) + 12 \zeta_2 \omm_A(0,3)  \label{Cauchy3} \\
(2\pi i \partial_\tau)^3 \Acyc{G111}&=  60 \GG{6}^0 - 36\zeta_2 \GG{4}^0  \,,\notag
\end{align}
see \eqn{GGk0} for our conventions for $\GG{k}^0$. In the previous subsection,
relations between $A$-cycle graph functions were found to only resemble those
of modular graph functions after dropping terms from the space $X$ defined in
\eqn{DefX}. Hence, we shall consider the simpler differential equations obeyed
by a hatted version of $A$-cycle graph functions, in which the terms projected
to zero by \eqn{adhoc} are omitted:
\begin{align}
 \hat{\text{\textbf{A}}}[\CG]&=   \text{\textbf{A}}[\CG]   \ {\rm mod} \ X  \,. \label{Cauchy6}
\end{align}
The simplest examples of $\hat{\text{\textbf{A}}}[\CG]$ can be expressed as:
\begin{align}
\hatAcyc{G2} &=  \omm_A(0,0,2)  \,, \ \ \ \ \ \   \hatAcyc{G111} = - \omm_A(0,0,0,3) + \frac{1}{6}\omm_A(0,3)  \notag \\
\hatAcyc{G1111} &=  \omm_A(0,0,0,0,4) -\frac{1}{6} \omm_A(0,0,4)    \label{Cauchy5} \\
\hatAcyc{G11111} &= - \omm_A(0,0,0,0,0,5) + \frac{1}{6} \omm_A(0,0,0,5) - \frac{7}{360} \omm_A(0,5)
\notag \\
\hatAcyc{G211} &= \frac{1}{2} \omm_A({0, 0, 2})^2 - \frac{1}{2} \omm_A({0, 0, 0, 0, 4}) -  \omm_A({0, 0, 0, 2, 2})   \, .
\notag
\end{align}
Writing the analogue of \eqn{Cauchy3} for $\hat{\text{\textbf{A}}}[\CG]$,
the Eisenstein series $\GG{4}^0$ in the last line is no longer existent.
Considering other simple graphs, one finds for instance
\begin{align}
(2\pi i \partial_\tau)^2 \hatAcyc{G2} &= 6 \GG{4}^0  \,, \ \ \ \  \ \ \ (2\pi i \partial_\tau)^3 \hatAcyc{G111}=  60 \GG{6}^0    \label{Cauchy4} 
\\
(2\pi i \partial_\tau)^4 \hatAcyc{G1111}&= 840 \GG{8}^0 \,, \ \ \ \ (2\pi i \partial_\tau)^5 \hatAcyc{G11111} = 15120 \GG{10}^0\, ,
\notag
\end{align}
which intriguingly resemble the following instances of \eqn{CR2}:
\begin{align}
(\pi \nabla)^2 \Dcyc{G2} &= 6 \, (\Im(\tau))^4 \GG{4}  \,, \ \ \ \  \ \  \ (\pi \nabla)^3 \Dcyc{G111}=  60 \, (\Im(\tau))^6 \GG{6}    \label{Cauchy4D} 
\\
(\pi \nabla)^4 \Dcyc{G1111}&= 840 \, (\Im(\tau))^8 \GG{8} \,, \ \ \ \ (\pi \nabla)^5 \Dcyc{G11111} = 15120 \, (\Im(\tau))^{10} \GG{10} \,.
\notag
\end{align}
A similar correspondence can be established for graphs with more than one loop:
For instance, the expression for $\hatAcyc{G211}$ in \eqn{Cauchy5} yields
\begin{align}
(2\pi i \partial_\tau)^3 \hatAcyc{G211} &= 12 \GG{4}^0 \omm_A(0,3) - 108 \omm_A(0,7) 
- 72 \zeta_4  \omm_A(0,3) 
 \label{Cauchy8}  \\
&= - 6 \GG{4}^0 2\pi i \partial_\tau \hatAcyc{G2} + \frac{9}{10} (2\pi i \partial_\tau)^3 \hatAcyc{G1111} 
\ {\rm mod} \ X \, ,
\notag
\end{align}
which resembles the differential equation (\ref{CR3}) among modular graph functions
\begin{equation}
(\pi \nabla)^3 \Dcyc{G211} = 
- 6 (\Im(\tau))^4 \GG{4}  \pi \nabla \Dcyc{G2} + \frac{9}{10} (\pi \nabla)^3 \Dcyc{G1111}  \,.
\label{Cauchy9}
\end{equation}
In passing to the second line of \eqn{Cauchy8}, we have identified $2\pi i
\partial_\tau  \hatAcyc{G2} = -2 \omm_A(0,3)$ as well as $(2\pi i
\partial_\tau)^3 \hatAcyc{G1111} = - 120 \omm_A(0,7)$ and dropped $- 72 \zeta_4
\omm_A(0,3)$ as it is contained in the space $X$ defined in
\eqn{DefX}. In a
similar way, discarding\footnote{Of course, we will as well discard terms like
$\GG{2k}^0\zeta_2 \omm_A(\underline{n})$ containing a factor from $X$.} terms
from $X$ in the third $\tau$-derivative of $\Acyc{G2111}$ gives rise to an
open-string counterpart of \eqn{CR4}. 

We infer the following general conjecture from the above examples:
Suppose that $A$-cycle graph functions associated with
some graphs $\CG$ satisfy the differential equation 
\begin{equation}
{\cal Q}(2\pi i \partial_\tau, \GG{2k}^0,  \text{\textbf{A}}[\CG])
= 0 \ {\rm mod} \ X
 \,,
\label{Cauchy10}
\end{equation}
with some polynomial ${\cal Q}$ in $\GG{2k}^0(2\pi i\partial_\tau)^n
\text{\textbf{A}}[\CG] $ where $ k,n \geq 0$. Then, one can coherently
replace $\text{\textbf{A}}[\CG] \rightarrow \text{\textbf{D}}[\CG]$
as well as $2\pi i\partial_\tau \rightarrow \pi \nabla$ and $\GG{2k}^0
\rightarrow (\Im(\tau))^{2k} \GG{2k}$ in that polynomial and obtain
a Cauchy--Riemann equation among modular graph functions
\begin{equation}
{\cal Q}(\pi \nabla, (\Im(\tau))^{2k} \GG{2k},  \text{\textbf{D}}[\CG])
= 0
 \,.
\label{Cauchy11}
\end{equation}
This procedure has been used at weight $w = 5,6$ to derive conjectural
Cauchy--Riemann differential equations for modular graph functions from
$A$-cycle graph functions and thus constitutes an alternative way compared to
the graphical manipulations of \rcites{DHoker:2016mwo, DHoker:2016quv}.  Our
method has been checked to either reproduce the Cauchy--Riemann equations in
the above reference or to yield expressions for modular graph functions that
satisfy the Laplace equations in \subsecref{ssec:modularB} as discussed in the
following section.


\subsubsection{Integrating Cauchy--Riemann equations}
\label{intproc}

We shall now describe techniques to convert Cauchy--Riemann equations derived
via \eqns{Cauchy10}{Cauchy11} into explicit representations of modular graph
functions. The idea is to solve the differential equations in terms of iterated
Eisenstein integrals \eqn{eqn:iteis0} along with integer powers of $\Im(\tau)$
and to fix the integration constants via modular invariance and reality of
$\text{\textbf{D}}[\CG]$. However, these constraints do not fix the last integration constant 
which amounts to adding MZVs of the appropriate weight to the modular graph function 
under investigations. This shortcoming can be fixed either by numerical evaluation or
by employing the alternative method described in \subsecref{sec:funrules}.

In case of one-loop graphs, \eqn{CR2} can be integrated to yield the
representation \eqn{nheis} of non-holomorphic Eisenstein series $\EE{k}$ up to
integration constants and antiholomorphic iterated Eisenstein integrals.  The
case $k=2$ in \eqn{CR2} reads
\begin{equation}
(\pi \nabla)^2 \EE{2} = 6(\Im(\tau))^4 \GG{4}
\end{equation}
which -- upon integration in $\tau$ -- yields
\begin{equation}
\pi \nabla \EE{2} = \frac{2 y^3 }{45} +  c_1  \zeta_3 + 24 y^2 \cez(4) + 12 y \cez(4,0) + 3 
 \cez(4,0,0) + c_2  \overline{\cez(4,0,0)} 
 \label{CR6}
\end{equation}
with rational constants $c_1, c_2$ and $y= \pi \Im(\tau)$. Then, a further
integration gives rise to
\begin{equation}
\EE{2} =  \frac{y^2}{45} - c_1 \frac{ \zeta_3}{y}   - 6  \cez({4, 0}) + c_3 \overline{ \cez({4, 0}) } - \frac{3}{y} \cez({4, 0, 0})  -  \frac{c_2}{y}  \overline{\cez({4, 0, 0}) } 
 \label{CR6p}
\end{equation}
with another rational constant $c_3$. While performing the above integrations,
we have used that Cauchy--Riemann derivatives act via
\begin{equation}
\pi \nabla(y^n) = n \, y^{n+1} \,, \ \ \ \ \ \ \pi \nabla ( \cez(k_1,k_2,\ldots,k_r)) = \frac{ 4y^2}{(2\pi i)^{k_r}} \, \GG{k_r}^0 \,  \cez(k_1,k_2,\ldots,k_{r-1}) \,,
 \label{CR6pp}
\end{equation}
and the integration constants $c_i \in \mathbb Q$ have been introduced
following two selection rules:
\begin{itemize}
\item[(i)] Let $\text{\textbf{D}}[\CG_w]$ denote a modular graph function of
  weight $w$, then the admissible integration constants in $(\pi \nabla)^n
  \text{\textbf{D}}[\CG_w]$ without any accompanying
  $\overline{\cez(\underline{k}) }$ are rational combinations of single-valued
  MZVs of weight $w{+}n$.
\item[(ii)] Whenever $(\pi \nabla)^n \text{\textbf{D}}[\CG_w]$ contains a term
  $\zeta^{\rm sv}_{\underline{m}} \cez(\underline{k})$, then rational multiples
  of its complex conjugate $\zeta^{\rm sv}_{\underline{m}} \overline{
  \cez(\underline{k})}$ have to be included in the integration constant.
\end{itemize}
Note that, as a consequence of (i), there is no rational multiple of $\zeta_2$
in \eqn{CR6p}.

The rational constants $c_i \in \mathbb Q$ in \eqn{CR6p} can be fixed by
imposing reality $\text{\textbf{D}}[\CG_w]=
\overline{\text{\textbf{D}}[\CG_w]}$ and modular invariance: Reality requires
the coefficients of $ \cez({4, 0}) $ and $ \overline{ \cez({4, 0}) }$ as well
as $ \cez({4, 0, 0}) $ and $\overline{\cez({4, 0, 0}) } $ to match, yielding
$c_2 = 3$ and $c_3 = -6$. Then, the modular transformations
\eqns{simex1}{simex2} of $\cez({4, 0}), \cez({4, 0, 0})$ and their complex
conjugates introduce $\zeta_3$ in a way such that \eqn{CR6p} can only be
modular invariant for $c_1=-1$. Hence, we arrive at
\begin{equation}
\EE{2} =  \frac{y^2}{45} + \frac{ \zeta_3}{y}   - 12 \Re[  \cez({4, 0}) ] - \frac{6}{y} \Re[\cez({4, 0, 0}) ] \,,
 \label{CR6ppp}
\end{equation}
which agrees with \eqn{nheis}. However, the criterion based on modular
invariance still leaves the freedom to add single-valued MZVs to
$\text{\textbf{D}}[\CG_w]$ which do not exist in the case at hand with $w=2$.
When applying the above integration procedure to obtain the expressions
\begin{align}
  \EE{3} &= \frac{2y^3}{945} + \frac{3 \zeta_5}{4y^2}  
  -120 \Re[ \cez({6, 0, 0}) ]- \frac{180 }{y} \Re[ \cez({6, 0, 0, 0})] - \frac{ 90}{y^2} \Re[ \cez({6, 0, 0, 0, 0}) ]
  \label{nholo3} \\
  \EE{4} &=  \frac{y^4}{4725} + \frac{5 \zeta_7}{8 y^3}
  -1680 \Re [\cez({8, 0, 0, 0}) ]- \frac{5040 }{y} \Re [\cez({8, 0, 0, 0, 0})] \notag \\
& \ \ \ \ \ - 
\frac{ 6300 }{y^2} \Re [ \cez({8, 0, 0, 0, 0, 0}) ] - \frac{ 3150 }{y^3} \Re [ \cez({8, 0, 0, 0, 0, 0, 0}) ]
  \label{nholo4} 
\end{align}
at weight $w=3,4$, the absence of $\zeta_3$ in $\EE{3}$ must be checked either
by numerical evaluation or by the methods of \secref{sec:funrules}.

Note that the task of integrating Cauchy--Riemann equations is completely
analogous to computing modular transformations of iterated Eisenstein integrals
from their differential equations, see \secref{ssec:Bdiff}. In particular, the
differential operator $\sim \tau^2 \partial_\tau$ for recursive computations of
$B$-cycle graph functions in eqs.\ (\ref{diffmod1}) to (\ref{diffmod3}) can be
mapped to the Cauchy--Riemann derivative \eqn{CR1} by replacing $\tau^2
\partial_\tau \rightarrow (\Im \tau)^2 \partial_\tau$. This is another reason
to expect strong parallels between $B$-cycle graph functions and modular graph
functions.


\subsubsection{Simplifying Cauchy--Riemann equations for multi-loop graphs}
\label{sec:service}

When applying the integration procedure of the previous subsection to modular
graph functions corresponding to graphs with more than one loop, it is useful
to disentangle iterated Eisenstein integrals with different types of entries.
For instance, the simplest irreducible two-loop modular graph function
$\Dcyc{G211}$ will comprise two kinds of iterated Eisenstein
integrals involving either two instances of $\GG{4}^0$ or a single integration
kernel $\GG{8}^0$. Any appearance of $\GG{8}^0$ in modular graph functions at
weight four can be captured via $\EE{4}$, so it is convenient to study the
linear combination
\begin{equation}
\EE{2,2} := \Dcyc{G211}  - \frac{9}{10} \EE{4}    \label{higher09a}
\end{equation}
for which the Cauchy--Riemann equation \eqref{Cauchy9} simplifies to
\begin{equation}
(\pi \nabla)^3  \EE{2,2}=  - 6 \Im(\tau)^4 \GG{4} \pi \nabla \EE{2} \,.
\label{newCR1}
\end{equation}
Then, starting from the representation \eqref{CR6ppp} of $\EE{2}$, integration
of \eqn{newCR1} yields depth-two iterated Eisenstein integrals with two entries
of $\GG{4}^0$. This observation motivates us to define the depth of a modular
graph function to be the minimum depth of the iterated Eisenstein integrals
required to represent it, see \secref{SectionItEisInt}.  Hence, the object
$\EE{2,2} $ in \eqn{higher09a} is our simplest example of a modular graph
function of depth two.

Similarly, Cauchy--Riemann equations at higher weight (which can be extracted
from \rcites{DHoker:2016mwo, DHoker:2016quv} and which we obtained from
employing the correspondence in \eqns{Cauchy10}{Cauchy11}) simplify when
considering the following combinations:
\begin{align}
\EE{2,3} &= \Dcyc{G2111} - \frac{43}{35} \text{E}_5   \label{higher09b} \\
\EE{3,3}&=   3 \Dcyc{GC321} + \Dcyc{GC222} - \frac{15}{14} \text{E}_6   \label{higher09c}  \\
\text{E}'_{3,3} &=   \Dcyc{GC321} + \frac{17}{60} \Dcyc{GC222} - \frac{59}{140} \text{E}_6
  \label{higher09d} \\
\EE{2,4} &=   9 \Dcyc{GC411} + 3 \Dcyc{GC321} + \Dcyc{GC222} - 13 \text{E}_6   \label{higher09e}  \\
\EE{2,2,2} &= - \Dcyc{GC2211}   + 
 \frac{232}{45} \Dcyc{GC222} + \frac{292}{15} \Dcyc{GC321} + \frac{2}{5} \Dcyc{GC411}+ 2 \text{E}_3^2 + \text{E}_2 \text{E}_4 - \frac{466}{45} \text{E}_6 \,.   \label{higher09f} 
\end{align}
The above combinations can be thought of as higher-depth generalizations of
non-holomorphic Eisenstein series. The benefit of the subtractions of
$\text{E}_k$ in \eqn{higher09b} to \eqn{higher09f} becomes
apparent\footnote{Note that these subtractions also simplify the respective
Laplace equations, e.g.\ we have $(\Delta-2)\EE{2,2} = - \EE{2}^2$ instead of
\eqn{unique11}.} in
\begin{align}
(\pi \nabla)^3 \EE{2,3}&= - 2 (\pi \nabla \text{E}_2) (\pi \nabla)^2 \text{E}_3
- 4 \Im(\tau)^4 \GG{4}  \pi \nabla \text{E}_3
\label{newCR2}\\
(\pi \nabla )^5 \EE{3,3} &= 180 \Im(\tau)^6 \GG{6}  (\pi \nabla)^2 \text{E}_3 \label{newCR3} \\
(\pi \nabla)^4\, \text{E}'_{3,3} &= - 12 \Im(\tau)^6 \GG{6} (\pi \nabla) \text{E}_3   
\label{newCR4}\\
(\pi \nabla)^3 \EE{2,4} &= - 27 \Im(\tau)^4 \GG{4} (\pi \nabla) \text{E}_4   + \text{R}_{2,4}  \label{newCR5} \\
\pi \nabla \, \text{R}_{2,4} &= - 81 \Im(\tau)^4 \GG{4} (\pi \nabla)^2 \text{E}_4  - 27 (\pi \nabla) \text{E}_2  (\pi \nabla)^3 \text{E}_4 \label{newCR6}
\end{align}
from which we can anticipate all of $\EE{2,3},\EE{3,3},\text{E}'_{3,3}$ and
$\EE{2,4}$ to be of depth two.  Finally, modular graph functions at weight six
contain one independent depth-three representative satisfying
\begin{align}
(\pi \nabla)^3 \EE{2,2,2} &=
-12 \Im(\tau)^4 \GG{4} \pi \nabla \EE{2,2} + (\pi \nabla \text{E}_2)^3  \,. \label{newCR7}
\end{align}
For all terms $\nabla^n \text{E}_{\underline{k}}$ on the right-hand side of the
above Cauchy--Riemann equations, a representation in terms of iterated
Eisenstein integrals $\cez$ can be found in \appref{app:service}.  We will now
proceed to solving \eqn{newCR1} and eqs.\ \eqref{newCR2} to \eqref{newCR7}
using the method in \subsecref{intproc}.


\subsubsection{Explicit solutions to Cauchy--Riemann equations at higher depth}
\label{sec:modezeroEX}

For the simplest modular graph function of depth two, $\EE{2,2}$, the
differential equation \eqn{newCR1} can be integrated to yield
\begin{align}
\EE{2,2} &= -\frac{ y^4}{20250} + \frac{y \zeta_3}{45}+ \frac{ 5 \zeta_5}{12 y}  - \frac{ \zeta_3^2}{4 y^2} -
\Big(\frac{ 2y}{15}  - \frac{ 3  \zeta_3}{y^2} \Big) \Re[ \cez(4, 0, 0) ]   \notag \\
&- \frac{9}{2y^2} \big( \Re[ \cez(4, 0, 0) ] ^2+\Im[ \cez(4, 0, 0) ] ^2 \big)
 - 72 \Re[ \cez(4, 4, 0, 0) ]    -  \frac{1}{5} \Re[ \cez(4, 0, 0, 0) ]   \notag \\
& - \frac{36 \Re[ \cez(4, 0, 4, 0, 0) ] }{y} 
 - \frac{ 108 \Re[ \cez(4, 4, 0, 0, 0) ] }{y} 
 - \frac{ \Re[ \cez(4, 0, 0, 0, 0) ] }{10 y}  \label{newCR8raw} \\
& - \frac{9 \Re[\cez(4, 0, 0, 4, 0, 0)] }{y^2} 
 - \frac{ 27 \Re[\cez(4, 0, 4, 0, 0, 0)] }{y^2} 
 - \frac{ 54 \Re[\cez(4, 4, 0, 0, 0, 0)] }{y^2}  \ , \notag
\end{align}
see \eqn{service1} for a convenient representation of the factor $\nabla
\text{E}_2$ therein.  Unlike the expression for $\EE{k}$, \eqn{newCR8raw}
contains products of holomorphic and antiholomorphic iterated Eisenstein
integrals, for example in 
\begin{equation}
\Re[ \cez(4, 0, 0) ] ^2 = \frac{1}{4}\cez(4, 0, 0)^2 +
\frac{1}{2}\cez(4, 0, 0) \overline{\cez(4, 0, 0)} + \frac{1}{4}\overline{\cez(4, 0, 0)}^2
\end{equation}
and $\Im[ \cez(4, 0, 0) ]^2$. The latter can be eliminated from 
\eqn{newCR8raw} by taking the real part of $\cez(4, 0, 0)^2$ and taking the shuffle relation 
\begin{equation}
\cez(4, 0, 0)^2= 2\cez(4, 0, 0,4,0,0) + 6 \cez(4, 0,
 4,0,0,0) + 12 \cez(4, 4,0,0,0, 0) \ ,
 \end{equation}
into account. This manipulation turns out to cancel all iterated
Eisenstein integrals of length six from \eqn{newCR8raw}:
\begin{align}
\EE{2,2} &= -\frac{ y^4}{20250} + \frac{y \zeta_3}{45}+ \frac{ 5 \zeta_5}{12 y}  - \frac{ \zeta_3^2}{4 y^2} -
\Big(\frac{ 2y}{15}  - \frac{ 3  \zeta_3}{y^2} \Big) \Re[ \cez(4, 0, 0) ]   \notag \\
&- \frac{9 \Re[ \cez(4, 0, 0) ] ^2}{y^2}
 - 72 \Re[ \cez(4, 4, 0, 0) ]    -  \frac{1}{5} \Re[ \cez(4, 0, 0, 0) ]   \label{newCR8}\\
& - \frac{36 \Re[ \cez(4, 0, 4, 0, 0) ] }{y} 
 - \frac{ 108 \Re[ \cez(4, 4, 0, 0, 0) ] }{y} 
 - \frac{ \Re[ \cez(4, 0, 0, 0, 0) ] }{10 y} \, . \notag
\end{align}
The coefficients of $\frac{\zeta_5}{y}$ and $\frac{\zeta_3^2}{y^2}$ in 
\eqns{newCR8raw}{newCR8} appear as integration
  constants in intermediate steps and can by fixed by imposing modular
  invariance\footnote{The modular transformations in eqs.\ \eqref{simex2},
  \eqref{modular91a} and \eqref{modular92} are sufficient to check this.} of
  \eqn{newCR8}. We have checked the resulting expression for $\Dcyc{G211}$ to
  satisfy the Laplace eigenvalue equation \eqref{unique11}, and its coefficient
  of $q^1 \overline{q}^0$ has been verified to agree with the results of
  \rcite{DHoker:2015gmr}. By inserting the $q$-expansion \eqn{qgamma1} of iterated 
  Eisenstein integrals, any term in the expansion \eqn{ModGraphExpansion}
  of modular graph functions around the cusp is readily available from \eqn{newCR8}
  and similar expressions below.
      
Similarly, the Cauchy--Riemann equation \eqref{newCR2} for the depth-two
modular graph function \eqn{higher09b} at weight five can be integrated to yield
\begin{align}
\EE{2,3} &= - \frac{ 4 y^5}{297675} + \frac{2y^2 \zeta_3}{945}  - \frac{\zeta_5}{180}
  - \frac{ \zeta_3 \zeta_5}{2 y^3} + \frac{ 7 \zeta_7}{16 y^2}
-  \Big( \frac{ 4 y^2}{315}   - \frac{  3  \zeta_5}{y^3} \Big) \Re[ \cez(4, 0, 0) ]  \notag \\
&
 - \Big( \frac{ 8y}{3} - \frac{  60  \zeta_3}{y^2} \Big)  \Re[ \cez(6, 0, 0, 0) ] 
 -\Big( 8   - \frac{  60 \zeta_3}{y^3} \Big) \Re[ \cez(6, 0, 0, 0, 0) ] \notag \\
&   - \frac{ 360 \Re[ \cez(4, 0, 0) ]  \Re[ \cez(6, 0, 0, 0) ] }{y^2}
  - \frac{  360 \Re[ \cez(4, 0, 0) ]  \Re[ \cez(6, 0, 0, 0, 0) ] }{y^3} \notag \\
&   - 1440 \Re[ \cez(4, 6, 0, 0, 0) ]   + \frac{ \Re[ \cez(4, 0, 0, 0, 0) ] }{21}
   -  480 \Re[ \cez(6, 0, 4, 0, 0) ]  - 1440 \Re[ \cez(6, 4, 0, 0, 0) ] 
\notag\\
&    - \frac{ 720 \Re[ \cez(4, 0, 6, 0, 0, 0) ] }{y} 
   - \frac{  4320 \Re[ \cez(4, 6, 0, 0, 0, 0) ] }{y}
  + \frac{  \Re[ \cez(4, 0, 0, 0, 0, 0) ] }{14 y}
  \label{newCR9} \\
& - \frac{  720 \Re[ \cez(6, 0, 0, 4, 0, 0) ] }{y} 
 - \frac{  2160 \Re[ \cez(6, 0, 4, 0, 0, 0) ] }{y} 
 - \frac{  4320 \Re[ \cez(6, 4, 0, 0, 0, 0) ] }{y} \notag \\
 &  - \frac{  10 \Re[ \cez(6, 0, 0, 0, 0, 0) ] }{y}
   - \frac{ 720 \Re[ \cez(4, 0, 6, 0, 0, 0, 0) ] }{y^2}
    - \frac{  3600 \Re[ \cez(4, 6, 0, 0, 0, 0, 0) ] }{y^2}  \notag \\
& +\frac{  \Re[ \cez(4, 0, 0, 0, 0, 0, 0) ] }{ 28 y^2}
 - \frac{  360 \Re[ \cez(6, 0, 0, 0, 4, 0, 0) ] }{y^2} 
 - \frac{  1080 \Re[ \cez(6, 0, 0, 4, 0, 0, 0) ] }{y^2} \notag \\
&  - \frac{  2160 \Re[ \cez(6, 0, 4, 0, 0, 0, 0) ] }{y^2}
   - \frac{  3600 \Re[ \cez(6, 4, 0, 0, 0, 0, 0) ] }{y^2}
    - \frac{  5 \Re[ \cez(6, 0, 0, 0, 0, 0, 0) ] }{y^2} \,, \notag
\end{align}
see \eqns{service2}{service3} for explicit expressions of $\nabla \text{E}_3$
and $\nabla^2 \text{E}_3$. Following the strategy of simplifying
$\text{E}_{2,2}$, we have eliminated the appearance of $\Im[ \cez(4, 0, 0) ]
\Im[ \cez(6, 0, 0, 0) ]$ and $\Im[ \cez(4, 0, 0) ]  \Im[ \cez(6, 0, 0, 0, 0) ]
$ in intermediate steps by taking the real part of appropriate shuffle
relations. These manipulations also remove all iterated Eisenstein integrals of
length $8$ from our final expression \eqn{newCR9}.  Hence, elimination of any
$\Im[ \cez(\ldots) ] $ via shuffle relations will be our guiding principle for
all subsequent cases which turns out to reduce the maximum length of the
iterated Eisenstein integrals appearing in a given $\text{E}_{\underline{k}}$.

The coefficient of $\zeta_5$ in $\text{E}_{2,3}$ is not fixed by modular
invariance and can be inferred by comparison with the results in the
literature, numerical evaluation or by the method discussed in
\subsecref{sec:funrules}. The expression for $\Dcyc{G2111}$ resulting from
\eqn{newCR9} has been  checked to satisfy the Laplace equation
\eqref{higher07}, and its coefficient of $q^1 \overline{q}^0$ agrees with the
results of \cite{DHoker:2015gmr}.

There are three independent modular graph functions at weight six and depth
two: $\EE{3,3},\EE{3,3}'$ as well as $\EE{2,4}$ defined in
eqs.~\eqref{higher09c} to \eqref{higher09e} are a convenient choice of basis.
Integrating the Cauchy--Riemann equation \eqref{newCR3} for $\EE{3,3}$ gives
rise to
\begin{align}
\EE{3,3} &= \frac{ 2 y^6}{6251175} + \frac{y \zeta_5 }{210} + \frac{ \zeta_7}{16 y} 
- \frac{7 \zeta_9}{64 y^3} + \frac{9 \zeta_5^2}{ 64 y^4} - \Big( \frac{4y}{7}  + \frac{ 135   \zeta_5}{4 y^4} \Big)  \Re[ \cez(6, 0, 0, 0, 0) ]\notag \\
&  + \frac{  2025 \Re[ \cez(6, 0, 0, 0, 0) ] ^2}{y^4 } + 21600 \Re[ \cez(6, 6, 0, 0, 0, 0) ]  -  \frac{20}{7} \Re[ \cez(6, 0, 0, 0, 0, 0) ]   \notag \\
&  + \frac{ 21600 \Re[ \cez(6, 0, 6, 0, 0, 0, 0) ] }{y} + \frac{
 108000 \Re[ \cez(6, 6, 0, 0, 0, 0, 0) ] }{y} \notag \\
 &   - \frac{ 45 \Re[ \cez(6, 0, 0, 0, 0, 0, 0) ] }{7 y}
  + \frac{ 16200 \Re[ \cez(6, 0, 0, 6, 0, 0, 0, 0) ] }{y^2}  \label{newCR10} \\
& + \frac{ 81000 \Re[ \cez(6, 0, 6, 0, 0, 0, 0, 0) ] }{y^2} + 
\frac{ 243000 \Re[ \cez(6, 6, 0, 0, 0, 0, 0, 0) ] }{y^2}  \notag \\
& - \frac{ 15 \Re[ \cez(6, 0, 0, 0, 0, 0, 0, 0) ] }{2 y^2}
+ \frac{ 8100 \Re[ \cez(6, 0, 0, 0, 6, 0, 0, 0, 0) ] }{y^3} \notag \\
& + \frac{ 40500 \Re[ \cez(6, 0, 0, 6, 0, 0, 0, 0, 0) ] }{y^3}
  + \frac{ 121500 \Re[ \cez(6, 0, 6, 0, 0, 0, 0, 0, 0) ] }{y^3} \notag \\
  & + \frac{ 283500 \Re[ \cez(6, 6, 0, 0, 0, 0, 0, 0, 0) ] }{y^3} 
  - \frac{ 15 \Re[ \cez(6, 0, 0, 0, 0, 0, 0, 0, 0) ] }{4 y^3}  \,, \notag
\end{align}
and similar expressions for $\text{E}'_{3,3}$ and $\EE{2,4}$ based on
eqs.~\eqref{newCR4} to \eqref{newCR6} are provided in \appref{app:explicitmod}.
The resulting expressions for $\Dcyc{GC411}, \Dcyc{GC321}$ and $\Dcyc{GC222}$
have been checked to satisfy the Laplace eigenvalue equations \eqref{higher10}.

Finally, there is a single irreducible modular graph function of
depth three at weight six: $\EE{2,2,2}$ defined in \eqn{higher09f}. Integrating its
Cauchy--Riemann equation \eqref{newCR7} (with $\nabla \EE{2,2}$ spelt out in
\eqn{service7}) yields 
\begin{align}
\EE{2,2,2} &=  \frac{ 4 y^6}{9568125}   - \frac{ 2 y^3 \zeta_3}{10125 }
+  \frac{ y \zeta_5 }{54}   +  \frac{ \zeta_3^2}{90}  + \frac{661 \zeta_7}{1800 y}
  - \frac{ 5 \zeta_3 \zeta_5}{12 y^2} + \frac{ \zeta_3^3}{6 y^3}  \notag \\
&\! \! \! + \Big( \frac{ 4 y^3  }{3375}  {-}  \frac{ 2 \zeta_3 }{15 }
 {+} \frac{ 5   \zeta_5}{2 y^2} {-} \frac{  3  \zeta_3^2}{y^3} \Big) \Re[ \cez(4, 0, 0) ] + \Big( \frac{ 2}{5}  {+} \frac{  18 \zeta_3}{y^3}  \Big) \Re[ \cez(4, 0, 0) ] ^2
  - \frac{ 36 \Re[ \cez(4, 0, 0) ] ^3}{y^3}  \notag \\
&\! \! \!    - 36   \Big( \frac{ 2y}{45} - \frac{ \zeta_3}{ y^2} + \frac{ 6 \Re[ \cez(4, 0, 0) ] }{y^2} \Big)
 \Re \! \Big[ \cez(4, 0, 4, 0, 0) ] + 3   \cez(4, 4, 0, 0, 0) 
  + \frac{    \cez(4, 0, 0, 0, 0) }{360} \Big]  \notag \\
&\! \! \! 
 - 864 \Re[ \cez(4, 4, 0, 4, 0, 0) ]  - 2592 \Re[ \cez(4, 4, 4, 0, 0, 0) ]
   - \frac{ 12}{5} \Re[ \cez(4, 0, 0, 4, 0, 0) ]  \notag \\
&\! \! \!  - \frac{  36}{5} \Re[ \cez(4, 0, 4, 0, 0, 0) ] 
    - \frac{ 84}{5} \Re[ \cez(4, 4, 0, 0, 0, 0) ] -  \frac{ 1}{150} \Re[ \cez(4, 0, 0, 0, 0, 0) ]   \label{newCR13} \\
 &\! \! \!  - \frac{  1296 \Re[ \cez(4, 4, 0, 0, 4, 0, 0) ] }{y}
 - \frac{  3888 \Re[ \cez(4, 4, 0, 4, 0, 0, 0) ] }{y}
  - \frac{  7776 \Re[ \cez(4, 4, 4, 0, 0, 0, 0) ] }{y} \notag \\
  &\! \! \!  - \frac{  432 \Re[ \cez(4, 0, 4, 0, 4, 0, 0) ] }{y}
   - \frac{  1296 \Re[ \cez(4, 0, 4, 4, 0, 0, 0) ] }{y}
  - \frac{ 6 \Re[ \cez(4, 0, 0, 0, 4, 0, 0) ] }{5 y} \notag \\
  &\! \! \!  - \frac{ 18 \Re[ \cez(4, 0, 0, 4, 0, 0, 0) ] }{5 y}
   - \frac{  42 \Re[ \cez(4, 0, 4, 0, 0, 0, 0) ] }{5 y} \notag \\
   &\! \! \!  - \frac{  18 \Re[ \cez(4, 4, 0, 0, 0, 0, 0) ] }{y} 
   - \frac{   \Re[ \cez(4, 0, 0, 0, 0, 0, 0) ] }{300 y}  \,,\notag
\end{align}
which, together with $\EE{3,3},\text{E}'_{3,3}$ and $\EE{2,4}$, completes the
basis of weight-six modular graph functions under the relations in
\appref{app:w6rel}. For all the above expressions for modular graph functions,
modular invariance has been confirmed numerically. 

All the above examples confirm our conjecture that the number of loops in a
graph is an upper bound for the depth of the associated modular graph function.
Said upper bound is saturated for the independent modular graph functions
$\Dcyc{G211},\Dcyc{G2111} ,\Dcyc{GC321},\Dcyc{GC222}, \Dcyc{GC411} $ and $
\Dcyc{GC2211}$ at weight $w \leq 6$. However, $\Dcyc{G3}$ being of depth one
(cf.~\eqn{eqn:weightthreerelation}) and $\Dcyc{G6}$ being of depth three
(cf.~\eqn{eqn:weightsixrelations}) are examples where the loop order exceeds
the depth.


\subsubsection{Laplace equation at weight six}
\label{sec:newlapl}

From their representations in terms of iterated Eisenstein integrals, we infer
the following Laplace equation among modular graph functions which has not yet
been spelt out in the literature: 
\begin{align}
&(\Delta-2)\big( \Dcyc{GC2211} - 2  \EE{3}^2 - \EE{2} \EE{4}  \big)  -\frac{14}{9} \Dcyc{GC222} + 
\frac{16}{3} \Dcyc{GC321} -4 \Dcyc{GC411} \notag \\
& \ \ \ \ \   + \frac{284}{9} \EE{6}   +  \frac{ 2}{3} \EE{2}^3 + 16 \EE{3}^2 + 
\frac{12}{5} \EE{2} \EE{4}  -4 \EE{2}  \EE{2,2} 
 = 0 \,.  \label{dpt3.5}
\end{align}
The combination $\Dcyc{GC2211} - 2  \EE{3}^2 - \EE{2} \EE{4}$ along with the
Laplacian is designed to absorb contributions $\sim \partial_\tau \EE{p}
\overline{ \partial_{\tau} \EE{q}  }$ in \eqn{dpt3.5} with $p+q=6$.  Moreover,
the combination $\Dcyc{GC2211} - 2  \EE{3}^2 - \EE{2} \EE{4}$ is selected by
the formalism of \rcite{DHoker:2016quv} to linearize the relations between
modular graph functions\footnote{The general formalism \rcite{DHoker:2016quv}
  assigns a so-called ``primitive'' version to each modular graph function
  which is observed to linearize all relations known up to date. We are grateful
to Eric D'Hoker and Justin Kaidi for bringing the connection between primitive
modular graph functions and the Laplace equation (\ref{dpt3.5}) to our
attention.}, as can be verified from the second equation from below in
\eqn{eqn:weightsixrelations}.


\subsubsection{Representations of modular graph functions in terms of $\ce$ rather than $\cez$?}

While all expressions for modular graph functions or their constituents have
been expressed in terms of iterated Eisenstein integrals $\cez$ defined in
\eqn{eqn:iteis0}, we conclude this subsection with expressions for modular
graph functions in terms of iterated Eisenstein integrals $\ce$ defined in
\eqn{eqn:iteis}, where the constant terms $2\zeta_k$ of the integrands $\GG{k}$
are not subtracted. At depth one, these $\ce$ appear to be the more suitable
language for modular graph functions than the $\cez$ since the polynomial term
$\EE{k} \sim y^k$ in \eqn{eisen1} is absorbed in this way:
\begin{align}
  \EE{k} &= \frac{ 4 \, (2k{-}3)! \, \zeta_{2k-1} \, (4y)^{1-k} }{(k{-}2)! \, (k{-}1)!  } 
  - 8y(2k{-}1)!\suml_{j=0}^{k-1}\binom{2k{-}2{-}j}{k{-}1}\frac{(4y)^{j-k}}{j!}\,\Re[\ce(2k,\underbrace{0,\ldots,0}_{2k-2-j};q)]\,. \label{CE1}
\end{align}
However, the analogous rearrangements at depth two convert \eqn{newCR8} into
\begin{align}
\EE{2,2} &= \frac{\zeta_3 \, |T|^2}{60 y} + \frac{ 5 \zeta_5}{12 y}  - \frac{ \zeta_3^2}{4 y^2}  + \frac{ 3  \zeta_3}{y^2}  \Re[ \ce(4, 0, 0) ] - \frac{9 \Re[ \ce(4, 0, 0) ] ^2}{y^2} \label{CE1prime} \\
& - 72 \Re[ \ce(4, 4, 0, 0) ]    - \frac{36 \Re[ \ce(4, 0, 4, 0, 0) ] }{y} 
 - \frac{ 108 \Re[ \ce(4, 4, 0, 0, 0) ] }{y} 
 \notag
\end{align}
and introduce an explicit appearance of $\Re \tau$ via $|T|^2 = \pi^2 ( (\Re
\tau)^2 + (\Im \tau)^2)$. Similar observations have been made for $\EE{2,3}$
and examples at higher weight, so it is not clear if representations in terms
of $\ce$ are preferable at generic depth.


\subsection{Modular graph functions from
\texorpdfstring{$B$}{$B$}-cycle graph functions}
\label{sec:funrules}

In this section, we suggest a mapping between $B$-cycle graph functions and the
corresponding modular graph functions which is based on their representations
via iterated Eisenstein integrals (see \subsecref{sec:evalBEX} and
\subsecref{sec:modezeroEX}, respectively).


\subsubsection{Depth one}

For illustrative purposes, we repeat the expressions
\begin{align}
\Dcyc{G2}  &=  \frac{y^2}{45} + \frac{ \zeta_3}{y}   - 12 \Re[  \cez({4, 0}) ] 
- \frac{6}{y} \Re[\cez({4, 0, 0}) ] 
\notag \\
\Dcyc{G111}  &=  
 \frac{2y^3}{945} + \frac{3 \zeta_5}{4y^2}  
  -120 \Re[ \cez({6, 0, 0}) ]- \frac{180 }{y} \Re[ \cez({6, 0, 0, 0})] - \frac{ 90}{y^2} \Re[ \cez({6, 0, 0, 0, 0}) ]
\notag \\
\Dcyc{G1111}  &=   \frac{y^4}{4725} + \frac{5 \zeta_7}{8 y^3}
  -1680 \Re [\cez({8, 0, 0, 0}) ]- \frac{5040 }{y} \Re [\cez({8, 0, 0, 0, 0})] \label{rep2}\\
& \ \ \ \ \ - 
\frac{ 6300 }{y^2} \Re [ \cez({8, 0, 0, 0, 0, 0}) ] - \frac{ 3150 }{y^3} \Re [ \cez({8, 0, 0, 0, 0, 0, 0}) ] \,,
\notag
\end{align}
for the simplest modular graph functions which agree with the all-weight
formula \eqn{nheis} for non-holomorphic Eisenstein series. These closed-string
expressions will be brought into correspondence with the analogous $B$-cycle
graph functions \eqn{bcyc1} modulo $\zeta_2$ on the open-string side,
\begin{align}
\Bcyc{G2} &=
\frac{ T^2}{180}   - \frac{ \zeta_3}{T}   - 6 \cez(4, 0) + \frac{6 \cez(4, 0, 0)}{T} \ {\rm mod} \ \zeta_2 \notag \\
\Bcyc{G111} &= -\frac{T^3}{3780}   + \frac{3 \zeta_5}{2 T^2}- 60 \cez(6, 0, 0) + \frac{180 \cez(6, 0, 0, 0)}{T} - \frac{
 180 \cez(6, 0, 0, 0, 0)}{T^2}  \ {\rm mod} \ \zeta_2
\notag \\
\Bcyc{G1111} &= \frac{T^4}{75600}  - \frac{5 \zeta_7}{ 2 T^3}  - 840 \cez(8, 0, 0, 0) + 
\frac{ 5040 \cez(8, 0, 0, 0, 0)}{T}
 \label{bcyc2} \\
 & \ \ \ \  
- \frac{ 12600 \cez(8, 0, 0, 0, 0, 0)}{T^2} 
+ \frac{ 12600 \cez(8, 0, 0, 0, 0, 0, 0)}{T^3}  \ {\rm mod} \ \zeta_2 \,.
\notag
 \end{align}
As in \eqn{bcyc5}, the notion of ``mod $\zeta_2$''  refers to
a representation of all the $\tau$-dependence via $T$ and $
\cez(\underline{k})$, where all terms of the form $\zeta_2^{n} T^m
\cez(\underline{k})$ with $n \geq 1$ and $m \in \ZZ$ are suppressed.

In comparing the above expressions for modular graph
functions and $B$-cycle graph functions, both iterated Eisenstein integrals
and Laurent polynomials in $y$ or $T$ exhibit striking similarities in their
coefficients: every single term in \eqn{bcyc2} will find a correspondent in
\eqn{rep2} once we replace
\begin{align}
T \rightarrow -2 y \,, \ \ \ \ 
\cez(2k, 0,\ldots)  \rightarrow 2 \Re[\cez(2k, 0,\ldots)] \,, \ \ \ \ 
\zeta_{2k+1} \rightarrow 2 \zeta_{2k+1} \,, \ \ \ \ 
\zeta_{2k} \rightarrow 0 \,.
\label{bcyc79}
\end{align} 
The $\tau$ dependence of the $\cez(\underline{k})$
through their $q$-series \eqn{qgamma1} is understood to be unaffected by
the prescription $T \rightarrow -2y$. 
The same correspondence has been verified between the depth-one modular graph
functions $\Dcyc{G11111} $, $\Dcyc{G111111}$ and their $B$-cycle counterparts
$\Bcyc{G11111} $, $\Bcyc{G111111}$, where the latter has been inferred from the 
$A$-cycle counterpart of the relations in \appref{app:w6rel}.


\subsubsection{General form}

Both the doubling of odd zeta-values in \eqn{bcyc79} and the suppression of
$\zeta_2$ in matching $B$-cycle graph functions with non-holomorphic Eisenstein
series are reminiscent of the single-valued projection of MZVs. From the above
examples associated with one-loop graphs, it is tempting to study the following
generalization of \eqn{bcyc79}\footnote{As pointed out in \appref{app:gamma0},
one can always write $\cez$'s in terms of powers of $T$ and $\cez$'s with
$k_1\neq 0$.} 
\begin{equation}
{\rm esv} : \ \ \ \left\{ 
\begin{aligned}[rcrcl]
(i)   &\hspace{8pt}&T                     &\rightarrow && \hspace{-5pt}-2 y \\
(ii)  &            &\cez(k_1,k_2,\ldots,k_r)   &\rightarrow && \hspace{-5pt}2 \Re [ \cez(k_1,k_2,\ldots,k_r) ],\quad k_1\neq 0 \\
(iii) &            &\zeta_{\underline{n}} &\rightarrow && \hspace{-5pt}\zeta^{\rm sv}_{\underline{n}}
\end{aligned} \right.
\label{eq:funrules}
\end{equation}
to arbitrary MZVs and iterated Eisenstein integrals. Note that part $(i)$ is in fact a special case of Brown's single valued map for multiple polylogarithms \cite{BrownSVMPL}, because $2T=\log q$ is sent to $-4y=\log q+\log\overline{q}$.\footnote{The reason why we define $T=\pi i\tau$ instead of $2\pi i\tau$ is that we want its image under the single-valued map to be consistent with the choice of variable $\mathbb{L}:=-2y$ operated in \rcites{Brown:2017qwo, Brown:2017qwo2}.} As before, part $(i)$ is understood to not act on the
$q$-series \eqn{qgamma1} of $\cez(k,\ldots)$ with $k\neq 0$.  Moreover, part
$(iii)$ motivates our earlier choices to occasionally display $B$-cycle graph
function modulo terms sent to zero by the esv-map such as $\zeta_2 T^m
\cez(\underline{k})$. As the key result of this section, we conjecture that,
once a $B$-cycle graph function is {\it suitably} expressed in terms of $T$ and
$\cez(\underline{k})$, the esv-map in \eqn{eq:funrules} yields the
corresponding modular graph function,
\begin{equation}
\esv \ \text{\textbf{B}}[\CG] = \text{\textbf{D}}[\CG]  \,.
\label{eq:funrules1}
\end{equation}
The notion of \textit{suitably} expressing $B$-cycle graph function in terms of
$T$ and $\cez(\underline{k})$ will be made more precise in the next
\subsecref{sec:solution} using examples at depth two and three. We must
introduce this notion, because the esv-map is a map on iterated integrals only
if we consider them as symbols and forget about the algebraic relations among
them.  The reason is, that these relations would not be respected by part
$(ii)$ of \eqn{eq:funrules}. 

Since open- and closed-string amplitudes comprise
generating functions of the respective graph functions, \eqn{eq:funrules1}
immediately implies the main result of this work -- the connection
\eqn{punchline} between the four-point open- and closed-string integrals
\eqns{Acyc4}{Acyc1}.

Let us already note here that the esv-map \eqn{eq:funrules} is consistent with
the truncation \eqn{Cauchy6} of $A$-cycle graph functions selected by the
replacement in \eqn{adhoc}: Using the result \eqn{Texpansion}, it follows that
adding any term $\zeta_{\underline{n}}\omm(\underline{m})$ contained in the
space $X$ defined in \secref{sec:ADrels} to an $A$-cycle graph function will
result in terms proportional to $\zeta_{\underline{n}} T^m \cez(\ldots)$ in the
corresponding $B$-cycle graph function\footnote{For instance, even though
  $\zeta_2 \omm_A(0,0,1,0\db\tau)= - \frac{\zeta_3}{8}+ \frac{3}{4} \cez(4, 0,
  0\db\tau) $ appears to introduce a term proportional to $\zeta_3$ which is
  preserved by the single-valued projection, the modular image $\zeta_2
  \omm_A(0,0,1,0\db\!- \tfrac{1}{\tau}) = \frac{\pi^6}{720 T^3} + \frac{\pi^4}{144 T} + \frac{\pi^2 T}{720}  + \frac{\pi^2 \zeta_3}{8 T^2} -\frac{
  3 \pi^2 \cez(4, 0, 0;\tau)}{4 T^2} $ and thereby the contribution to a
$B$-cycle graph function is sent to zero by the esv-map \eqn{eq:funrules} term
by term.}, which are in turn annihilated by part $(iii)$ of \eqn{eq:funrules}.
In other words, all terms contained in $X$ will yield zero upon taking their
modular transformation and applying the rules \eqn{eq:funrules} afterwards.
Accordingly, the observation of \subsecref{sec:ADrels} that $A$-cycle graph
functions -- after omission of terms from the space $X$ and replacing
$\zeta_{\underline{n}} \rightarrow \zeta^{\rm sv}_{\underline{n}} $ -- satisfy
the relations of modular graph functions, is made plausible by
\eqn{eq:funrules1}. 


\subsubsection{Higher depth}
\label{sec:solution}

We shall now discuss the representations of $B$-cycle graph functions in which
the esv-map \eqn{eq:funrules1} to modular graph functions is applicable. 
 At depth two, it is instructive to compare the expression \eqn{newCR8} for
$\EE{2,2}= \Dcyc{G211}  -\tfrac{9}{10}\Dcyc{G1111}$ with the analogous
$B$-cycle graph function
 \begin{align}
&\Bcyc{G211}  - \tfrac{9}{10}\Bcyc{G1111}  = - \frac{T^4}{324000} 
  - \frac{T \zeta_3}{180}  
- \frac{ 5 \zeta_5}{12 T} 
 - \frac{ \zeta_3^2}{4 T^2}  \notag \\
&
 +\Big(\frac{T}{30}  + \frac{3   \zeta_3}{T^2}   \Big) \cez(4, 0, 0)
  - \frac{9 \cez(4, 0, 0)^2}{T^2}  -  36 \cez(4, 4, 0, 0)  - \frac{1}{10} \cez(4, 0, 0, 0) 
 \label{bcyc4} \\
  &  + \frac{36 \cez(4, 0, 4, 0, 0)}{T} 
   + \frac{108 \cez(4, 4, 0, 0, 0)}{T}  
+ \frac{ \cez(4, 0, 0, 0, 0)}{10 T}
\ {\rm mod} \ \zeta_2   \,,  \notag
\end{align}
see \eqn{bcyc3} for the terms $\sim \zeta_2$ suppressed by the esv-map. In the
present form of \eqn{bcyc4}, the esv-map in \eqn{eq:funrules} correctly
reproduces the corresponding modular graph function in \eqn{newCR8}.  However,
as already anticipated in the previous section, a major shortcoming is that the
esv-map in \eqn{eq:funrules} is not compatible with shuffle multiplication:
Rewriting \eqn{bcyc4} via $\cez(4, 0, 0)^2=2\cez(4, 0, 0,4,0,0) + 6 \cez(4, 0,
4,0,0,0) + 12 \cez(4, 4,0,0,0, 0)$ results in a different image under the
esv-map. When performing this shuffle multiplication, one could at best hope to
make contact with the more cumbersome representation of $\EE{2,2} $ in
\eqn{newCR8raw} with spurious iterated Eisenstein integrals of length six, but
it is not clear how to extend the esv-map such as to generate $\Im [
\cez(\underline{k})]$.

In the depth-two case at hand, one can still argue that the representation in
\eqn{bcyc4} is optimized with respect to the length of the iterated Eisenstein
integrals and therefore particularly canonical: There is currently no
$\cez(\ldots)$ at length six, provided that the shuffle multiplication of
$\cez(4, 0, 0)^2$ is not performed.

At weight five, the expression \eqn{newCR9} for the two-loop modular graph
function $\EE{2,3}$ can be reached by applying \eqn{eq:funrules} to the
representation \eqn{bcyc5} of the corresponding $B$-cycle graph function.  For
the first non-trivial product $\cez(4, 0, 0) \cez(6, 0, 0, 0, 0)$ of iterated
Eisenstein integrals in \eqn{bcyc5}, the absence of $\cez(\underline{k})$ at
length eight in the remaining equation suggests to not perform this shuffle
multiplication. However, the other product $\cez(4, 0, 0) \cez(6, 0, 0, 0)$ in
\eqn{bcyc5} does not admit a comparable argument to leave it inert: Said
$B$-cycle graph function inevitably contains $\cez(\underline{k})$ at length
seven, independent on the treatment of $\cez(4, 0, 0) \cez(6, 0, 0, 0)$. 
 
We expect that each $B$-cycle graph function admits a scheme of performing
selected shuffle multiplications such that the esv-map \eqn{eq:funrules}
converts it to the corresponding modular graph function via \eqn{eq:funrules1}.
In absence of MZVs of depth $\geq 2$, one may reformulate 
the conjecture \eqn{eq:funrules1} in a way that is insensitive to the issues 
with the shuffle multiplication: given a representation of $\text{\textbf{D}}[\CG]$
in terms of $y,\zeta_{2k+1}$ and $\Re [ \cez(\underline{k}) ]$, reversing the
rules of \eqn{eq:funrules} via $y \rightarrow-\frac{ T}{2}, \ \zeta_{2k+1}\rightarrow \frac{\zeta_{2k+1}}{2}$ and
$\Re [ \cez(\underline{k}) ] \rightarrow \frac{  \cez(\underline{k}) }{2}$ is claimed
to yield the corresponding $ \text{\textbf{B}}[\CG] $ modulo $\zeta_2$. 

It would be desirable to identify a general criterion on the representations of
$B$-cycle graph functions that are tailored to the esv-map. It is encouraging to
observe that inappropriate ways of performing shuffle multiplications before
applying the esv-map seem to always result in a breakdown of modular
invariance.  Also, note that esv is well defined at depth one since any use of
shuffle multiplication would necessarily introduce cases $\cez(k_1,\ldots)$
with $k_1=0$ that are explicitly excluded from \eqn{eq:funrules}.

We have checked that the independent modular graph functions $\EE{3,3},
\text{E}'_{3,3},\EE{2,4}$ and $\EE{2,2,2}$ at weight six can be obtained
through the esv-map from suitable representations of the corresponding
$B$-cycle graph functions. This adds a depth-three example to support the
general conjecture \eqn{eq:funrules1}.


\subsubsection{Expressing esv rules in terms of \texorpdfstring{$\cez$}{$\cez$} versus
\texorpdfstring{$\ce$}{$\ce$}}

One very obvious question is whether one can find a formulation of the esv-map
in \eqn{eq:funrules} which applies to iterated Eisenstein integrals $\ce$
rather than $\cez$.  It has been noted in \eqn{CE1} that the leading term in
the Laurent polynomial of non-holomorphic Eisenstein series cancels when $\cez$
are collectively traded for $\ce$.  Indeed, inserting $\cez(4,0;\tau) =
\ce(4,0;\tau) -\frac{\pi^2 \tau^2}{360}$ and $\cez(4,0,0;\tau) =
\ce(4,0,0;\tau) - \frac{i\pi^3 \tau^3}{540}$ into the $B$-cycle graph function
\eqn{bcyc2} gives rise to the analogous cancellation of the term $T^2$,
\begin{equation}
\Bcyc{G2}  =  -\frac{ \zeta_3}{T}   - 6 \ce(4, 0) + \frac{6 \ce(4, 0, 0)}{T}  \ {\rm mod} \ \zeta_2 \,.
\end{equation}
Given that this effect persists in one-loop graph functions of higher weight,
it is conceivable that replacing the esv-rule $(ii)$ by ${\cal
E}(\underline{k}) \rightarrow 2 \Re [ \ce(\underline{k})]$ correctly reproduces
all $\text{E}_n$ from $B$-cycle graph functions in terms of iterated Eisenstein
integrals $\ce$.

At depth two, however, there is a discouraging example: When replacing $\cez$
by combinations of $\ce$ in \eqn{bcyc4}, one arrives at a shorter expression
 \begin{align}
\Bcyc{G211}  - \tfrac{9}{10}\Bcyc{G1111}  &=  -\frac{5 \zeta_5}{12 T}  - \frac{ \zeta_3^2}{4 T^2}  +  \frac{ 3 \zeta_3 \ce(4, 0, 0)}{T^2} - 
 \frac{  9 \ce(4, 0, 0)^2}{T^2}  - 36 \ce(4, 4, 0, 0) \notag \\
 &  \ \ \   + \frac{ 36}{T}
( \ce(4, 0, 4, 0, 0)+3 \ce(4, 4, 0, 0, 0))  \ {\rm mod} \ \zeta_2 \,,
\end{align}
which should be compared with the representation \eqn{CE1prime} of the modular
graph function $\EE{2,2}$. It turns out that there is no $B$-cycle analogue of
the term $\EE{2,2} = \frac{ \zeta_3 |T|^2}{60y}+\ldots$ in \eqn{CE1prime},
which was already mentioned as a drawback of representations in terms of $\ce$.
We hope that this particular term in the expression for $\EE{2,2}$ will shed
light on a reformulation of the esv-map which respects shuffle
multiplication.


\subsubsection{Zero modes}

We recall that (by abuse of nomenclature) the zero mode of a modular graph
function (see \eqn{ModGraphExpansion})
\begin{equation}
\DcycLetter[\CG]= \sum_{m,n=0}^{\infty} c^{{\cal
G}}_{m,n}(y)q^m \bar q^n,
\end{equation}
is defined to be $\smallDcycLetter[\CG]:=c^{\CG}_{0,0}(y)$. 
The analogue of the zero mode for a $B$-cycle graph function 
\begin{equation}
\BcycLetter[\CG]= \sum_{m=0}^{\infty} b^{{\cal
G}}_{m}(T)q^m 
\end{equation}
will be denoted as $\smallBcycLetter[\CG]:=b^{\CG}_{0}(T)$ and is also referred
to as a {\it constant term} in \appref{app:constantB}.  The map esv is well
defined on zero modes, as it does not present the problem of being dependent on
the way we write $\BcycLetter[\CG]$ in terms of iterated Eisenstein integrals
$\cez(\underline{k};q) = {\cal O}(q)$.  Hence, the conjectural formula
\begin{equation}
\esv \ \smallBcycLetter[\CG] = \smallDcycLetter[\CG]
\label{thisone}
\end{equation}
is well defined, and it has been verified on all examples up to weight six.
Moreover, in order to confirm part $(iii)$ of the esv-map for an MZV of depth
$3$ and weight $11$, where the sv-map acts in a non-trivial way, \eqn{thisone}
has been checked to hold for the weight-seven examples $\smallBcyc{G511}$ and
$\smallDcyc{G511}$ spelt out in \eqns{b511}{laurnewsv}.

Hence, we propose \eqn{thisone} as a conjectural method to compute the zero
modes $\text{\textbf{d}}[{\cal G}] $ of modular graph functions. Expressions
for the constant terms $\text{\textbf{b}}[\CG]$ of $B$-cycle graph functions
can be calculated using the methods described in \appref{app:constantB}. There
is no conceptual bottleneck to addressing graphs of arbitrary complexity in
this way, though the amount of data in intermediate steps of the calculations
in \appref{app:constantB} imposes practical limitations\footnote{The expression
  for $\smallBcyc{G511}$ in \eqn{b511} has been obtained by modular
transformation of the $A$-cycle graph function $\Acyc{G511}$, involving
numerical evaluations of multiple modular values, see \secref{ssec:Bcyc3}.} for
weights larger than six. 
As pointed out in \subsecref{sec:evalB}, the analytic computation of
$\text{\textbf{b}}[\CG]$ bypasses the necessity to numerically determine the
multiple modular values arising in the modular transformations of iterated
Eisenstein integrals described in \subsecref{ssec:Bcyc3} as well as the
integration constants in the method outlined in \subsecref{ssec:Bdiff}. 


\section{Non-planar \texorpdfstring{$A$}{$A$}-cycle graph function}
\label{sec:npcyc}

As will be demonstrated in this section, the graphical organization of
open-string $\ap$-expansions is not tied to planar one-loop amplitudes. Even
for non-abelian open-string states, the $\ap$-expansions of the non-planar
open-string amplitudes can be conveniently expressed via mild generalizations
of $A$-cycle graph functions which we will call {\it non-planar $A$-cycle graph
functions}. As shown in \rcite{Broedel:2017jdo}, non-planar $\ap$-expansions
are composed of $A$-cycle eMZVs, and one-particle reducible graphs will be
shown to again decouple once one employs a suitable choice of the Green
function.  Most surprisingly, planar and non-planar $A$-cycle graph functions
turn out to be indistinguishable under the esv-map \eqn{eq:funrules}, i.e.\
under the correspondence \eqn{tree3141} between open-string graph functions and
modular graph functions for the closed string.  As will be detailed in
\secref{sec:npcycC}, this gives rise to expect that planar open-string
amplitudes carry the complete information on the closed string, without any
need for non-planar input.


\subsection{Non-planar open-string integrals}
\label{sec:npcycA}
 
Non-planar one-loop amplitudes of both abelian and non-abelian open-string states comprise 
the integrals \cite{Green:1987mn} 
\begin{align}
M^{\rm open}_{12|34}(s_{ij}\db\tau) \, &:= \,  \int^1_{0} \dd z_2 \int^1_{0} \dd z_3 \int^1_{0} \dd z_4 \ \exp \left(s_{12}\PP_{12}+s_{34} \PP_{34}+ \sum_{i=1}^2 \sum_{j=3}^4 s_{ij}   Q_{ij} \right) 
\label{npln4}
\\
M^{\rm open}_{123|4}(s_{ij}\db\tau) \, &:= \,  \int^1_{0} \dd z_2 \int^1_{0} \dd z_3 \int^1_{0} \dd z_4 \ \exp \left( \sum_{i<j}^3 s_{ij} \PP_{ij}  +\sum_{j=1}^3 s_{j4} Q_{j4} \right) \,,
\label{npln5}
\end{align}
which remain to be integrated over the modular parameter $\tau \in i \ZR$ of
the respective cylinder worldsheet. The subscripts $12|34$ and $123|4$ refer to
the distribution of the open-string states over the two boundaries of the
cylinder. When performing the $\tau$-integral for $M^{\rm
open}_{12|34}(s_{ij}\db\tau)$ an additional factor of $q^{\ap k_1\cdot k_2/2}$
needs to be taken into account\footnote{The integral $I_{12|34}$ expanded in
\rcite{Broedel:2017jdo} is related to \eqn{npln4} via $I_{12|34} = q^{\ap
k_1\cdot k_2/2} M^{\rm open}_{12|34}$.}.
Given that none of the worldsheet boundaries in \eqns{npln4}{npln5} comprises
more than three punctures, the integrals are universal to both abelian and
non-abelian open-string states. However, generic non-planar integrals at $n\geq
5$ points will necessitate a distinction between abelian and non-abelian
states.

In contrast to the integrals \eqns{Acyc4}{Acyc12op} from the planar abelian
sector, \eqns{npln4}{npln5} contain a second species of propagators: $Q_{ij}$.
This propagator connects punctures on different boundaries of the cylinder, and
its representation following from $P_{ij}$ in \eqn{Acyc5} reads
\begin{equation}
Q_{1j}    =  \omwb{1, 0}{\tauh,0} - \Gamma\left( \begin{smallmatrix}
1 \\ \tauh
\end{smallmatrix} ;  z_{j}\right)
\,, \ \ \ \ \ \ 
Q_{ij}    = 
\omwb{1, 0}{\tauh,0}
-  \Gamma\left( \begin{smallmatrix}1 \\z_j+\tauh
\end{smallmatrix} ;  z_{i}\right) - \Gamma\left( \begin{smallmatrix}
1 \\ \tauh
\end{smallmatrix} ;  z_{j}\right)
 \,.
\label{Acyc5np}
\end{equation}
The planar and non-planar propagators $P_{ij}$ and $Q_{ij}$ given by
\eqns{Acyc5}{Acyc5np} are related to the Green functions employed in section 4
of \rcite{Broedel:2017jdo} through a shift by $\omm_A(1,0)$.  Momentum
conservation $\sum^4_{i<j} s_{ij}=0$ again guarantees that the Green functions of
the reference and the present expressions for $P_{ij}$ and $Q_{ij}$ yield the
same integrand in \eqns{npln4}{npln5}.  As a key benefit of the
representations \eqns{Acyc5}{Acyc5np} of $P_{ij}$ and $Q_{ij}$, they satisfy
\eqn{Acyc7a} and
\begin{equation}
\int^1_0 \dd z_i \ Q_{ij}  \ = \ 0 \,,
\label{Acyc7ano}
\end{equation} 
which furnish a suitable starting point to again organize the $\ap$-expansion
of \eqns{npln4}{npln5} in terms of one-particle irreducible graphs. As
elaborated on in \rcite{Broedel:2017jdo}, the non-planar propagator introduces
{\it twisted eMZVs} in intermediate steps,
\begin{equation}
\omwb{n_1,n_2,\ldots,n_r}{\nbeta_1,\nbeta_2,\ldots,\nbeta_r}  := \GLarg{n_r & n_{r-1} & \ldots & n_1}{\nbeta_r & \nbeta_{r-1} & \ldots & \nbeta_1}{1} \,,
\label{twemzv}
\end{equation} 
which drop out from the final expressions such as\footnote{The $M^{\rm
  open}_{123|4}$ is defined as twice the integral $I_{123|4}$ expanded in
  \rcite{Broedel:2017jdo} in order to illustrate the parallels to the
  $\ap$-expansion of $M^{\rm open}_{12|34}$, see in particular
  \secref{sec:npcycC}. Also note the relative minus sign in the definition of
  the Mandelstam variables in this work and ref.\ \cite{Broedel:2017jdo} which
  affects the third order in $\alpha'$ in the subsequent expansions.}
\begin{align}
M^{\rm open}_{12|34}(s_{ij}\db\tau) \ &= \   1+ s_{12}^2  \Big( \frac{ 7\zeta_2 }{6} + 2\omm_A(0,0,2)  \Big)
-2s_{13}s_{23} \Big( \frac{ \zeta_2 }{3} + \omm_A(0,0,2)  \Big) \label{npln44} \\
 &\! \! \! \! \! \!\! \! \! \! \! - 4 \, \zeta_2 \, \omm_A(0,1,0,0) \, s_{12}^3 + s_{12}s_{13}s_{23} \Big(  \frac{ \zeta_3}{2} -
 \frac{5}{3}\omm_A(0,3,0,0) - 4 \, \zeta_2  \omm_A(0,1,0,0) 
 \Big) + {\cal O}(\ap^4)  
\notag
\\
M^{\rm open}_{123|4}(s_{ij}\db\tau) \ &= \  1 +(s_{12}^2 + s_{12}s_{23} + s_{23}^2)  \Big( \frac{ 7\zeta_2 }{6} +2 \omm_A(0,0,2)  \Big) \label{npln55} \\
&\! \! \! \! \! \!\! \! \! \! \!+
s_{12}s_{23}s_{13} \Big( \frac{\zeta_3}{2}+4\, \zeta_2 \omm_A(0,1,0,0) - \frac{5}{3} \omm_A(0,3,0,0)
\Big) + {\cal O}(\ap^4) \,,
\notag
\end{align}
also see \cite{Hohenegger:2017kqy} for the linear orders in $s_{ij}$. Moreover,
all-order expressions for the $\tau \rightarrow i\infty$ limit of \eqn{npln4}
and \eqn{npln5} have been given in \cite{Hohenegger:2017kqy} and
\cite{Green:1981ya}, respectively.


\subsection{Examples of non-planar \texorpdfstring{$A$}{$A$}-cycle
graph functions}
\label{sec:npcycB}

The definition of planar $A$-cycle graph functions in
\secref{ssec:graphfunctions} has a natural extension to the non-planar setup.
Monomials in $P_{ij}$ and $Q_{ij}$ from the expansion of the integrand in
\eqns{npln4}{npln5} are translated into graphs according to the following
rules: Each integration variable in the open-string measure \eqn{Acyc12op} is
again represented by a vertex, and the two kinds of propagators $P_{ij}$ and
$Q_{ij}$ in \eqns{Acyc5}{Acyc5np} between vertices $i$ and $j$ are visualized
by two types of undirected edges. 

\begin{figure}
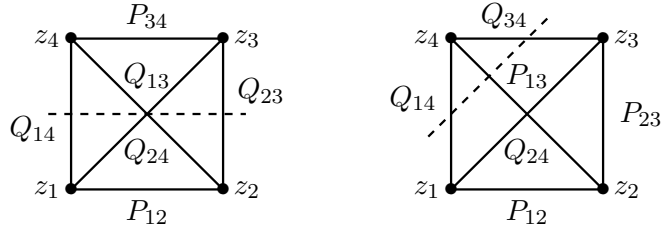

\begin{center}
\tikzpicture[line width=0.3mm]
\draw (-1,-1)node{$\bullet$} node[left]{$z_1$};
\draw (1,-1)node{$\bullet$} node[right]{$z_2$};
\draw (1,1)node{$\bullet$}node[right]{$z_3$} ;
\draw (-1,1)node{$\bullet$}node[left]{$z_4$} ;
\draw(-1,-1)--(-1,1);
\draw(-1,-1)--(1,-1);
\draw(1,1)--(-1,1);
\draw(1,1)--(1,-1);
\draw(-1,-1)--(1,1);
\draw(-1,1)--(1,-1);
\draw(0,-1.3)node{$P_{12}$};
\draw(1.5,0.3)node{$Q_{23}$};
\draw(0,1.3)node{$P_{34}$};
\draw(-1.5,-0.2)node{$Q_{14}$};
\draw(0,0.5)node{$Q_{13}$};
\draw(0,-0.5)node{$Q_{24}$};
\draw[dashed](-1.3,0) --(1.3,0);
\scope[xshift=5cm]
\draw (-1,-1)node{$\bullet$} node[left]{$z_1$};
\draw (1,-1)node{$\bullet$} node[right]{$z_2$};
\draw (1,1)node{$\bullet$}node[right]{$z_3$} ;
\draw (-1,1)node{$\bullet$}node[left]{$z_4$} ;
\draw(-1,-1)--(-1,1);
\draw(-1,-1)--(1,-1);
\draw(1,1)--(-1,1);
\draw(1,1)--(1,-1);
\draw(-1,-1)--(1,1);
\draw(-1,1)--(1,-1);
\draw(0,-1.3)node{$P_{12}$};
\draw(1.5,0)node{$P_{23}$};
\draw(-0.3,1.3)node{$Q_{34}$};
\draw(-1.5,0.2)node{$Q_{14}$};
\draw(0,0.5)node{$P_{13}$};
\draw(0,-0.5)node{$Q_{24}$};
\draw[dashed](-1.3,-0.3) --(0.3,1.3);
\endscope
\endtikzpicture
\caption{In the framework of non-planar $A$-cycle graph functions, the two
  types of propagators $Q_{ij}$ and $P_{ij}$ are represented by edges (drawn as
  solid lines) which do and do not cross the dashed line, respectively. Said
  dashed line tracks the distribution of the punctures in non-planar
  open-string amplitudes over two cylinder boundaries. The situations for the
non-planar integrals $M^{\rm open}_{12|34}$ and $M^{\rm open}_{123|4}$ are
depicted in the left and right panel, respectively.}
\label{fig:npl}
\end{center}
\end{figure}

A convenient way to track the two types of edges stems from the distribution of
the punctures in $M^{\rm open}_{12|34}(s_{ij}\db\tau)$ and $M^{\rm
open}_{123|4}(s_{ij}\db\tau) $ into two sets according to the vertical-bar
notation. These two sets correspond to the location of the punctures on two
different boundaries of a cylindrical worldsheet, and the separation of the
boundaries can be incorporated into the graphs through the dashed line in
\figref{fig:npl}. Then, propagators $Q_{ij}$ and $P_{ij}$ correspond to edges
that cross and do not cross the dashed line, respectively. The generalization
of \eqn{mgf1} then reads

\begin{equation} 
  P_{ij} \, = \,\,\mpostuse[align=c]{G1label} \ , \ \ \ \ \ \ 
    Q_{ij} \, = \,\,\mpostuse[align=c]{Q1label}  \,.
\label{dotline}
\end{equation}
At weight two and three, for instance, one will have to evaluate the following
non-planar $A$-cycle graph functions
\begin{align}
\Acyc{G2np} = \! \int \dd \mu_2^{\rm open} \ Q_{12}^2  \ ,
 \ \ \ \ 
\Acyc{G3np} =  \! \int  \dd \mu_2^{\rm open} \, Q_{12}^3 \ ,
 \ \ \ \ 
\Acyc{G111np} = \!  \int  \dd \mu_3^{\rm open}\, P_{12} Q_{13} Q_{23}   
 \label{npln7}
\end{align}
in addition to their planar counterparts $\Acyc{G2}, \Acyc{G3}$ and
$\Acyc{G111}$. The essential steps of their computation can be assembled from
\rcite{Broedel:2017jdo}, see in particular appendix I.1 of the reference, with
the following results in terms of \textit{untwisted} $A$-cycle eMZVs:
\begin{align}
\Acyc{G2np}   \ &= \ \omm_A(0,0,2) + \frac{\zeta_2}{3}
\label{npln6a} \\
\Acyc{G3np} \ &= \ \frac{\zeta_3}{2}  -\frac{1}{3}  \omm_A(0,3,0,0) - 4 \zm_2  \omm_A(0,1,0,0)  \label{npln7a} \\ 
\Acyc{G111np}  \ &= \  - \frac{1}{3}  \omm_A(0,3,0,0) \,.
\label{npln8a}
\end{align}
From weight four on, certain graph topologies admit several non-planar
$A$-cycle graph functions which correspond to different distributions
of punctures over two boundaries or different numbers of $Q_{ij}$ propagators. 
For instance, there are two and three non-planar analogues to $\Acyc{G211}$ 
and $\Acyc{G1111}$, respectively,
\begin{align}
\Acyc{G4np} \ &= \ \int  \dd \mu_2^{\rm open} \, Q_{12}^4 \label{npln9} \\ 
\Acyc{G211np1} \ &= \ \int  \dd \mu_3^{\rm open}\, P_{12}^2 Q_{13} Q_{23}    \,, \ \ \ \ 
\Acyc{G211np2} \ = \ \int  \dd \mu_3^{\rm open}\, Q_{12}^2 P_{13} Q_{23}   
\label{npln10} \\
\Acyc{G1111np1} \ &= \ \int  \dd \mu_4^{\rm open}\, P_{12} Q_{23} P_{34} Q_{41}  \,, \ \ \ \
\Acyc{G1111np2} \ = \ \int  \dd \mu_4^{\rm open}\, P_{12} P_{23} Q_{34} Q_{41}    \notag \\
\Acyc{G1111np3} \ &= \ \int  \dd \mu_4^{\rm open}\, Q_{12} Q_{23} Q_{34} Q_{41} \,.
\label{npln11} 
\end{align}


\subsection{Comparing \texorpdfstring{$\ap$}{$\ap$}-expansions of
planar and non-planar integrals}
\label{sec:npcycC}

By means of momentum conservation, the $\ap$-expansion \eqns{npln44}{npln55} of
the non-planar integrals \eqns{npln4}{npln5} can be recovered from the
following planar and non-planar $A$-cycle graph functions:
\begin{align}
M^{\rm open}_{12|34}(s_{ij}\db\tau) \ &= \  
1 + s_{12}^2 ( \Acyc{G2} + \Acyc{G2np}) - 2 s_{13} s_{23} \Acyc{G2np} \notag \\
& + \frac{s_{12}^3}{3} (\Acyc{G3} - \Acyc{G3np}) + s_{12} s_{13} s_{23} (\Acyc{G3np}+ 4 \Acyc{G111np}) + {\cal O}(\ap^4)\,,
 \label{npln12} \\
M^{\rm open}_{123|4}(s_{ij}\db\tau) \ &= \  1 + (s_{12}^2+s_{12}s_{23}+s_{23}^2) (\Acyc{G2} + \Acyc{G2np})  \notag \\
&+ s_{12}s_{13}s_{23} \Big(
\frac{1}{2} \Acyc{G3} + \frac{1}{2} \Acyc{G3np}+ \Acyc{G111} + 3 \Acyc{G111np} 
\Big)
+ {\cal O}(\ap^4)\,. \label{npln13} 
\end{align}
Clearly, when identifying the two boundaries and formally sending $\Acyc{G2np}
\rightarrow \Acyc{G2}$, $\Acyc{G3np} \rightarrow \Acyc{G3}$ and
$\Acyc{G111np}\rightarrow \Acyc{G111}$, both \eqn{npln12} and \eqn{npln13}
reduce to the integral \eqn{AAstructure} of the abelian planar amplitude. 
Nevertheless, the expressions \eqns{npln12}{npln13} for non-planar 
integrals also apply to non-abelian open-string amplitudes.


It is tempting to compare the eMZV representation of non-planar $A$-cycle graph
functions with their planar counterparts.  The above examples in eqs.\
(\ref{npln6a}) to (\ref{npln8a}),
\begin{align}
\Acyc{G2} - \Acyc{G2np} &= \frac{1}{2}\zm_2
\notag \\
\Acyc{G3} - \Acyc{G3np} &= 12 \zm_2 \omm_A(0,1,0,0)
\label{npln14} \\
\Acyc{G111} - \Acyc{G111np}&= 2 \zm_2 \omm_A(0,1,0,0)
\notag 
\end{align}
give rise to the following observation: to the orders considered, planar and
non-planar $A$-cycle graph functions associated with the same graph differ by
terms in the space $X$ defined in \eqn{DefX}
and thus lead to identical results after applying \eqn{adhoc}. This is
furthermore supported by the weight-four example
\begin{equation}
\Acyc{G4np} =  216 \cez(4,0,4,0) - 432  \cez(4,4,0,0) - 3024  \cez(8,0,0,0)  \ {\rm mod} \ \zeta_2
\end{equation}
which again matches the expression for $\Acyc{G4}$ in \eqn{eis7} modulo terms
in $X$ and validates the observation beyond one-loop graphs. We therefore
expect the matching of planar and non-planar $A$-cycle graph functions modulo
terms in $X$ to persist at higher weight and depth,
\begin{equation}
 \text{\textbf{A}}[\CG]  =  \text{\textbf{A}}[ \, \ \CG \! \! \! \! \! \! \! \cdots\! \! \! \! \! \! \! \cdots]   \ {\rm mod} \ X \ ,
 \label{persist}
\end{equation}
where $\ \, \CG \! \! \! \! \! \! \! \cdots\! \! \! \! \! \! \! \cdots$
represents an arbitrary non-planar generalization\footnote{For instance, $\ \,
\CG \! \! \! \! \! \! \! \cdots\! \! \! \! \! \! \! \cdots$ can be either
$\mpostuse[width=0.5cm,align=c]{G211np1}$ or
$\mpostuse[width=0.5cm,align=c]{G211np2}$ when $\CG$ is taken to be
$\mpostuse[width=0.5cm,align=c]{G211}$.} of the graph $\CG$. This conjecture
also applies to $A$-cycle graph functions \eqn{defAcyc} and their non-planar generalizations
with $n\geq 5$ vertices. Still, this class of integrals will not capture all the terms
in the $\ap$-expansions of five-point and higher-multiplicity 
amplitudes of the open superstring.

As explained in the paragraph below \eqn{eq:funrules1}, the modular
$S$-transformation maps terms contained in the space $X$ to terms sent to zero
by esv in \eqn{eq:funrules}. Hence, given the definition of non-planar
$B$-cycle graph functions
\begin{equation}
 \text{\textbf{B}}[ \, \ \CG \! \! \! \! \! \! \! \cdots\! \! \! \! \! \! \! \cdots] :=  \text{\textbf{A}}[ \, \ \CG \! \! \! \! \! \! \! \cdots\! \! \! \! \! \! \! \cdots]  \, \big|_{ \tau \rightarrow -\frac{1}{\tau} }\,,
\label{bgraph2}
\end{equation} 
in direct analogy with the planar ones \eqn{bgraph}, our conjecture
\eqn{persist} implies that modular graph functions can also be constructed from
non-planar open-string graph functions
\begin{equation}
\esv \ \text{\textbf{B}}[ \, \ \CG \! \! \! \! \! \! \! \cdots\! \! \! \! \! \! \! \cdots]  =
\esv \ \text{\textbf{B}}[\CG] = \text{\textbf{D}}[\CG] \ .
\label{bgraph3}
\end{equation} 
At the level of their generating functions, this leads to the following
conclusion: When the closed-string four-point amplitude is obtained from
open-string input through the esv-projection as in \eqn{punchline}, then the
non-planar sectors do not carry any additional information beyond the planar
sector for abelian open-string states:
\begin{equation}
\esv \ M^{\rm open}_{4}(s_{ij}\db\!-\tfrac{1}{\tau}) = \esv \ M^{\rm open}_{12|34}(s_{ij}\db\!-\tfrac{1}{\tau}) = \esv \ M^{\rm open}_{123|4}(s_{ij}\db\!-\tfrac{1}{\tau})  \ .
\label{bgraph4}
\end{equation}
In other words -- the esv-map identifies non-planar open-string integrals
with planar abelian ones!  It would be interesting to understand this in the
light of monodromy relations among one-loop open-string amplitudes
\cite{Tourkine:2016bak, Hohenegger:2017kqy}.


\section{Conclusions}
\label{sec:conclusion}

In this work, we have identified new connections between building blocks of
open- and closed-string one-loop amplitudes at the level of their
$\ap$-expansions. In view of the relation between the respective tree-level
amplitudes through the single-valued projection of MZVs, we have proposed an
elliptic version of a single-valued map called ``esv''.  The latter acts on the
eMZVs in symmetrized one-loop open-string integrals and yields the
corresponding integrals of the closed string.  This connection between open and
closed strings through the esv-map has been explicitly verified at the leading
seven orders in $\ap$ and suggests to envision the following scenario in the
long run: Closed-string $\ap$-expansions at generic multiplicity and loop order
might be entirely derivable from open-string data using suitable operations.

Our construction is based on a graphical organization of the $\ap$-expansions
of planar and non-planar open-string amplitudes: Convenient arrangements of the
genus-one Green function cancel the contributions from one-particle reducible
graphs which has already been used to simplify closed-string $\ap$-expansions.
For each one-particle irreducible graph, we have defined a meromorphic
$A$-cycle graph function comprising eMZVs, its modular $S$-transformation
called $B$-cycle graph function as well as non-planar generalizations.
Representing these open-string constituents in terms of iterated Eisenstein
integrals leads to a straightforward identification through the esv-map with
the modular graph functions governing the closed-string $\ap$-expansion.

Expressing modular graph functions in terms of iterated Eisenstein integrals gives
rise to new results on their Fourier expansions beyond the simplest cases of
non-holomorphic Eisenstein series. Furthermore, our
iterated-Eisenstein-integral representations automatically manifest 
all relations between modular graph functions and
their Laplace equations at the weights under consideration. We expect that this
language is suitable to represent the general systematics of indecomposable
modular graph functions and their network of Laplace eigenvalue equations.

Having applied methods from the open string to modular graph functions on the
closed-string side, it would be interesting to try the converse: The
representation of modular graph functions in terms of nested lattice sums,
which is immediately accessible from their definition through the genus-one
Green function, should have an echo for eMZVs. In particular, tentative
lattice-sum representations of $A$-cycle and $B$-cycle eMZVs are likely to 
offer new perspectives on their algebraic and differential relations
and new insights on the esv-map.

Moreover, it would be desirable to connect the present proposal for the esv-map
with the framework developed by Brown in \rcites{Brown:mmv,
Brown:2017qwo, Brown:2017qwo2}. This would make clear whether our observations
hold true for any graph or should be corrected at higher depth. 

While the present results are restricted to scattering amplitudes of four
external states, a natural follow-up question concerns the generalization to
$n$-point one-loop amplitudes. Since the coefficients of the
Kronecker--Eisenstein series capture the all-multiplicity integrands
\cite{Broedel:2014vla}, the $n$-point $\alpha'$-expansion for both open and
closed strings is expressible in terms of iterated Eisenstein integrals and
therefore accessible to the proposed esv-map.  However, it remains to identify
the correspondence between cyclic orderings in higher-point open-string
amplitudes and the additional functions of the punctures in closed-string
integrands at five and more points \cite{Richards:2008jg, Green:2013bza,
Mafra:2016nwr, Basu:2016mmk}. The recent double-copy representation
\cite{Mafra:2017ioj} of open-string integrands is expected to play a key role
in this endeavor.

Relations between open- and closed-string amplitudes at higher genus should be
encoded in a similar organization scheme of the integrals over the punctures. A
strategic path at genus two would be to express the moduli-space integrand for
the Zhang--Kawazumi invariant \cite{D'Hoker:2013eea, D'Hoker:2014gfa} and its
recent generalizations to higher orders in $\ap$ \cite{DHoker:2017pvk} in terms
of open-string quantities. For this purpose, higher-genus generalizations of
eMZVs along with the appropriate analogues of iterated Eisenstein integrals
seem to be a suitable framework.

Finally, both the $\ap$-expansion of closed-string one-loop amplitudes and the
iterated-integral description of modular graph functions have important
implications for the non-perturbative $S$-duality of type-IIB superstrings
\cite{Hull:1994ys}: this duality symmetry connects amplitudes of different
loop orders and incorporates their non-perturbative completion. It would be
desirable to express the underlying modular invariant functions of the
axion-dilaton field -- non-holomorphic Eisenstein series at half-odd integer
arguments \cite{Green:1997tv, Green:1999pu} and beyond \cite{Green:2005ba,
Green:2014yxa} -- via esv-projected open-string quantities. This would set the
stage for taking maximal advantage of $S$-duality to infer exact and
non-perturbative results on the low-energy regime of type-II superstrings at
unprecedented orders in $\ap$.


\subsection*{Acknowledgments}

First of all, we are very grateful to Nils Matthes for collaboration in
early stages of the project. We would like to thank Francis Brown, Axel
Kleinschmidt, Eric D'Hoker, Nils Matthes and Pierre Vanhove for comments on the
draft and various helpful discussions. In addition, we are grateful to Eric
D'Hoker and Justin Kaidi for several email exchanges. We would like to thank
the Kolleg Mathematik und Physik Berlin for support in various ways and the
Hausdorff Research Institute for Mathematics for hospitality while finalizing
this article. 

JB and OS would like to thank the Munich Institute for Astro- and Particle
Physics for hospitality and providing a stimulating atmosphere during a
workshop ``Mathematics and Physics of Scattering Amplitudes'' in August and
September 2017, where a substantial part of this project was realized.  This
research was supported in part by the National Science Foundation under Grant
No.~NSF PHY17-48958, and we are grateful to the KITP Santa Barbara for
providing a vibrant research environment during the workshop ``Scattering
Amplitudes and Beyond''.

The research of OS was supported in part by Perimeter Institute for Theoretical
Physics. Research at Perimeter Institute is supported by the Government of
Canada through the Department of Innovation, Science and Economic Development
Canada and by the Province of Ontario through the Ministry of Research,
Innovation and Science.

The research of FZ was supported by the Max Planck Institute for Mathematics,
by a French public grant as part of the Investissement d'avenir project,
reference ANR-11-LABX-0056-LMH, LabEx LMH, and by the People Programme (Marie
Curie Actions) of the European Union’s Seventh Framework Programme
(FP7/2007-2013) under REA grant agreement n.\ PCOFUND-GA-2013-609102, through
the PRESTIGE programme coordinated by Campus France.


\newpage

\section*{Appendix}
\appendix


\section{Translating between graphs and notations for modular graph functions}
\label{app:translate}

In this appendix, we spell out the dictionary between the graphical notation $\text{\textbf{D}}[\CG]$ 
for modular graph functions used in this work and the $D$- and $C$-notations 
introduced \rcite{Green:2008uj} (see also appendix C of \cite{Green:2013bza}).

\begin{table}[h]
\begin{minipage}[t]{0.45\textwidth}
\setlength\tabcolsep{4pt}
  \begin{tabular}{c|c|c}
   Graph                                                 &$D$-notation       &$C$-notation        \\\hline
  \mpostuse[align=c]{G2}                                 &$D_2=\EE{2}$       &$C_{1,1}$           \\\hline
  \mpostuse[align=c]{G3}                                 &$D_3$              &$C_{1,1,1}$         \\\hline
  \mpostuse[align=c]{G111}                               &$D_{111}=\EE{3}$   &$C_{2,1}$           \\\hline
  \mpostuse[align=c]{G4}                                 &$D_4$              &$C_{1,1,1,1}$       \\\hline
  \mpostuse[align=c]{G211}                               &$D_{211}$          &$C_{2,1,1}$         \\\hline
  \mpostuse[align=c]{G1111}                              &$D_{1111}=\EE{4}$  &$C_{3,1}$           \\\hline
  \mpostuse[align=c]{G5}                                 &$D_5$              &$C_{1,1,1,1,1}$     \\\hline
  \mpostuse[align=c]{G221}                               &$D_{221}$          & -                  \\\hline
  \mpostuse[align=c]{G311}                               &$D_{311}$          &$C_{2,1,1,1}$       \\\hline
  \mpostuse[align=c]{G2111}                              &$D_{2111}$         &$C_{3,1,1}$         \\\hline
  \mpostuse[align=c]{Gp1111} = \mpostuse[align=c]{GC221} &$D'_{1111}$        &$C_{2,2,1}$         \\\hline
  \mpostuse[align=c]{G11111}                             &$D_{11111}=\EE{5}$ &$C_{4,1}$           \\\hline 
  \mpostuse[align=c]{G511}				 &$D_{511}$              &$C_{2,1,1,1,1,1}$                    
\end{tabular}
\end{minipage}\hspace{5mm}
\begin{minipage}[t]{0.45\textwidth}
\setlength\tabcolsep{4pt}
  \begin{tabular}{c|c|c}
  Graph                                                  &$D$-notation         &$C$-notation            \\\hline
  \mpostuse[align=c]{G6}                                 &$D_6$                &$C_{1,1,1,1,1,1}$       \\\hline
  \mpostuse[align=c]{G411}                               &$D_{411}$            &$C_{2,1,1,1,1}$         \\\hline
  \mpostuse[align=c]{G321}                               &$D_{321}$            & -                      \\\hline
  \mpostuse[align=c]{G222}                               &$D_{222}$            & -                      \\\hline
  \mpostuse[align=c]{G3111}                              &$D_{3111}$           &$C_{3,1,1,1}$           \\\hline
  \mpostuse[align=c]{G2211}                              &$D_{2211}$           & -                      \\\hline
  \mpostuse[align=c]{Gp2111}                             &$D'_{2111}$          & -                      \\\hline
  \mpostuse[align=c]{Gpp1111}                            &$D''_{1111}$         &$C_{2,2,1,1}$           \\\hline
  \mpostuse[align=c]{Gx1111}                             &$D^\times_{1111}$    & -                      \\\hline
  \mpostuse[align=c]{G21111}                             &$D_{21111}$          &$C_{4,1,1}$             \\\hline
  \mpostuse[align=c]{Gp11111} = \mpostuse[align=c]{GC321}&$D'_{11111}$         &$C_{3,2,1}$             \\\hline
  \mpostuse[align=c]{GC222}                              &$D_{11,11,11}$       &$C_{2,2,2}$             \\\hline 
  \mpostuse[align=c]{G111111}                            &$D_{111111}=\EE{6}$  &$C_{5,1}$               
  \end{tabular}
\end{minipage}%
\caption{Different notations for modular graph functions in various publications}
\label{tab:graphs}
\end{table}


\section{Constant term of \texorpdfstring{$B$}{$B$}-cycle eMZVs}
\label{app:constantB}

As explained in detail in subsection 2.3 in \rcite{Broedel:2015hia}, the
constant term of an $A$-cycle eMZV can be calculated using a method developed
in \rcites{KZB, Enriquez:Emzv}. In short, the construction relies on
comparing the properly regulated generating series of $A$-cycle eMZVs, the
elliptic KZB associator $A(\tau)$ 
\begin{equation}
e^{\pi i[y,x]}A(\tau) := \sum_{r \geq 0} (-1)^r \sum_{n_1,n_2,\ldots,n_r\geq 0} \omega_A(n_1,n_2,\ldots,n_r\db\tau)\ad^{n_r}_x(y)\ldots \ad^{n_2}_x(y) \ad^{n_1}_x(y) \ ,
\label{Agenerating}
\end{equation}
to its asymptotic expansion as $\tau \to i\infty$ \cite{KZB}
\begin{equation}
  A(\tau)=\Phi(\tilde{y},t)\,e^{2\pi i\tilde{y}}\,\Phi(\tilde{y},t)^{-1}+{\cal{O}}(e^{2\pi i\tau}) \ .
  \label{AAssociator}
\end{equation}
Taking the limit $\tau\to i\infty$ in \eqn{Agenerating} amounts to replacing
the full eMZV $\omega_A(n_1,\ldots,n_r\db\tau)$ with its constant part 
$\omega_{A,0}(n_1,\ldots,n_r):= \lim_{\tau \rightarrow i \infty} \omega_A(n_1,\ldots,n_r\db\tau)$, 
which is the quantity of interest here.  

Comparison between \eqns{Agenerating}{AAssociator} is done for coefficients of
words built from the letters $x$ and $y$, which in turn denote generators of a
complete and free algebra $\ZC \langle \!\langle x,y \rangle \! \rangle$, whose
multiplication is concatenation and $\ad_x(y) := [x,y]$. \Eqn{AAssociator}
takes its concise and short form only after defining additional auxiliary
letters 
\begin{equation}
t = [y,x] \co \tilde{y} = -\frac{\ad_x}{e^{2\pi i\ad_x}-1}(y) \ .
\label{tyt}
\end{equation}
Finally, $\Phi$ in \eqn{AAssociator} denotes the Drinfeld associator
\cite{Drinfeld:1989st, Drinfeld2, Le} 
\begin{equation}
\Phi(e_0,e_1) := \sum_{W \in \langle e_0,e_1 \rangle} \zeta^{\shuffle}(\hat W) \cdot W \ .
\end{equation}
The sum over $W \in \langle e_0,e_1 \rangle$ runs over all non-commutative
words built from letters $e_0$ and $e_1$. The operation $\hat{\ }$ acts on a word
$W$ by replacing letters $e_0$ and $e_1$ by $0$ and $1$, respectively.  
The notion $\zeta^{\shuffle}(\hat W)$ refers to shuffle-regularized MZVs \cite{Rac}
which are uniquely determined from \eqn{tree05}, the shuffle product and 
the definition $\zeta^{\shuffle}(0)=\zeta^{\shuffle}(1)=0$.  Accordingly, the first couple of
terms of $\Phi(e_0,e_1)$ are given by 
\begin{equation}
\Phi(e_0,e_1)=1-\zm_2 [e_0,e_1]-\zm_3 [e_0+e_1,[e_0,e_1]] + \ldots \ .
\end{equation}
Numerous results for constant terms have been calculated and noted in
\rcite{Broedel:2015hia}.

For $B$-cycle eMZVs, an analogous construction does exist. Considering the
expansion of $B$-cycle eMZVs, this time it is not a constant term, but rather a
Laurent polynomial $\omega_{B,0}(n_1,\ldots,n_r\db\tau)$ in $\tau$ which comes
with the term $q^0$ in the expansion
\begin{equation}
  \omega_{B}(n_1,\ldots,n_r\db\tau) = \omega_{B,0}(n_1,\ldots,n_r\db\tau) +\sum_{k=1}^\infty\omega_{B,k}(n_1,\ldots,n_r\db\tau)\,q^k\,,
  \label{omegaBexpansion} 
\end{equation}
see \eqn{B-cycleAsymptExp}. The $B$-cycle analogue of the $A$-cycle 
associator in \eqn{Agenerating} reads \cite{Enriquez:Emzv}
\begin{equation}
e^{\pi i[y,x]}B(\tau):= \sum_{r \geq 0} (-1)^r \sum_{n_1,n_2,\ldots,n_r\geq 0} \omega_B(n_1,n_2,\ldots,n_r\db\tau)\ad^{n_r}_x(y)\ldots \ad^{n_2}_x(y) \ad^{n_1}_x(y) \ .
\label{Bgenerating}
\end{equation}
While taking the limit $\tau\to i\infty$ in the above equation will again
replace $\omega_B$ by $\omega_{B,0}$, obtaining the $B$-cycle analogue of
\eqn{AAssociator} takes a little more effort. In \rcite{Enriquez:Emzv}, it was shown that
the comparison ought to be done between the $(\tau\to i \infty)$-limit of
\eqn{Bgenerating} and
\begin{equation}\label{EnrConstBassociator}
B(\tau) = \exp \left( -\frac{2\pi i }{\tau } \, e_+ \right) \Phi(-\tilde y - t,t)e^{2\pi i x} e^{2\pi i \tilde y \tau} \Phi^{-1}(\tilde y,t)
+ {\cal O}(e^{2\pi i(1-\epsilon)\tau}) \, ,
\end{equation} 
where the introduction of an arbitrary  $\epsilon >0$ is needed 
to account for the suppressed terms of the form $\tau^l q^k$ with $k,l\geq 1$.
The new ingredient in comparison to \eqn{AAssociator} is the 
derivation $e_+$, which acts on algebra generators $x$ and $y$ via 
\begin{equation*}
  e_+(x)=0 \ \ \text{ and } \ \ e_+(y)=x\,. 
\end{equation*}
The term $\omega_{B,0}(n_1,\ldots,n_r\db\tau)$ in the expansion \eqn{omegaBexpansion} of $B$-cycle 
eMZVs can then be obtained by equating \eqns{Bgenerating}{EnrConstBassociator} and isolating 
the coefficients of a given word in $\ad^{n_i}_x(y)$.

For instance, applying this procedure to the simplest $B$-cycle graph function
\begin{align}
\Bcyc{G211} &= \frac{1}{2} \frac{\omm_B(0, 0, 2)^2}{\tau^2} - \frac{1}{2} \frac{\omm_B(0, 0, 0, 0, 4)}{\tau} - \frac{\omm_B(0, 0, 0, 2, 2)}{\tau} \notag \\
& + 
  \frac{7}{3} \zeta_2\frac{ \omm_B(0, 0, 2)}{\tau} - 
  14 \zeta_2 \frac{\omm_B(0, 0, 0, 0, 2)}{\tau^3} + \frac{ 301 \zeta_4}{180}
\end{align}
involving eMZVs of depth two, we arrive at the
constant term
\begin{align}
\smallBcyc{G211}  &= \frac{T^4}{113400} +  \frac{ T^2 \zeta_2 }{540} - \frac{ T \zeta_3}{180} + \frac{37 \zeta_4}{180} - \frac{ 5 \zeta_5}{12 T} \label{Bconst11} \\
&+ \frac{29 \zeta_6}{16 T^2} - \frac{\zeta_3^2}{
 4 T^2}   - \frac{9 \zeta_7}{4 T^3}   + \frac{7 \zeta_2 \zeta_5}{T^3} - \frac{3 \zeta_3 \zeta_4}{2 T^3} + \frac{28 \zeta_8}{3 T^4}  \ ,\notag
\end{align}
and the same can be repeated at higher weight.


\section{Expanding \texorpdfstring{$\boldsymbol{S}$}{$S$}-transformed
\texorpdfstring{$A$}{$A$}-cycle eMZVs}
\label{app:prf}

This appendix is dedicated to a proof of our observation on the expansion
\eqn{Texpansion} of $S$-transformed $A$-cycle eMZVs: the coefficients of a
given $(2\pi i \tau)^l q^k$ (with $l \in \ZZ$ and $k\geq 0$) in the expansion
of $\omm_A(n_1,n_2,\ldots,n_r\db\!-\tfrac{1}{\tau})$ around the cusp are
claimed to be $\ZQ$-linear combinations of MZVs\footnote{We suppose that
$n_1,n_r\neq 1$. Otherwise, the asymptotic expansion of
$\omm_A(\underline{n}\db\!-\tfrac{1}{\tau})$ would in general involve terms
proportional to $\log(\tau)$ \cite{ZerbiniThesis}, but this is never the case
in our context.}. This claim is essential in \subsecref{sec:funrules}, where we
show that terms from the space $X$ defined in \eqn{DefX} are annihilated by the
esv-map after a modular transformation.

Using \eqns{eqn:fromAtoB}{B-cycleAsymptExp}, we can write
\begin{equation}
\omm_A(n_1,n_2,\ldots,n_r\db\!-\tfrac{1}{\tau})=\sum_{l=1-r}^{n_1+\cdots +n_r}\tau^l\sum_{k= 0}^\infty \tilde{b}_{k,l}(n_1,n_2,\ldots ,n_r)q^k \ ,
\label{klseries}
\end{equation}
where the coefficients $\tilde{b}_{k,l}$ are $\ZQ[(2\pi i)^{\pm 1}]$-linear
combinations of MZVs and related to the coefficients in \eqn{B-cycleAsymptExp}
via $b_{k,l}=\tilde{b}_{k,l+n_1+\cdots n_r-r}$. What we need to prove is
that $c_{k,l}:=\tilde{b}_{k,l}/(2\pi i)^l$ are $\ZQ$-linear
combinations of MZVs.  The proof is divided into two parts. In
\appref{app:prf.1}, the setup of \appref{app:constantB} will be used to prove
that the $c_{k,l}$ at $k=0$ are $\ZQ$-linear combinations of
MZVs. Then, the analogous statement for $c_{k,l}$ at $k>0$ will be deduced from
the differential equation of eMZVs in \appref{app:prf.2}, which together with
the previous step implies our claim for all $k$.


\subsection{The Laurent polynomial}
\label{app:prf.1}

For the coefficients $b_{k,l}(n_1,n_2,\ldots,n_r)$ of the $B$-cycle eMZV
$\omm_B(n_1,n_2,\ldots,n_r\db\tau)$ in \eqn{B-cycleAsymptExp}, it will now be
shown that $b_{0,l}/(2\pi i)^{l+n_1+\cdots +n_r-r}$ is a $\ZQ$-linear combination of MZVs. This implies the above claim at $k=0$. 

The $B$-cycle eMZV
$\omm_B(n_1,n_2,\ldots,n_r\db\tau)$ can be obtained as the coefficient of the
word $\mbox{ad}^{n_r}_{x}(y)\ldots \mbox{ad}^{n_1}_{x}(y)$ in the
$B$-elliptic associator \eqn{Bgenerating}. In the degeneration
\eqn{EnrConstBassociator} of the associator where all the coefficients
$b_{k>0,l}$ of the $B$-cycle eMZVs \eqn{B-cycleAsymptExp} are suppressed, each
instance of the letter $x$ is accompanied by $2\pi i$,
each instance of $y$ by $\frac{1}{2\pi
i}$ and each instance of $\tau$ by $2\pi i$. This follows from the following trivial remarks about \eqn{EnrConstBassociator}:
\begin{itemize}
\item The letter $t$ can be written
  as $t=[\frac{y}{2\pi i},2\pi i x]$.
\item The expansion of the letter $\tilde y$ in \eqn{tyt} and the factor of $\exp(2\pi i\tilde{y}\tau)=\exp((2\pi i\tau)\tilde{y})$ in \eqn{EnrConstBassociator} can be written as
\begin{equation}
\tilde{y} = -\sum_{n\geq 0}\frac{B_n}{n!} \mbox{ad}^n_{2\pi ix}\Big(\frac{y}{2\pi i}\Big)  \ .
\label{Blettersnew}
\end{equation}
\end{itemize}
Moreover, the factor of $\exp \big( -\frac{2\pi i }{\tau} e_+ \big)$ in
\eqn{EnrConstBassociator} does not alter the argument, because when it acts on
$\frac{y}{2\pi i}$ it gives back $\frac{x}{\tau}=\frac{2\pi ix}{2\pi i\tau}$.  Hence, the fact that the
$\tau \rightarrow i\infty$ asymptotics of $\omm_B(n_1,n_2,\ldots,n_r\db\tau)$
enter \eqn{EnrConstBassociator} along with $n_1{+}n_2{+}\ldots{+}n_r$ letters $x$ and
$r$ letters $y$ implies that the coefficient $b_{0,l}$ can be written as $(2\pi
i)^{l+n_1+n_2+\ldots+n_r - r}$ times a $\ZQ$-linear combination of
MZVs, where $(2\pi i)^{l}$ comes from the fact that each $\tau$ is accompanied by $2\pi i$. This means that all $c_{0,l}=\tilde{b}_{0,l}/(2\pi i)^l$ are $\ZQ$-linear
combinations of MZVs, and concludes the first part of the proof.


\subsection{The \texorpdfstring{$q$}{$q$}-expansion}
\label{app:prf.2}

The second part of the proof is based on Enriquez's differential equation
satisfied by the generating series of $A$-cycle eMZVs \cite{Enriquez:Emzv}. For
any fixed $A$-cycle eMZV, it implies that 
\begin{align}
2\pi i & \frac{ \partial }{\partial \tau} \omm_A(n_1,\ldots,n_r \db \tau) =  n_1 \GG{n_1+1} \omm_A(n_2,\ldots,n_r \db \tau) - n_r \GG{n_r+1} \omm_A(n_1,\ldots,n_{r-1} \db \tau)\notag \\
&+ \sum_{i=2}^r \Big\{ (-1)^{n_i} (n_{i-1}+n_i)   \GG{n_{i-1}+n_i+1} \omm_A(n_1,\ldots,n_{i-2},0,n_{i+1},\ldots,n_r \db \tau) \label{eqn:tauder} \\
& \ \ \ \ \ \ -  \sum_{k=0}^{n_{i-1}+1} (n_{i-1}-k) { n_i+k-1 \choose k }  \GG{n_{i-1}-k+1} \omm_A(n_1,\ldots,n_{i-2},k+n_i,n_{i+1},\ldots,n_r \db \tau) \notag \\
& \ \ \ \ \ \ + \sum_{k=0}^{n_{i}+1} (n_{i}-k) { n_{i-1}+k-1 \choose k }  \GG{n_{i}-k+1} \omm_A(n_1,\ldots,n_{i-2},k+n_{i-1},n_{i+1},\ldots,n_r \db \tau) \Big\} \,.
\notag 
\end{align}
One can straightforwardly deduce the differential equation w.r.t.~the variable
$2\pi i\tau$ for the modular images (which up to powers of $\tau$ coincide with
  $B$-cycle MZVs, see
\eqn{eqn:fromAtoB})
\begin{align}
&\frac{1}{2\pi i}\frac{\partial}{\partial \tau} \omm_A(n_1,\ldots,n_r\db\!-\tfrac{1}{\tau}) = n_1 \bigg( (2\pi i \tau)^{n_1-1}\hat{\rm G}_{n_1+1}(\tau) -\frac{\delta_{n_1,1}}{2\pi i\tau}\bigg) \omm_A(n_2,\ldots,n_r\db\!-\tfrac{1}{\tau}) \label{DiffEqB}\\
&- n_r \bigg( (2\pi i \tau)^{n_r-1}\hat{\rm G}_{n_r+1}(\tau)-\frac{ \delta_{n_r,1}}{2\pi i\tau}\bigg)\omm_A(n_1,\ldots,n_{r-1}\db\!-\tfrac{1}{\tau})\notag \\
&+ \sum_{i=2}^r \Big\{ (-1)^{n_i} (n_{i-1}{+}n_i)(2\pi i \tau)^{n_{i-1}+n_i-1} \hat{\rm G}_{n_{i-1}+n_i+1}(\tau)  \omm_A(n_1,\ldots,n_{i-2},0,n_{i+1},\ldots,n_r\db\!-\tfrac{1}{\tau}) \notag\\
&- \! \! \sum_{k=0}^{n_{i-1}+1} \! (n_{i-1}{-}k) { n_i{+}k{-}1 \choose k } (2\pi i \tau)^{n_{i-1}-k-1} \hat{\rm G}_{n_{i-1}-k+1}(\tau) \omm_A(n_1,\ldots,n_{i-2},k{+}n_i,n_{i+1},\ldots,n_r\db\!-\tfrac{1}{\tau}) \notag \\
&+ \sum_{k=0}^{n_{i}+1} (n_{i}{-}k) { n_{i-1}{+}k{-}1 \choose k } (2\pi i \tau)^{n_{i}-k-1} \hat{\rm G}_{n_{i}-k+1}(\tau) \omm_A(n_1,\ldots,n_{i-2},k{+}n_{i-1},n_{i+1},\ldots,n_r\db\!-\tfrac{1}{\tau}) \Big\}  \notag\ ,
\end{align}
where the Kronecker-delta terms in the first and second line follow from the
exceptional modular transformation $\GG{2}(-\frac{1}{\tau}) = \tau^2
\GG{2}(\tau) - 2\pi i \tau$. We are using the normalization conventions
\begin{equation}
\hat{\rm G}_{2k}(\tau):=\frac{\GG{2k}(\tau)}{(2\pi i)^{2k}}=\frac{2\zeta_{2k}}{(2\pi i)^{2k}}+\frac{2}{(2k{-}1)!}\sum_{m\geq 1}\frac{m^{2k-1}q^m}{1-q^m} 
\end{equation}
for the Eisenstein series in \eqn{DiffEqB}.

The right-hand side of \eqn{DiffEqB} involves $A$-cycle eMZVs of smaller length
$r{-}1$ multiplied by MZV-linear combinations\footnote{We mean that the coefficients lie in the $\mathbb{Q}$-algebra generated by MZVs.} of $(2\pi i \tau)^l q^k$ with $l \in \mathbb Z$ and
$k\geq 0$. When $r=1$ it is easy to see that
$\omm_A(2m\db\!-\tfrac{1}{\tau})=-2\zeta_{2m}$ and
$\omm_A(2m{+}1\db\!-\tfrac{1}{\tau})=0$ with $m \geq 0$ are compatible with our
claim. Hence, \eqn{DiffEqB} implies by induction in $r$ that  that the $\tilde
b_{k,l}(\underline{n})$ in \eqn{klseries} with $k>0$ can be written as
$(2\pi i)^l$ times a $\ZQ$-linear combinations of MZVs.

In view of the discussion in \appref{app:prf.1}, there is no need to revisit
the constant term of the Laurent polynomial $\tilde b_{0,l}(\underline{n})$
which is annihilated by $\partial_\tau$, so this concludes the proof.


\section{Different flavors of iterated Eisenstein integrals}

\subsection{Another convention for iterated Eisenstein integrals}
\label{app:conv}
In this appendix we make precise how to convert the iterated Eisenstein
integrals $\ce(\underline{k})$ considered in the present work to the
differently normalized iterated integrals $\gamma(\underline{k})$ considered in
refs. \cite{Broedel:2014vla, Broedel:2015hia}, defined for
\begin{equation}
  \GG{k}(\tau):=\suml_{(m,n)\neq(0,0)}\frac{1}{(m+\tau n)^k} = 2\bigg(\zeta_k+\frac{(2\pi i)^k}{(k-1)!} \suml_{m,n=1}^\infty m^{k-1}q^{mn}\bigg)=2\zeta_k+\GG{k}^0(\tau)
\end{equation}
when $k\geq 2$ (even) and $\GG{0}(\tau)=\GG{0}^0(\tau)=-1$ as
\begin{align}
\gamma(k_1,k_2,\ldots,k_r;\tau)  &:= \frac{1}{2\pi i} \int^{i\infty}_{\tau} \dd \tau_r \ \GG{k_r}(\tau_r) \, \gamma(k_1,k_2,\ldots,k_{r-1};\tau_r) 
\notag \\
&\phantom{:}= \frac{1}{4\pi^2} \int^q_0 \dlog q_r \, \GG{k_r}(q_r) \, \gamma(k_1,k_2,\ldots,k_{r-1};q_r) \\
&\phantom{:}=  \frac{1}{(4\pi^2)^r}\! \! \! \! \! \! \!  \int \limits_{0\leq q_1\leq q_2 \leq \ldots \leq q_r\leq q} \! \! \! \! \! \! \! \! \!  \dlog q_1 \, \GG{k_1}(q_1) \ 
 \dlog q_2 \, \GG{k_2}(q_2) \, \ldots \, 
  \dlog q_r \, \GG{k_r}(q_r) \,,\nnl
\gamma_0(k_1,k_2,\ldots,k_r;\tau)  &:=
 \frac{1}{2\pi i} \int^{i\infty}_{\tau} \dd \tau_r \ \GGz{k_r}(\tau_r) \, \gamma_0(k_1,k_2,\ldots,k_{r-1};\tau_r)
\notag \\
&\phantom{:}=  \frac{1}{4\pi^2} \int^q_0 \dd \log q_r \, \GG{k_r}^0(q_r) \, \gamma_0(k_1,k_2,\ldots,k_{r-1};q_r) \\
&\phantom{:}=  \frac{1}{(4\pi^2)^r}\! \! \! \! \! \! \!  \int \limits_{0\leq q_1\leq q_2\leq \ldots \leq q_r\leq q} \! \! \! \! \! \! \! \! \!  \dlog q_1 \, \GG{k_1}^0(q_1) \ 
 \dlog q_2 \, \GG{k_2}^0(q_2) \, \ldots \, 
  \dlog q_r \, \GG{k_r}^0(q_r) \,.
\notag
\end{align}
The conversion reads
\begin{align}
  \gm(k_1,k_2,\ldots,k_r;\tau) &= (2 \pi i)^{k_1+k_2+\cdots+k_r-2r}\ce(k_1,k_2,\ldots,k_r;\tau) \, ,
  \\
  \gmz(k_1,k_2,\ldots,k_r;\tau) &= (2 \pi i)^{k_1+k_2+\cdots+k_r-2r}\cez(k_1,k_2,\ldots,k_r;\tau) \, .
\end{align}
%


\subsection{Conversion between \texorpdfstring{$\cez$}{$\cez$} and \texorpdfstring{$\ce$}{$\ce$}}
\label{app:gamma0}

Recall the generating series
\begin{align}
\ZE_{\underline{k}}(Y_0,Y_1,\ldots ,Y_r;\tau)&:= 
\sum_{p_0,p_1,\ldots ,p_r\geq 0}\frac{1}{(2\pi i)^{2p_0}} \bigg[ \prod_{i=1}^r(2\pi i)^{k_i-2p_i-1} \bigg]  \label{withzero}\\
&\times \ce(0^{p_0},k_1,0^{p_1},\ldots ,k_r,0^{p_r};\tau)Y_0^{p_0}Y_1^{p_1}\cdots Y_r^{p_r} \, . \notag 
\end{align}
Let us introduce a lighter notation for iterated integrals: we denote the
iterated integral of the differential forms $\omega_1(t)\dd t,\ldots ,\omega_r(t)\dd t$,
integrated from $\omega_r$ to $\omega_1$ along a path $\gamma\subset\mathbb{C}$
as $\int_\gamma\omega_1\cdots \omega_r$. For instance, on the straight path
$[0,1]$ we have 
\begin{equation}
\int_{[0,1]}\omega_1\cdots \omega_r=\int_0^1 \omega_1(t_1) \, \dd t_1 \int_0^{t_1}\omega_2(t_2) \, \dd t_2 \cdots \int_0^{t_{r-1}}\omega_r(t_r) \, \dd t_r\, ,
\end{equation}
while on a path $[\tau,i\infty]$ in the upper half plane we have
\begin{equation}
\int_{[\tau,i\infty]}\omega_1\cdots \omega_r=\int_\tau^{i\infty} \omega_r(t_1) \, \dd t_1\int_{t_1}^{i\infty}\omega_{r-1}(t_2) \, \dd t_2 \cdots \int_{t_{r-1}}^{i\infty}\omega_1(t_r) \, \dd t_r \, .
\end{equation}
Let us then rewrite our generating series as
\begin{align}
\ZE_{\underline{k}}(Y_0,Y_1,\ldots ,Y_r;\tau) &= 
\sum_{p_0,p_1,\ldots ,p_r\geq 0}  \frac{ Y_0^{p_0} Y_1^{p_1} Y_2^{p_2}\cdots Y_r^{p_r} }{ (2\pi i)^{p_0+p_1+p_2+\ldots + p_r}} \notag\\
& \ \ \times\Big[\int_{[\tau,i\infty]}\underbrace{\GG{0}\cdots \GG{0}}_{p_0} \GG{k_1}\underbrace{\GG{0}\cdots \GG{0}}_{p_1} \GG{k_2} \cdots \GG{k_r}\underbrace{\GG{0}\cdots \GG{0}}_{p_r}\Big] \label{modular71}  \\
&=
\int_{[\tau,i\infty]}\exp\bigg(\frac{t_1Y_0}{2\pi i}\bigg)\GG{k_1}(t_1)\exp\bigg(\frac{(t_2{-}t_1)Y_1}{2\pi i}\bigg) \GG{k_2}(t_2) \cdots  \notag\\
& \ \ \times \cdots \GG{k_{r-1}}(t_{r-1}) \exp\bigg(\frac{(t_r{-}t_{r-1})Y_{r-1}}{2\pi i}\bigg) \GG{k_r}(t_r)\exp\bigg(\frac{(\tau -t_{r})Y_{r}}{2\pi i}\bigg) \,, \notag 
\end{align}
where in the last step we have used that 
\begin{equation}
\int_{[t_i,t_j]}\underbrace{\GG{0}\GG{0}\cdots \GG{0}}_{p}=\frac{(t_i-t_j)^p}{p!}
\end{equation}
and that (according to the regularization introduced in \cite{Brown:mmv}) 
\begin{equation}
\int_{[t,i\infty]}\underbrace{\GG{0}\cdots \GG{0}}_{p}=(-1)^p\int_{[0,t]}\underbrace{\GG{0}\cdots \GG{0}}_{p}=\frac{t^p}{p!} \, .
\end{equation}
Let us also introduce a modified generating series
\begin{align}
\hat{\ZE}_{\underline{k}}(Y_1,\ldots ,Y_r;\tau):= 
\sum_{p_1,\ldots ,p_r\geq 0} \bigg[ \prod_{i=1}^r(2\pi i)^{k_i-2p_i-1} \bigg] \ce(k_1,0^{p_1},\ldots ,k_r,0^{p_r};\tau)Y_1^{p_1}\cdots Y_r^{p_r} \, ,
\end{align}
where, in comparison to \eqn{withzero}, the iterated Eisenstein integrals
$\ce(0,\ldots)$ with $0$ in the first entry are suppressed.  It is easy to
check repeating for $\hat{\ZE}_{\underline{k}}$ the same steps of
(\ref{modular71}) that
\begin{equation}
\ZE_{\underline{k}}(Y_0,Y_1,\ldots ,Y_r;\tau)=\exp\bigg(\frac{\tau Y_0}{2\pi i}\bigg) \, \hat{\ZE}_{\underline{k}}(Y_1,\ldots ,Y_r;\tau) \, ,
\end{equation}
which leads to the explicit formula
\begin{align}\label{ceToceZero}
&\ce(0^{p_0},k_1,0^{p_1},\ldots ,k_r,0^{p_r};\tau)= \sum_{t+s_1+\cdots +s_r=p_0} (-1)^{s_1+\cdots +s_r}\\
& \ \ \ \ \times
\bigg(\prod_{j=1}^r\binomial{p_j+s_j}{s_j}\bigg)  \frac{ (2\pi i\tau)^t}{t!}\ce(k_1,0^{p_1+s_1},\ldots ,k_r,0^{p_r+s_r};\tau)  \, .\notag
\end{align}
Iterated Eisenstein integrals of the kind $\ce(k_1,0^{p_1},\ldots
,k_r,0^{p_r};\tau)$ can be written as the sum of $\cez(k_1,0^{p_1},\ldots
,k_r,0^{p_r};\tau)$ and other iterated Eisenstein integrals $\ce$'s of strictly
lower depth\footnote{Recall that this was defined as the number of non-zero
entries of $\ce$.}, therefore one can iteratively make use of \eqn{ceToceZero}
and write any $\ce$ in terms of the $\cez$'s. For instance, one easily gets
\begin{equation}
\ce(k_1,0^{p_1};\tau)=\cez(k_1,0^{p_1};\tau)-\frac{2\zeta_{k_1}}{(2\pi i)^{k_1}}\cez(0^{p_1+1};\tau)
\end{equation}
and
\begin{align}
&\ce(k_1,0^{p_1},k_2,0^{p_2};\tau)=\cez(k_1,0^{p_1},k_2,0^{p_2};\tau)-\frac{2\zeta_{k_2}}{(2\pi i)^{k_2}}\cez(k_1,0^{p_1+p_2+1};\tau)  \\
& \ \ \ -\frac{2\zeta_{k_1}}{(2\pi i)^{k_1}}\sum_{s+t=p_1+1}\binomial{p_2+s}{s}\frac{(-1)^s(2\pi i)^t}{t!}\Big(\cez(k_2,0^{p_2+s};\tau)-\frac{2\zeta_{k_2}}{(2\pi i)^{k_2}}\cez(0^{p_2+s+1};\tau)\Big) \, .\notag
\end{align}
The motivation to do this conversion comes from the fact that we know
explicitly\footnote{This is a consequence of \eqn{qgamma1} and the fact that
our regularization of iterated Eisenstein integrals yields
$\cez(0^p;\tau)=\ce(0^p;\tau)=(2\pi i\tau)^p/p!$.} the \mbox{$q$-expansion} of any
$\cez$, which therefore allowed us to exploit the very fast convergence of
these series and verify numerically our results and conjectures to arbitrary
precision.


\subsection{Conversion between \texorpdfstring{$\ce$}{$\ce$} and \texorpdfstring{$\CG$}{$\CG$}}
\label{app:GvsE}

The relations \eqn{modular42} between the two generating series \eqn{modular41}
of iterated Eisenstein integrals $\ce$ and $\CG$ can be proven by writing, as
in the previous section,
\begin{align}
\ZE_{\underline{k}}(Y_0,Y_1,\ldots ,Y_r;\tau) &= \int_{[\tau,i\infty]}\exp\bigg(\frac{t_1Y_0}{2\pi i}\bigg)\GG{k_1}(t_1)\exp\bigg(\frac{(t_2{-}t_1)Y_1}{2\pi i}\bigg) \GG{k_2}(t_2) \cdots   \\
& \ \ \times \cdots \GG{k_{r-1}}(t_{r-1}) \exp\bigg(\frac{(t_r{-}t_{r-1})Y_{r-1}}{2\pi i}\bigg) \GG{k_r}(t_r)\exp\bigg(\frac{(\tau -t_{r})Y_{r}}{2\pi i}\bigg)  \notag 
\end{align}
as well as writing the generating series $\mathbb{G}_{\underline{k}}$ in \eqn{modular41} as
\begin{equation}
\mathbb{G}_{\underline{k}}(T_1,\ldots ,T_r;\tau)= \int_{[\tau,i\infty]}\exp\bigg(\frac{t_1T_1}{2\pi i}\bigg)\GG{k_1}(t_1)
\exp\bigg(\frac{t_2T_2}{2\pi i}\bigg)\GG{k_2}(t_2)
\cdots\exp\bigg(\frac{t_rT_r}{2\pi i}\bigg) \GG{k_r}(t_r) \, 
\end{equation}
by a completely similar computation.


\subsection{Examples of modular transformations}
\label{moremod}

We give here one more example in depth two of a special linear combination of
iterated Eisenstein integrals which can be $S$-transformed with our methods:
\begin{align}
&3\ce(4,6,0,0,0;-\tfrac{1}{\tau})+\ce(6,0,4,0,0;-\tfrac{1}{\tau})+3\ce(6,4,0,0,0;-\tfrac{1}{\tau}) = \frac{\zeta_4}{240}\ce(4;\tau)-\frac{\zeta_3}{6}\ce(0,6;\tau)\notag \\
&+\ce(0,0,4,0,6;\tau)+3\ce(0,0,0,4,6;\tau)+3\ce(0,0,0,6,4;\tau)-\frac{1}{T}\Big(-\frac{\zeta_5}{20}\ce(4;\tau)+\frac{\zeta_4}{240}\ce(0,4;\tau)\notag \\
&-\frac{\zeta_3}{2}\ce(0,0,6;\tau)+3\ce(0,0,4,0,0,6;\tau)+9\ce(0,0,0,4,0,6;\tau)+18\ce(0,0,0,0,4,6;\tau)\notag \\
&+3\ce(0,0,0,6,4;\tau)+18\ce(0,0,0,0,6,4;\tau)+ \frac{143 \zeta_6 }{241920} \Big)-\frac{1}{T^2}\Big(\frac{\zeta_5}{20}\ce(0,4;\tau)\notag \\
&-\frac{\zeta_4}{480}\ce(0,0,4;\tau)+\frac{3\zeta_3}{4}\ce(0,0,0,6;\tau)-\frac{9}{2}\ce(0,0,4,0,0,0,6;\tau)-\frac{27}{2}\ce(0,0,0,4,0,0,6;\tau)\notag \\
&-27\ce(0,0,0,0,4,0,6;\tau)-45\ce(0,0,0,0,0,4,6;\tau)-\frac{3}{2}\ce(0,0,0,6,0,0,4;\tau)\notag \\
&-12\ce(0,0,0,0,6,0,4;\tau)-45\ce(0,0,0,0,0,6,4;\tau)-\frac{13\zeta_4\zeta_3}{20160}+\frac{7\zeta_7}{1920}\Big)\notag \\
&+\frac{1}{T^3}\Big(\frac{\zeta_5}{40}\ce(0,0,4;\tau)+\frac{\zeta_3}{2}\ce(0,0,0,0,6;\tau)
-3\ce(0,0,4;\tau)\ce(0,0,0,0,6;\tau)
-\frac{\zeta_3\zeta_5}{240}\Big)  \ .
 \label{modular98}
\end{align}
Moreover, the following iterated Eisenstein integral of depth three gives rise to the depth-three MZV
$\zeta_{3,5,3}$ upon modular $S$-transformation:
\begin{align}
&\ce(6, 4, 4;-\tfrac{1}{\tau}) = \frac{1}{8(2\pi i)^{8}}\bigg[
\frac{533  \zeta_3 \zeta_8}{4050}
-\frac{4\zeta_3\zeta_{3,5}}{225}-\frac{4\zeta_5\zeta_3^2}{45}-\frac{4\zeta_{3,5,3}}{225} -\frac{221\zeta_{11}}{5400}\notag \\
&+\bigg(\frac{16\zeta_5\zeta_3}{15}
-\frac{ 503 \zeta_8}{675}
+\frac{16\zeta_{3,5}}{75}\bigg)\ce(0^2,4;\tau)
-\frac{16 \zeta_5}{5} \big[ \ce(0^2,4;\tau) \,\big]^2
\notag \\
&+80640\ce(0^7,6,4,0,4;\tau)+69120\ce(0^6,6,0^2,4,4;\tau)+23040\ce(0^5,6,0^3,4,4;\tau)\notag \\
&+768\ce(0^4,6,0^2,4,0^2,4;\tau)+161280\ce(0^7,6,0,4,4;\tau)+11520\ce(0^6,6,4,0^2,4;\tau)\notag \\
&+34560\ce(0^6,6,0,4,0,4;\tau)+11520\ce(0^5,6,0^2,4,0,4;\tau)+2304\ce(0^4,6,0^3,4,0,4;\tau)\notag \\
&+4608\ce(0^4,6,0^4,4,4;\tau)+322560\ce(0^8,6,4,4;\tau)+3840\ce(0^5,6,0,4,0^2,4;\tau)\bigg] \ .
 \label{modular98a}
\end{align}


\section{\texorpdfstring{$A$}{$A$}-cycle graph functions at weight five}
\label{app:higher5}

Among the six $A$-cycle graph functions at weight five, two examples have
been spelt out in terms of eMZVs in \eqns{w5D}{w5F}, and the remaining
four are given by
\begin{align}
\Acyc{G5}  &=   8 \zeta_5  + \frac{8}{5} \omm_A(0, 5) + 32 \omm_A(0, 0, 0, 5) - 20 \omm_A(0, 0, 2, 3) + 
 \frac{10}{3} \omm_A(0, 0, 2) \omm_A(0, 0, 3, 0) \notag \\
 & + 120 \omm_A(0, 3) \omm_A(0, 0, 0, 0, 2) - 312 \omm_A(0, 0, 0, 0, 0, 5) +
 120 \omm_A(0, 0, 0, 0, 1, 4)   \notag \\
 &+ 5  \zeta_3\omm_A(0, 0, 2)  - 12 \zeta_2 \omm_A(0, 3)  -
 80  \zeta_2  \omm_A(0, 0, 2) \omm_A(0, 0, 1, 0)+ 
 \frac{25}{9} \zeta_2  \omm_A(0, 0, 3, 0)  \notag \\
 &+ 480 \zeta_2 \omm_A(0, 0, 0, 0, 0, 3)  - 
 960   \zeta_2\omm_A(0, 0, 0, 0, 1, 2)  + \frac{265}{6} \zeta_2 \zeta_3  \notag \\
 &+ \frac{412}{3} \zeta_4 \omm_A(0, 0, 1, 0) +  192  \zeta_4  \omm_A(0, 0, 0, 1, 0, 0)
 \label{w5A}\\
\Acyc{G311} &= \frac{
 7 \zeta_5}{80}+ \frac{87}{400} \omm_A(0, 5) + \frac{77}{20} \omm_A(0, 0, 0, 5) - 
 \frac{5}{2} \omm_A(0, 0, 2, 3) + \omm_A(0, 0, 2) \omm_A(0, 0, 3, 0) \notag \\
 & + 15 \omm_A(0, 3) \omm_A(0, 0, 0, 0, 2) - \frac{381}{10} \omm_A(0, 0, 0, 0, 0, 5) + 15 \omm_A(0, 0, 0, 0, 1, 4)    \notag \\
 & -  \frac{7}{10} \zeta_2 \omm_A(0, 3)  - 
 6  \zeta_2\omm_A(0, 0, 2) \omm_A(0, 0, 1, 0)  + 
 \frac{3}{2} \zeta_2  \omm_A(0, 0, 3, 0) + 
 36  \zeta_2\omm_A(0, 0, 0, 0, 0, 3)   \notag \\
 & - 48 \zeta_2  \omm_A(0, 0, 0, 0, 1, 2)  + 2 \zeta_2 \zeta_3  + 
 \frac{47}{10} \zeta_4 \omm_A(0, 0, 1, 0)  +
 \frac{48}{5}  \zeta_4\omm_A(0, 0, 0, 1, 0, 0)  \label{w5B}\\
\Acyc{G221}  &=  \frac{ 3 \zeta_5}{20}+ \frac{3}{100} \omm_A(0, 5) + \frac{19}{15} \omm_A(0, 0, 0, 5) - 
 \frac{2}{3} \omm_A(0, 0, 2, 3) + 4 \omm_A(0, 3) \omm_A(0, 0, 0, 0, 2) \notag \\
& -  \frac{58}{5} \omm_A(0, 0, 0, 0, 0, 5) + 4 \omm_A(0, 0, 0, 0, 1, 4)  - 
 \frac{1}{3}  \zeta_2 \omm_A(0, 3) + \frac{14}{9} \zeta_2  \omm_A(0, 0, 3, 0) \notag \\
 & + 
 32 \zeta_2 \omm_A(0, 0, 0, 0, 0, 3)  -
 8 \zeta_2 \omm_A(0, 0, 0, 0, 1, 2)  +\frac{1}{3} \zeta_2 \zeta_3\notag \\
 &  -  \frac{4}{5} \zeta_4 \omm_A(0, 0, 1, 0)  -
 \frac{352}{5} \zeta_4   \omm_A(0, 0, 0, 1, 0, 0) 
 \label{w5C} \\
\Acyc{GC221}&=  \frac{\zeta_5}{60} - \frac{7}{900} \omm_A({0, 5}) + \frac{1}{15} \omm_A({0, 0, 0, 5}) - 
 \frac{2}{5} \omm_A({0, 0, 0, 0, 0, 5}) + \frac{1}{10} \zeta_2 \omm_A({0, 3})   \notag \\ 
 & + \frac{1}{3}  \zeta_2  \omm_A({0, 0, 3, 0})+ 
 12  \zeta_2 \omm_A({0, 0, 0, 0, 1, 2}) - \frac{1}{2} \zeta_2 \zeta_3 - \frac{103}{15}  \zeta_4 \omm_A({0, 0, 1, 0}) \notag \\
 & + \frac{132}{5} \zeta_4 \omm_A({0, 0, 0, 1, 0, 0})   \,.
  \label{w5E}
 \end{align}


\section{Relations between modular graph functions at weight six}
\label{app:w6rel}

In this appendix, we collect the complete set of relations among modular graph
functions of weight six as given in \cite{DHoker:2016quv}:
%
%
\begin{align}
0 &= \Dcyc{G6} - 15 \Dcyc{G2} \Dcyc{G4} + 30 \Dcyc{G2}^3 - 10 \Dcyc{G3}^2 - 60 \Dcyc{G411} + 720 \Dcyc{GC2211}    \notag \\
  &\quad+ 240 \Dcyc{G111} \Dcyc{G3}-720 \Dcyc{G2} \Dcyc{G1111} -1440 \Dcyc{G111}^2 - 5280 \Dcyc{GC321} \nnl 
  &\quad + 360 \Dcyc{G2} \Dcyc{G211} - 1280 \Dcyc{GC222} + 3380 \Dcyc{G111111} \notag \\
0 &= 2 \Dcyc{GC3111} + 3 \Dcyc{GC2211} - 9 \Dcyc{G2} \Dcyc{G1111} - 6 \Dcyc{G111}^2 - 18 \Dcyc{GC411} \notag \\ 
  &\quad- 24 \Dcyc{GC321} - 2\Dcyc{GC222}  + 32 \Dcyc{G111111} \notag  \\
0 &= - 3 \Dcyc{G411} + 109 \Dcyc{GC222} + 408 \Dcyc{GC321} + 36 \Dcyc{GC411} + 18 \Dcyc{G2} \Dcyc{G211} \nnl 
  & \quad+ 12 \Dcyc{G111} \Dcyc{G3} - 211 \Dcyc{G111111} \notag  \\
0 &= 3 \Dcyc{G222} - 18 \Dcyc{GC2211} - 58 \Dcyc{GC222} - 192 \Dcyc{GC321} - 3 \Dcyc{G2}^3 + 24 \Dcyc{G111}^2 \nnl
  &\quad+ 18 \Dcyc{G2} \Dcyc{G1111} + 46 \Dcyc{G111111} \label{eqn:weightsixrelations}  \\
0 &= 2 \Dcyc{G321} + 18 \Dcyc{GC2211} - 36 \Dcyc{GC411} - 69 \Dcyc{GC222} - 288 \Dcyc{GC321} - 6 \Dcyc{G2} \Dcyc{G211}  \notag \\
&- 18 \Dcyc{G2} \Dcyc{G1111} - 36 \Dcyc{G111}^2 + 183 \Dcyc{G111111} \notag  \\
0 &= 3 \Dcyc{G2211} + 6 \Dcyc{GC2211} - 10 \Dcyc{GC222} - 48 \Dcyc{GC321} - 12 \Dcyc{GC411} \nnl
  &\quad - 6 \Dcyc{G2} \Dcyc{G1111} - 12 \Dcyc{G111}^2 + 40 \Dcyc{G111111} \notag  \\
0 &= 18 \Dcyc{Gp2111} - 9 \Dcyc{GC2211}  - 20 \Dcyc{GC222} - 60 \Dcyc{GC321} + 9 \Dcyc{G2} \Dcyc{G1111} \nnl
  &\quad + 18 \Dcyc{G111}^2 - 10 \Dcyc{G111111} \notag  \\
0 &= 3 \Dcyc{Gx1111} - \Dcyc{GC222} - 12 \Dcyc{GC321} + 4 \Dcyc{G111111} \,. 
\notag
\end{align}


\section{Explicit modular graph forms and modular graph functions}

In this appendix we gather explicit representations of all the modular graph
forms and modular graph functions which appear in the Cauchy--Riemann equations
up to weight six and have not been spelt out in the main text.


\subsection{Cauchy Riemann derivatives}
\label{app:service}

In order to supplement the discussion of the Cauchy--Riemann equations in
\secref{sec:service}, all the modular graph forms on their right-hand side will
be spelt out in this subsection. Starting from the expression \eqn{nheis} for
non-holomorphic Eisenstein series, repeated action of the Cauchy--Riemann
derivative \eqn{CR1} gives rise to
\begin{align}
\pi \nabla \EE{2} &=  \frac{ 2 y^3}{45} -
\zeta_3 + 24 y^2 \cez(4) + 12 y \cez(4, 0) + 6 \Re[ \cez(4, 0, 0) ] 
\label{service1} \\
\pi \nabla \EE{3} &= \frac{2 y^4}{315} - \frac{3 \zeta_5}{2 y} + 240 y^2 \cez(6, 0) + 360 y \cez(6, 0, 0) + 
 180 \cez(6, 0, 0, 0)   \label{service2}  \\
 &+ 180 \Re[ \cez(6, 0, 0, 0) ] + 
 \frac{180 \Re[ \cez(6, 0, 0, 0, 0) ]}{y}\notag \\
(\pi \nabla)^2 \EE{3} &= \frac{8 y^5}{315} + \frac{3 \zeta_5}{2} - 960 y^4 \cez(6) - 960 y^3 \cez(6, 0) - 
 720 y^2 \cez(6, 0, 0) \label{service3} \\
 &- 360 y \cez(6, 0, 0, 0) - 
 180 \Re[ \cez(6, 0, 0, 0, 0) ]\notag  \\
\pi \nabla \EE{4} &= \frac{4 y^5}{4725}- \frac{15 \zeta_7}{8 y^2} + 3360 y^2 \cez(8, 0^2) + 10080 y \cez(8, 0^3) + 
 12600 \cez(8, 0^4)  \label{service4}  \\
 &+ \frac{6300 \cez(8, 0^5)}{y} + 
 5040 \Re[ \cez(8, 0^4) ] + \frac{12600 \Re[ \cez(8, 0^5) ]}{y} + 
 \frac{9450 \Re[ \cez(8, 0^6) ]}{y^2} \notag \\
(\pi \nabla)^2 \EE{4} &= \frac{4 y^6}{945} + \frac{15 \zeta_7}{4 y} - 13440 y^4 \cez(8, 0) - 33600 y^3 \cez(8, 0^2) - 
 50400 y^2 \cez(8, 0^3)  \label{service5}  \\
 &- 50400 y \cez(8, 0^4) - 
 25200 \cez(8, 0^5) - 12600 \Re[ \cez(8, 0^5) ] - 
 \frac{18900 \Re[ \cez(8, 0^6 )]}{y}  \notag \\
(\pi \nabla)^3 \EE{4} &= \frac{8 y^7}{315} - \frac{15 \zeta_7}{4} + 53760 y^6 \cez(8) + 80640 y^5 \cez(8, 0) + 
 100800 y^4 \cez(8, 0^2)  \label{service6}  \\
 &+ 100800 y^3 \cez(8, 0^3) + 
 75600 y^2 \cez(8, 0^4) + 37800 y \cez(8, 0^5) + 
 18900 \Re[ \cez(8, 0^6) ]  \,. \notag
\end{align}
At depth two, the Cauchy--Riemann derivative of the modular graph function
$\EE{2,2}$ in \eqn{newCR8} is given by
\begin{align}
\pi \nabla \EE{2,2} &=
- \frac{2 y^5}{10125} +  \frac{y^2 \zeta_3 }{45} - \frac{5 \zeta_5}{12}  + \frac{\zeta_3^2}{2 y}
+\Big( \frac{ 4  y^3}{15}   - 6   \zeta_3 \Big)\cez(4, 0) -  \Big( \frac{2y^2 }{15}    + \frac{ 6 \zeta_3}{y} \Big) \Re[ \cez(4, 0, 0)]  \notag \\
&  +  \frac{ 2 y^2}{5}  \cez(4, 0, 0) 
+ 36 y \cez(4, 0)^2 + 36 \cez(4, 0) \Re[ \cez(4, 0, 0)] + \frac{18 \Re[ \cez(4, 0, 0)]^2}{y}  \label{service7}  \\
& + 144 y^2 \cez(4, 4, 0) + 72 y \big( \cez(4, 4, 0, 0) + \tfrac{1}{360} \cez(4, 0, 0, 0)  \big) \notag \\
&  + 36 \Re[ \cez(4, 0, 4, 0, 0) + 3 \cez(4, 4, 0, 0, 0) + \tfrac{1}{360} \cez(4, 0, 0, 0, 0) ] \,,\notag
 \end{align}
which enters on the right-hand side of the differential equation (\ref{newCR7}) for $\EE{2,2,2}$.


\subsection{Modular graph functions at weight six}
\label{app:explicitmod}

Using the method in \secref{intproc}, the Cauchy--Riemann equations
(\ref{newCR4}) to (\ref{newCR6}) give rise to the following expressions for
$\text{E}'_{3,3}$ and $\EE{2,4}$, respectively.
\begin{align}
\text{E}'_{3,3} &=  -\frac{y^6}{18753525} +\frac{ y \zeta_5}{630}   + \frac{3 \zeta_7}{160 y} - \frac{ 7 \zeta_9}{480 y^3}
 - \Big( \frac{4y^2}{105}   - \frac{ 9   \zeta_5}{y^3}  \Big)\Re[ \cez(6, 0, 0, 0) ] 
 \notag \\
& - \frac{ 540 \Re[ \cez(6, 0, 0, 0) ] ^2}{y^2}  
 - \frac{ 1080 \Re[ \cez(6, 0, 0, 0) ]  \Re[ \cez(6, 0, 0, 0, 0) ] }{y^3} 
 \notag \\
 & - \frac{4y}{21}  \Re[ \cez(6, 0, 0, 0, 0) ] - 1440 \Re[ \cez(6, 0, 6, 0, 0, 0) ] - 
\frac{ 11}{21} \Re[ \cez(6, 0, 0, 0, 0, 0) ]  \notag \\
& 
- \frac{2160 \Re[ \cez(6, 0, 0, 6, 0, 0, 0) ] }{y} 
- \frac{ 4320 \Re[ \cez(6, 0, 6, 0, 0, 0, 0) ] }{y}  \notag \\ 
& - \frac{13 \Re[ \cez(6, 0, 0, 0, 0, 0, 0) ] }{14 y}  
  - \frac{1080 \Re[ \cez(6, 0, 0, 0, 6, 0, 0, 0) ] }{y^2}  \label{newCR11} \\
& - \frac{ 2160 \Re[ \cez(6, 0, 0, 6, 0, 0, 0, 0) ] }{y^2} 
+ \frac{ 10800 \Re[ \cez(6, 6, 0, 0, 0, 0, 0, 0) ] }{y^2} \notag \\
& - \frac{ \Re[ \cez(6, 0, 0, 0, 0, 0, 0, 0) ] }{y^2}
 + \frac{ 1080 \Re[ \cez(6, 0, 0, 0, 6, 0, 0, 0, 0) ] }{y^3}  \notag \\
&+ \frac{5400 \Re[ \cez(6, 0, 0, 6, 0, 0, 0, 0, 0) ] }{y^3} + 
\frac{ 16200 \Re[ \cez(6, 0, 6, 0, 0, 0, 0, 0, 0) ] }{y^3} \notag \\
 & + \frac{37800 \Re[ \cez(6, 6, 0, 0, 0, 0, 0, 0, 0) ] }{y^3}
  - \frac{ \Re[ \cez(6, 0, 0, 0, 0, 0, 0, 0, 0) ] }{2 y^3} \notag
\end{align}
\begin{align}
\EE{2,4} &=  - \frac{ y^6}{70875} +   \frac{  y^3 \zeta_3 }{525} +  \frac{ 3 \zeta_7 }{ 40 y} 
 + \frac{ 25 \zeta_9}{8 y^3} - \frac{  135 \zeta_3 \zeta_7}{32 y^4}
  - \Big( \frac{2y^3}{175} - \frac{405   \zeta_7}{16 y^4} \Big)  \Re[ \cez(4, 0, 0) ]   \notag\\
&- \Big(  504 y-  \frac{11340  \zeta_3}{y^2} \Big)  \Re[ \cez(8, 0, 0, 0, 0) ]  
- \Big( 2520 - \frac{ 28350  \zeta_3}{y^3} \Big) \Re[ \cez(8, 0, 0, 0, 0, 0) ]   \notag \\
& - \Big( \frac{ 5670 }{y} - \frac{42525   \zeta_3}{2 y^4}  \Big) \Re[ \cez(8, 0, 0, 0, 0, 0, 0) ] 
 - \frac{68040 \Re[ \cez(4, 0, 0) ]  \Re[ \cez(8, 0, 0, 0, 0) ] }{y^2} \notag \\
 & - \frac{170100 \Re[ \cez(4, 0, 0) ]  \Re[ \cez(8, 0, 0, 0, 0, 0) ] }{y^3}
 - \frac{ 127575 \Re[ \cez(4, 0, 0) ]  \Re[ \cez(8, 0, 0, 0, 0, 0, 0) ] }{y^4} \notag \\
&  -  45360 \Re[ \cez(8, 0, 0, 4, 0, 0) ]  - 136080 \Re[ \cez(8, 0, 4, 0, 0, 0) ]  -  272160 \Re[ \cez(8, 4, 0, 0, 0, 0) ]  
 \notag \\
 & - 272160 \Re[ \cez(4, 8, 0, 0, 0, 0) ] -  \frac{3}{20} \Re[ \cez(4, 0, 0, 0, 0, 0) ]    - \frac{136080 \Re[ \cez(4, 0, 8, 0^4) ] }{y} 
\notag \\
&  - \frac{1360800 \Re[ \cez(4, 8, 0^5) ] }{y}  - \frac{9 \Re[ \cez(4, 0^6) ] }{ 20 y}   - \frac{136080 \Re[ \cez(8, 0^3, 4, 0^2) ]}{y}   
 \label{newCR12} \\
&  - \frac{ 408240 \Re[ \cez(8, 0^2, 4, 0^3) ] }{y}  - \frac{ 816480 \Re[ \cez(8, 0, 4, 0^4) ] }{y} - \frac{ 1360800 \Re[ \cez(8, 4, 0^5) ] }{y} 
 \notag \\
 &  - \frac{ 340200 \Re[ \cez(4, 0, 8, 0^5) ] }{y^2} - \frac{ 2551500 \Re[ \cez(4, 8, 0^6) ] }{y^2}
  - \frac{ 9 \Re[ \cez(4, 0^7) ] }{16 y^2} \notag \\
& - \frac{170100 \Re[ \cez(8, 0^4, 4, 0^2) ] }{y^2} 
- \frac{ 510300 \Re[ \cez(8, 0^3, 4, 0^3) ] }{y^2}
 - \frac{ 1020600 \Re[ \cez(8, 0^2, 4, 0^4) ] }{y^2} \notag \\
 & - \frac{1701000 \Re[ \cez(8, 0, 4, 0^5) ] }{y^2}
  - \frac{ 2551500 \Re[ \cez(8, 4, 0^6) ] }{y^2}
   - \frac{ 6615 \Re[ \cez(8, 0^7) ] }{y^2} - \frac{ 9 \Re[ \cez(4, 0^8) ] }{32 y^3}\notag \\
&- \frac{ 255150 \Re[ \cez(4, 0, 8, 0^6) ] }{y^3}
 - \frac{ 1786050 \Re[ \cez(4, 8, 0^7) ] }{y^3}
   - \frac{ 85050 \Re[ \cez(8, 0^5, 4, 0^2) ] }{y^3}   \notag \\
& - \frac{ 255150 \Re[ \cez(8, 0^4, 4, 0^3) ] }{y^3}
  - \frac{ 510300 \Re[ \cez(8, 0^3, 4, 0^4) ] }{y^3}
   - \frac{ 850500 \Re[ \cez(8, 0^2, 4, 0^5) ] }{y^3} \notag \\
 &
  - \frac{ 1275750 \Re[ \cez(8, 0, 4, 0^6) ] }{y^3}
   - \frac{ 1786050 \Re[ \cez(8, 4, 0^7) ] }{y^3}
     - \frac{ 6615 \Re[ \cez(8, 0^8) ] }{2 y^3} \notag
\end{align}
The contributing expressions for $\nabla \text{E}_3$ and $\nabla^{j=1,2,3} \text{E}_4$
can be found in the previous subsection.

\bibliographystyle{sv}
\bibliography{sv}

\end{document}